\documentclass[prd,twocolumn,showpacs,preprintnumbers,floatfix,superscriptaddress,10pt]{revtex4}

\usepackage[dvips]{color}
\usepackage[normalem]{ulem}

\usepackage[mathscr]{eucal}
\usepackage{amssymb}
\usepackage{amsmath}

\usepackage{graphicx}
\usepackage[dvips]{color}
\usepackage{epsf}
\usepackage{bm}

\newcommand{\be}{\begin{equation}}

\newcommand{\ee}{\end{equation}}
\newcommand{\bea}{\begin{eqnarray}}
\newcommand{\eea}{\end{eqnarray}}

\newcommand{\de}{\partial}

\newcommand{\De}{\Delta}

\newcommand{\fourier}[1]{\int \frac{d^4#1}{(2\pi)^4}}
\newcommand{\intspace}[1]{\int d^4 {#1}\,}
\newcommand{\ha}{\frac{1}{2}}
\newcommand{\rr}{{\bf r}}
\newcommand{\q}[2]{ {\bf q}_{#1}^{#2}}
\newcommand{\setq}[2]{\{{\bf q}_{#1}^{#2}\}}
\newcommand{\coleps}{\epsilon_{I\alpha\beta}}
\newcommand{\flaeps}{\epsilon_{Iij}}
\newcommand{\cross}[1]{#1\!\!\!/}
\newcommand{\cf}{\alpha i,\beta j}
\newcommand{\bp}{\bar{\psi}}
\newcommand{\bF}{\bar{F}}
\newcommand{\bG}{\bar{G}^{(0)}}
\newcommand{\bD}{\bar{\Delta}}
\newcommand{\bbta}{\bar{\beta}}
\newcommand{\bgma}{\bar{\gamma}}

\newcommand{\x}[1]{\frac{M_s^2}{{#1}\mu}}

\newcommand{\dmu}{\delta\mu}
\newcommand{\ssq}{\sin^2 \frac{\phi}{2} }

\begin{document}

\title{The crystallography of three-flavor quark matter}
\date{\today}
\author{Krishna~Rajagopal}
\email{krishna@lns.mit.edu}
\affiliation{Center for Theoretical Physics, Massachusetts Institute
of Technology, Cambridge, MA 02139}
\affiliation{Nuclear Science Division, MS 70R319,
Lawrence Berkeley National Laboratory, Berkeley, CA 94720}
\author{Rishi~Sharma}
\email{sharma@mit.edu}
\affiliation{Center for Theoretical Physics, Massachusetts Institute
of Technology, Cambridge, MA 02139}
\preprint{MIT-CTP-3749}

\begin{abstract}
We analyze and compare candidate crystal structures for the
 crystalline color superconducting phase that may arise in cold,
 dense but not asymptotically dense, three-flavor quark matter.
 We determine the gap parameter $\Delta$ and free energy $\Omega(\Delta)$
 for many
 possible crystal structures within a Ginzburg-Landau approximation,
evaluating $\Omega(\Delta)$  to order $\Delta^6$.
In contrast to the two-flavor case, we find a positive $\Delta^6$ term
and hence an $\Omega(\Delta)$ that is bounded from below for all the
structures that we analyze.  This means that we are able to evaluate $\Delta$
and $\Omega$ as a function of the splitting between Fermi surfaces
for all the structures we consider.  We find two structures with particularly
robust values of $\Delta$ and the condensation energy, within a factor of 
two of those for the CFL phase which is known to characterize 
QCD at asymptotically large densities.  The robustness of these phases
results in their being favored over wide ranges of density.  However, it
also implies that the Ginzburg-Landau approximation is not quantitatively
reliable.  
We develop qualitative
insights into what makes a crystal structure favorable, and use these to winnow
the possibilities.  The two structures that
we find to be most favorable are both built from condensates with face-centered
cubic symmetry: in one case, the $\langle ud \rangle$ and $\langle us \rangle$
condensates are separately face centered cubic; in the other case $\langle ud\rangle$
and $\langle us\rangle$
combined make up a face centered cube.
\end{abstract}
\maketitle

\section{Introduction}

Quantum chromodynamics predicts that
at densities that are high enough that baryons are crushed into
quark matter, the quark matter that results features pairing
between quarks at low enough
temperatures, meaning that it is in one of a family of
possible color superconducting phases~\cite{reviews}.
The essence of color superconductivity is quark pairing driven by
the BCS mechanism, which operates whenever there are attractive
interactions between fermions at a Fermi surface~\cite{BCS}.
The interaction between quarks in QCD is strong and is attractive
between quarks that are antisymmetric in color, so we expect
cold dense quark matter to exhibit color superconductivity.
If color superconducting quark matter occurs in nature, it
lies within compact stars.   
Except during the first few seconds after their birth in
supernovae, these stars have temperatures
well below the tens of MeV. This implies that if these stars
feature quark matter cores,  these cores will be color superconductors,
and justifies us in restricting our investigation to $T=0$ throughout
this paper.  

We shall only consider Cooper pairs whose
pair wave function is antisymmetric in Dirac indices --- the relativistic
generalization of zero total spin. (Other
possibilities have been
investigated~\cite{reviews,Iwasaki:1994ij,Alford:1997zt,Alford:1998mk,Alford:2002kj}
and found to be less favorable.)
This in turn requires antisymmetry  in flavor, meaning in particular
that the two quarks in a Cooper pair must have different flavor.

It is by now well-established that at sufficiently high densities,
where the up, down and strange quarks can be treated
on an equal footing and the disruptive effects of the
strange quark mass can be neglected, quark matter
is in the color-flavor locked (CFL) phase, in which
quarks of all three colors and all three flavors form
conventional Cooper pairs with zero total momentum,
and all fermionic
excitations are gapped, with the gap parameter 
$\Delta_0\sim 10-100$~MeV~\cite{Alford:1998mk,reviews}. However, even at the very center
of a compact star the quark number chemical potential
$\mu$ cannot be much larger than 500 MeV, meaning
that the strange quark mass $M_s$ (which is density dependent, lying
somewhere between its vacuum current mass of about 100 MeV and constituent
mass of about 500 MeV) cannot be neglected.
Furthermore, bulk matter, as relevant for a compact star, must be in weak equilibrium
and must be electrically and color
neutral~\cite{Iida:2000ha,Amore:2001uf,Alford:2002kj,Steiner:2002gx,Huang:2002zd}.
All these factors work to separate the Fermi momenta of the three different
flavors of quarks, and thus disfavor the cross-species BCS pairing
that characterizes the CFL phase.
If we imagine beginning at asymptotically high densities and reducing
the density, and suppose that CFL pairing is disrupted by the heaviness
of the strange quark before color superconducting quark matter is superseded
by baryonic matter, the CFL phase must be replaced by some
phase of quark matter in which there is less, and less symmetric, pairing.

Within a spatially homogeneous ansatz, the next phase down in
density is the gapless CFL (gCFL)
phase~\cite{Alford:2003fq,Alford:2004hz,Alford:2004nf,Ruster:2004eg,Fukushima:2004zq,Alford:2004zr,Abuki:2004zk,Ruster:2005jc}.
In this phase, quarks of all three colors and all three flavors
still form ordinary Cooper pairs, with each pair having zero total
momentum, but there are regions of momentum space in which certain
quarks do not succeed in pairing, and these regions are bounded by
momenta at which certain fermionic quasiparticles are gapless. This
variation on BCS pairing --- in which the same species of fermions
that pair feature gapless quasiparticles --- was first proposed for
two flavor quark matter~\cite{Shovkovy:2003uu} and in an atomic
physics context~\cite{Gubankova:2003uj}.  In all these contexts,
however, the gapless paired state turns out in general to suffer
from a ``magnetic instability'': it can lower its energy by the
formation of counter-propagating
currents~\cite{Huang:2004bg,Casalbuoni:2004tb,Fukushima:2006su}. In the atomic
physics context, the resolution of the instability is phase
separation, into macroscopic regions of two phases in one of which
standard BCS pairing occurs and in the other of which no pairing
occurs~\cite{Bedaque:2003hi,KetterleImbalancedSpin,HuletPhaseSeparation}.
In three-flavor quark matter, where the instability of the gCFL
phase has been established in Refs.~\cite{Casalbuoni:2004tb}, phase
coexistence would require coexisting components with opposite color
charges, in addition to opposite electric charges, making it very
unlikely that a phase separated solution can have lower energy than
the gCFL phase~\cite{Alford:2004hz,Alford:2004nf}. Furthermore,
color superconducting phases which are less symmetric than the CFL
phase but still involve only conventional BCS pairing, for example
the much-studied 2SC phase in which only two colors of up and down
quarks pair~\cite{Bailin:1983bm,Alford:1997zt,Rapp:1997zu} but
including also many other possibilities~\cite{Rajagopal:2005dg},
cannot be the resolution of the gCFL
instability~\cite{Alford:2002kj,Rajagopal:2005dg}. It seems likely,
therefore, that a ground state with counter-propagating currents is
required.  This could take the form of a crystalline color
superconductor~\cite{Alford:2000ze,Bowers:2001ip,Casalbuoni:2001gt,Leibovich:2001xr,Kundu:2001tt,Bowers:2002xr,Casalbuoni:2003wh,Casalbuoni:2003sa,Casalbuoni:2004wm,Casalbuoni:2005zp,Ciminale:2006sm,Mannarelli:2006fy}
--- the QCD analogue of a form of non-BCS pairing first considered
by Larkin, Ovchinnikov, Fulde and Ferrell~\cite{LOFF}. Or, given
that the CFL phase itself is likely augmented by kaon
condensation~\cite{Bedaque:2001je,Kryjevski:2004jw}, it could take
the form of a phase in which a CFL kaon condensate carries a current
in one direction balanced by a counter-propagating current in the
opposite direction carried by gapless quark
quasiparticles~\cite{Kryjevski:2005qq,Gerhold:2006dt}. This meson supercurrent
phase has been shown to have a lower free energy than the gCFL
phase. 

Our purpose in this paper is to analyze and compare candidate crystal
structures for three-flavor crystalline color superconductivity.
The investigation of crystalline color superconductivity in
three-flavor QCD was initiated in Ref.~\cite{Casalbuoni:2005zp}.
Although such phases seem to be free from magnetic
instability~\cite{Ciminale:2006sm}, it remains to be seen whether
such a phase can have a lower free energy than the meson current
phase, making it a possible resolution to the gCFL instability.  The
simplest ``crystal'' structures do not 
suffice~\cite{Casalbuoni:2005zp,Mannarelli:2006fy}, but experience in
the two-flavor context~\cite{Bowers:2002xr} suggests that realistic
crystal structures constructed from more plane waves will prove to
be qualitatively more robust.   Our results confirm this expectation.

Determining the favored crystal structure(s) in the
crystalline color superconducting phase(s) of three-flavor QCD requires
determining the gaps and comparing the free energies for very many candidate structures,
as there are even more possibilities than the many that were investigated in
the two-flavor context~\cite{Bowers:2002xr}.   As there, we shall make
a Ginzburg-Landau approximation.
This approximation is controlled if $\Delta \ll \Delta_0$,
where $\Delta$ is the gap parameter of the crystalline color superconducting
phase itself and $\Delta_0$ is the gap parameter in the CFL phase that would
occur if $M_s$ were zero.    We shall find that the most favored crystal
structures can have $\Delta/\Delta_0$ as large as $\sim 1/2$, meaning
that we are pushing the approximation hard and so should not trust it quantitatively.
In earlier work with Mannarelli~\cite{Mannarelli:2006fy},  we analyzed
a particularly simple one parameter family of ``crystal'' structures in
three-flavor quark matter, simple enough that we were able to do
do the analysis both with and without the Ginzburg-Landau approximation.
We found that the approximation works when it should and that, at least
for the simple crystal structures we analyzed in Ref.~\cite{Mannarelli:2006fy},
when it breaks down it always underestimates the gap $\Delta$ and the
condensation energy.  Furthermore, we found that the Ginzburg-Landau approximation
correctly determines which crystal structure among the one parameter family
that we analyzed in Ref.~\cite{Mannarelli:2006fy}  
has the largest gap and lowest free energy.  

We shall work throughout in a Nambu--Jona-Lasinio (NJL)
model in which the QCD interaction between
quarks is replaced by a point-like four-quark interaction, with the quantum
numbers of single-gluon exchange, analyzed in mean field theory.
This is not a controlled approximation.
However, it suffices for our purposes: because this model has  attraction
in the same channels as in QCD, its high density phase is the CFL phase; and, the
Fermi surface splitting effects whose
qualitative consequences we wish to study can all be built
into the model.  Note that we
shall assume throughout that $\Delta_0\ll \mu$.  This weak coupling assumption
means that the pairing is dominated by modes near the Fermi surfaces.
Quantitatively,  this means that results for the gaps and condensation energies
of candidate crystalline phases are independent of the cutoff in the NJL model
when expressed in terms of the CFL gap $\Delta_0$: if the cutoff is changed with
the NJL coupling constant adjusted so that $\Delta_0$ stays fixed, the gaps and
condensation energies for the candidate crystalline phases also stay fixed.
This makes the NJL model valuable for making the comparisons
that are our goal.

We shall consider crystal structures in which there are two condensates
\begin{eqnarray}
\langle ud \rangle &\sim& \Delta_3 \sum_{a} \exp\left(2i\q{3}{a} \cdot {\bf r}\right)\nonumber\\
\langle us \rangle &\sim& \Delta_2 \sum_{a} \exp\left(2i\q{2}{a}\cdot {\bf r} \right)\ .
\label{udanduscondensates}
\end{eqnarray}
As in Refs.~\cite{Casalbuoni:2005zp,Mannarelli:2006fy}, and as we explain
in Section II, we neglect $\langle ds \rangle$ pairing because the $d$ and $s$
Fermi surfaces  are twice as far apart from each other as each is from
the intervening $u$ Fermi surface.  Were we to set $\Delta_2$ to zero, treating
only $\langle ud \rangle$ pairing, we would recover the two-flavor Ginzburg-Landau
analysis of Ref.~\cite{Bowers:2002xr}.   There, it was found that the best
choice of crystal structure was one in which pairing occurs for a set
of eight ${\bf q}_3^a$'s pointing at the corners of a cube in momentum space,
yielding a condensate with face-centered cubic symmetry.  The 
analyses of three-flavor crystalline color superconductivity in 
Refs.~\cite{Casalbuoni:2005zp,Mannarelli:2006fy} introduce nonzero $\Delta_2$,
but made the simplifying ansatz that pairing occurs only for a single ${\bf q}_3$
and a single ${\bf q}_2$. We consider crystal structures with up to eight
${\bf q}_3^a$'s and up to eight ${\bf q}_2^a$'s.  

We shall evaluate the
free energy $\Omega(\Delta_2,\Delta_3)$ for each crystal structure, in
a Ginzburg-Landau expansion in powers of the $\Delta$'s. We work
up to order $\Delta_2^p\Delta_3^q$ with $p+q=6$.  At sextic order, 
we find that $\Omega(\Delta,\Delta)$ is positive for large $\Delta$
for all the crystal structures that we investigate.  This is in marked
contrast to the results of Ref.~\cite{Bowers:2002xr}, which showed that
many two-flavor crystal structures have negative sextic terms,
with free energies that are unbounded from below when the 
Ginzburg-Landau expansion is stopped at sextic order.   Because we 
find positive sextic terms, we are able to use our sextic Ginzburg-Landau expansion
to evaluate $\Delta$ and $\Omega(\Delta,\Delta)$ for all the structures that we analyze.

The two crystal structures that
we argue are most favorable are both related to the face-centered cube
of Ref.~\cite{Bowers:2002xr}, but in different ways.  In the first, which
we denote ``CubeX'' in Section VI, there are
four ${\bf q}_3^a$'s and four ${\bf q}_2^a$'s which {\it together} point
at the eight corners of a cube in momentum space. In the second, 
denoted ``2Cube45z'' in Section VI, there
are eight ${\bf q}_3^a$'s and eight ${\bf q}_2^a$'s which {\it each}
point at the eight corners of a cube in momentum space, the two cubes rotated
relative to each other by 45 degrees about an axis perpendicular to their faces.
To a large degree, our argument that these two structures are the most
favorable relies only on two qualitative inputs.  First, if either the set
of $\setq{2}{a}$'s or the set of $\setq{3}{a}$'s yields a $\langle us \rangle$
or a $\langle ud \rangle$ condensate whose free energy, viewed in isolation
as a two-flavor problem and evaluated as in Ref.~\cite{Bowers:2002xr}, is unfavorable,
then the three-flavor condensate is unfavorable.  Thus, we can use all the qualitative
results of Ref.~\cite{Bowers:2002xr}.  Second, the free energy of a candidate 
three-flavor crystal structure becomes less favorable the closer any 
$\q{2}{a}$ comes to the antipodes of any $\q{3}{a}$.    This second result
is foreshadowed in the results of Refs.~\cite{Casalbuoni:2005zp,Mannarelli:2006fy},
and the results of  Ref.~\cite{Mannarelli:2006fy} indicate that it is valid beyond
the Ginzburg-Landau approximation.  We shall see in Section VI 
that these two qualitative lessons
are sufficient to winnow the space of candidate crystal structures down to the
two that our calculational results, also described in Section VI, demonstrate
are indeed the most favorable.

We find that several of the crystal structures that we consider have gap parameters
that can be as large as $\Delta_0/3$, and that one of them (the CubeX structure)
has $\Delta/\Delta_0$ that reaches 1/2.  The robustness of these crystalline
condensates thus pushes the Ginzburg-Landau approximation that we have used in the
derivation of our results to the edge of its regime of quantitative reliability.  
As we discussed above, the analysis of
Ref.~\cite{Mannarelli:2006fy} shows that for simpler crystal structures 
qualitative results obtained within this approximation remain valid when the
approximation has broken down quantitatively.   We expect this to be so
also for the more realistic, and complicated, crystal structures that we have
constructed, but a demonstration would require their analysis without
making a Ginzburg-Landau approximation, something we do not attempt here.

We find that the two crystal structures which we argue are most favorable
have large condensation energies, easily 1/3 to 1/2 of that in the CFL phase
with $M_s=0$, which is  $3\Delta_0^2\mu^2/\pi^2$.  This is remarkable, given
the only quarks that pair are those lying on (admittedly many) rings on
the Fermi surfaces, whereas in the CFL phase with $M_s=0$ pairing
occurs over the entire $u$, $d$ and $s$ Fermi surfaces.  

The gapless CFL (gCFL)
phase provides a useful
comparison at nonzero $M_s$. For $2 \Delta_0 < M_s^2/\mu < 5.2 \Delta_0$,
model analyses that are restricted to isotropic phases
predict a gCFL phase~\cite{Alford:2003fq,Alford:2004hz,Fukushima:2004zq}, finding
this phase to have lower free energy than either the CFL phase or unpaired quark 
matter.
However, this phase
is unstable to the formation of current-carrying
condensates~\cite{Casalbuoni:2004tb,Huang:2004bg,Kryjevski:2005qq,Fukushima:2006su,Gerhold:2006dt} and so it cannot be the ground state.  The true ground state must have lower
free energy than that of the gCFL phase, 
and for this reason the gCFL free energy provides a useful benchmark.
We find that three-flavor crystalline color superconducting quark matter 
has a lower free energy than both gCFL quark matter and unpaired quark matter
within a wide regime of density. For
\be
2.9 \Delta_0 < \frac{M_s^2}{\mu} < 10.4 \Delta_0
\label{WideLOFFWindow}
\ee
the crystalline phase with one or other of the two crystal structures that we argue
are most favorable has lower free energy (greater condensation energy)
than CFL quark matter, gCFL quark matter, and unpaired quark matter.
(See Fig.~\ref{omegavsx} in Section VI.)
This window in parameter space is in no sense narrow.
Our results therefore indicate that three-flavor crystalline quark matter
will occur over a wide range of densities, unless, that is, the pairing between
quarks is so strong (that is, $\Delta_0$ is so large making $M_s^2/\Delta_0$ so
small) that quark matter is in the 
CFL phase all the way down to the density at which quark matter
is superseded by nuclear matter.  

However, our results  also indicate
that unless the Ginzburg-Landau approximation is underestimating
the condensation energy of the
crystalline phase by about a factor of two, there is a fraction
of the ``gCFL window'' (with $2 \Delta_0 < M_s^2/\mu < 2.9 \Delta_0$, in
the Ginzburg-Landau approximation) in which no crystalline phase
has lower free energy than the gCFL phase.  This is thus the most
likely regime in which to find the current-carrying meson condensates
of Refs.~\cite{Kryjevski:2005qq,Gerhold:2006dt}.

Our paper is organized as follows.  In Section~\ref{section:model}, we shall specify the model 
we use and the simplifying assumptions we make, valid for $\Delta \ll \Delta_0$. Along
the way we review relevant aspects of two-flavor color superconductivity.
We shall also define our ansatz for the crystalline condensates more precisely
than in Eq.~(\ref{udanduscondensates}).
Much of Section~\ref{section:model} closely follows our earlier paper
in collaboration with Mannarelli~\cite{Mannarelli:2006fy}.
One simplifying  assumption that we make is that $\Delta_2$
and $\Delta_3$ are equal in magnitude, an assumption which is related to how
electric neutrality is maintained. 
In Appendix A, we use our results to confirm the validity of this assumption.
In Section III 
we introduce the Ginzburg-Landau expansion of the free energy, 
deferring the derivation
of the expressions for the Ginzburg-Landau
coefficients to Section IV and their evaluation to Section V.  
We give our
results in Section VI, and discuss their implications for future work in Section VII.

\section{Model, simplifications and ansatz}
\label{section:model}

\subsection{Neutral unpaired three-flavor quark matter}

We shall analyze quark matter containing massless $u$ and $d$ quarks and $s$ quarks with
an effective mass $M_s$.  (Although the strange quark mass can be
determined self-consistently by solving for an $\langle \bar s s\rangle$
condensate~\cite{Steiner:2002gx,Abuki:2004zk,Ruster:2005jc}, we shall leave this to future
work and treat $M_s$ as a
parameter.) The Lagrangian density describing this system in the absence
of interactions is given by
\begin{equation}
{\cal L}_0=\bar{\psi}_{i\alpha}\,\left(i\,\de\!\!\!
/^{\,\,\alpha\beta}_{\,\,ij} -M_{ij}^{\alpha\beta}+
\mu^{\alpha\beta}_{ij} \,\gamma_0\right)\,\psi_{\beta j}
\label{lagr1}\ \,,
\end{equation}
where $i,j=1,2,3$ are flavor indices and $\alpha,\beta=1,2,3$ are
color indices and we  have suppressed the Dirac indices,
where $M_{ij}^{\alpha\beta} =\delta^{\alpha\beta}\,{\rm
diag}(0,0,M_s)_{ij} $ is the mass matrix, where
$\de^{\alpha\beta}_{ij}=\partial\delta^{\alpha\beta}\delta_{ij}$ and where
the quark chemical potential matrix is  given by
\begin{equation}\mu^{\alpha\beta}_{ij}=(\mu\delta_{ij}-\mu_e
Q_{ij})\delta^{\alpha\beta} + \delta_{ij} \left(\mu_3
T_3^{\alpha\beta}+\frac{2}{\sqrt 3}\mu_8 T_8^{\alpha\beta}\right) \,
, \label{mu}
\end{equation} with  $Q = {\rm
diag}(2/3,-1/3,-1/3)_{ij} $ the quark electric-charge matrix and
$T_3$ and $T_8$ the Gell-Mann matrices in color space. We shall
quote results at quark number chemical potential $\mu=500$~MeV
throughout.

In QCD, $\mu_e$, $\mu_3$ and $\mu_8$ are the zeroth components of
electromagnetic and color gauge fields, and the gauge field dynamics
ensure that they take on values such that the matter is
neutral~\cite{Alford:2002kj,Gerhold:2003js}, satisfying
\be
\label{neutrality}
\frac{\partial \Omega}{\partial\mu_e} =
\frac{\partial \Omega}{\partial\mu_3} =
\frac{\partial \Omega}{\partial\mu_8} = 0\ ,
\ee
with $\Omega$ the free energy density of the system.
In the NJL model that we shall employ, in which quarks interact
via four-fermion interactions and there are no gauge fields, we introduce
$\mu_e$, $\mu_3$ and $\mu_8$ by hand, and choose them to satisfy
the neutrality constraints (\ref{neutrality}).  The assumption of weak equilibrium
is built into the calculation via the fact that the only flavor-dependent chemical
potential is $\mu_e$, ensuring for example that the chemical potentials of
$d$ and $s$ quarks with the same color must be equal.  Because
the strange quarks have greater mass, the equality
of their chemical potentials implies that the $s$ quarks  have smaller
Fermi momenta than the $d$ quarks in the absence
of BCS pairing.  In the absence of pairing, then, because weak equilibrium
drives the massive strange quarks to be less numerous than
the down quarks, electrical neutrality
requires a $\mu_e>0$, which makes the up quarks less numerous than the
down quarks and introduces some electrons into the system.
In the absence of pairing, color neutrality is obtained with $\mu_3=\mu_8=0.$

The Fermi momenta of the quarks and electrons in quark matter
that is electrically and color neutral and in weak equilibrium
are given in the absence of pairing by
\begin{eqnarray}
p_F^d &=& \mu+\frac{\mu_e}{3}\nonumber\\
p_F^u &=& \mu-\frac{2 \mu_e}{3}\nonumber\\
p_F^s &=& \sqrt{\left(\mu+\frac{\mu_e}{3}\right)^2- M_s^2} \approx \mu + \frac{\mu_e}{3}
-\frac{M_s^2}{2\mu}\nonumber\\
p_F^e &=& \mu_e\ ,
\label{pF1}
\end{eqnarray}
where we have simplified $p_F^s$ upon assuming that $M_s$ and $\mu_e$ are
small compared to $\mu$ by working only to linear order in $\mu_e$ and $M_s^2$.
The free energy of the noninteracting quarks and electrons is given by
\bea
\Omega_{\rm unpaired} &=& -\frac{ 
3 \left( p_F^u\right)^4 + 3 \left(p_F^d\right)^4 + \left(p_F^e\right)^4 }{12\pi^2}\nonumber\\
&~& + \frac{3}{\pi^2}\int_0^{p_F^s}p^2 dp \left(\sqrt{p^2+M_s^2}-\mu-\frac{\mu_e}{3}\right)
\nonumber\\
&\approx & -\frac{3}{4\pi^2}\left(\mu^4 -\mu^2 M_s^2\right)\nonumber\\
&~&\quad + \frac{1}{2\pi^2}\mu M_s^2 \mu_e -
\frac{1}{\pi^2}\mu^2\mu_e^2+\ldots
\label{OmegaUnpaired}
\eea
To this order, electric neutrality requires
\be
\mu_e=\frac{M_s^2}{4\mu}\ ,
\label{mueneutral}
\ee
yielding
\begin{eqnarray}
p_F^d &=& \mu+\frac{M_s^2}{12\mu}=p_F^u+\frac{M_s^2}{4\mu}\nonumber\\
p_F^u &=& \mu-\frac{M_s^2}{6\mu}\nonumber\\
p_F^s &=& \mu-\frac{5 M_s^2}{12 \mu} =p_F^u-\frac{M_s^2}{4\mu}\nonumber\\
p_F^e &=& \frac{M_s^2}{4\mu}\ .
\label{pF2}
\end{eqnarray}
We see from (\ref{pF1}) that to leading order in $M_s^2$ and $\mu_e$, the
effect of the strange quark mass on unpaired quark matter is as if instead
one reduced the strange quark chemical potential by $M_s^2/(2\mu)$.
We shall make this approximation throughout.
The corrections to this approximation in
an NJL analysis of a two-flavor crystalline
color superconductor have been evaluated and found
to be small~\cite{Kundu:2001tt}, and we expect the same to be true here.
Upon making this assumption, we need no longer be careful about the
distinction between $p_F$'s and $\mu$'s, as we can simply think of the three
flavors of quarks as if they have chemical potentials
\begin{eqnarray}
\mu_d &=& \mu_u + 2 \delta\mu_3 \nonumber\\
\mu_u &=&p_F^u \nonumber\\
\mu_s &=& \mu_u - 2 \delta\mu_2
\label{pF3}
\end{eqnarray}
with
\be
\delta\mu_3 = \delta\mu_2 = \frac{M_s^2}{8\mu}\equiv \delta\mu \ ,
\ee
where the choice of subscripts indicates that
$2\delta\mu_2$ is the
splitting between the Fermi surfaces for quarks 1 and 3 and
$2\delta\mu_3$ is that between the Fermi surfaces for quarks 1 and 2,
identifying $u,d,s$ with $1,2,3$.  (The prefactor $2$ in the equations defining
the $\delta\mu$'s is chosen to agree with the notation used in the analysis
of crystalline color superconductivity in a two flavor model~\cite{Alford:2000ze}, in which the
two Fermi surfaces were denoted by $\mu\pm\delta\mu$ meaning that they were
separated by $2\delta\mu$.)

Note that the equality of $\delta\mu_2$ and $\delta\mu_3$ is only valid to leading
order in $M_s^2$; at the next order, $\mu_e=M_s^2/(4\mu)-M_s^4/(48\mu^3)$ and
$\delta\mu_3=\mu_e/2$ while $\delta\mu_2=\delta\mu_3+M_s^4/(16\mu^3)$. In Section V,
we will utilize the fact that $\delta\mu_2$ and $\delta\mu_3$ are close to equal, but
not precisely equal.

\subsection{BCS pairing and neutrality}

As described in Refs. \cite{Rajagopal:2000ff,Alford:2002kj,Steiner:2002gx,Alford:2003fq},
BCS pairing introduces qualitative changes into the analysis of neutrality.  For example,
in the CFL phase $\mu_e=0$ and $\mu_8$ is nonzero and of order $M_s^2/\mu$.
This arises because the construction of a phase in which BCS pairing occurs between
fermions whose Fermi surface would be split in the absence of pairing can be
described as follows. 
First, adjust the Fermi surfaces of those fermions that pair to make
them equal. This costs a free energy price of order $\delta\mu^2\mu^2$.  And, it
changes the relation between the chemical potentials and the particle numbers,
meaning that the $\mu$'s required for neutrality can change qualitatively as
happens in the CFL example. Second, pair.
This yields a free energy benefit of order $\Delta_0^2\mu^2$, where $\Delta_0$
is the gap parameter describing the BCS pairing.
Hence, BCS pairing will only
occur if the attraction between  the fermions is large enough that
$\Delta_0 \gtrsim \delta\mu$.  In the CFL context, in which $\langle ud \rangle$,
$\langle us \rangle$ and $\langle ds \rangle$ pairing is fighting against the
splitting between the $d$, $u$ and $s$ Fermi surfaces described above, it
turns out that CFL pairing can occur if
$\Delta_0>4\delta\mu=M_s^2/(2\mu)$~\cite{Alford:2003fq},
a criterion that is
reduced somewhat by kaon condensation which
acts to stabilize CFL pairing~\cite{Kryjevski:2004jw}.

In this paper we are considering quark matter at densities that are
low enough ($\mu<M_s^2/(2\Delta_0)$) that CFL pairing is not possible.
The gap parameter $\Delta_0$ that would characterize the CFL
phase if $M_s^2$ and $\delta\mu$ were zero
is nevertheless an important scale in our problem, as it
quantifies the strength of the attraction between quarks.
Estimates of the magnitude of $\Delta_0$ are typically in the tens of MeV,
perhaps as large as 100 MeV~\cite{reviews}.  We shall treat $\Delta_0$ as a parameter,
and quote results for $\Delta_0=25$~MeV, although as we shall show
in Section VI.E our results can easily be scaled to any value of $\Delta_0$ 
as long as the weak-coupling approximation $\Delta_0\ll\mu$ is respected.

\subsection{Crystalline color superconductivity in two-flavor quark matter}

Crystalline color superconductivity can be thought of as the answer
to the question: ``Is there a way to pair quarks at differing Fermi
surfaces without first equalizing their Fermi momenta, given that
doing so exacts a cost?" The answer is ``Yes, but it requires Cooper
pairs with nonzero total momentum."  Ordinary BCS pairing pairs
quarks with momenta ${\bf p}$ and $-{\bf p}$, meaning that if the
Fermi surfaces are split at most one member of a pair can be at its
Fermi surface.  In the crystalline color superconducting phase,
pairs with total momentum $2\bf q$ condense, meaning that one member
of the pair has momentum ${\bf p}+{\bf q}$ and the other has
momentum $-{\bf p}+{\bf q}$ for some $\bf
p$~\cite{LOFF,Alford:2000ze}.  Suppose for a moment that only $u$
and $d$ quarks pair, making the analyses of a two-flavor model found
in
Refs.~\cite{Alford:2000ze,Bowers:2001ip,Casalbuoni:2001gt,Leibovich:2001xr,Kundu:2001tt,Bowers:2002xr,Casalbuoni:2003wh,Casalbuoni:2003sa,Casalbuoni:2004wm}
(and really going back to Ref.~\cite{LOFF}) valid. We sketch the
results of this analysis in this subsection.

The simplest ``crystalline'' phase
is one in which only pairs with a single $\bf q$ condense, yielding a condensate
\be
\langle \psi_u(x) C \gamma_5 \psi_d(x) \rangle \propto \Delta \exp(2i {\bf q}\cdot {\bf r})
\label{singleplanewave}
\ee
that is modulated in space like a plane wave.   (Here and throughout, we shall denote
by ${\bf r}$ the spatial three-vector corresponding to the Lorentz four-vector $x$.)
Assuming that $\Delta\ll \delta\mu \ll \mu$,
the energetically favored value of $|{\bf q}|\equiv q$ turns out to be $q=\eta \delta\mu$, where
the proportionality constant $\eta$ is given by $\eta=1.1997$~\cite{LOFF,Alford:2000ze}.
If $\eta$ were 1, then the only choice of $\bf p$ for which a Cooper pair
with momenta $(-\bf p+\bf q,\bf p+\bf q)$ would describe two quarks each on
their respective Fermi surfaces would correspond to a quark on
the north pole of one Fermi surface and a quark on the south pole of the other.
Instead, with $\eta>1$, the quarks on each Fermi surface that can pair lie
on one ring on each Fermi surface, the rings having opening angle
$\psi_0=2\cos^{-1}(1/\eta)=67.1^\circ$.  The energetic calculation that determines $\eta$
can be thought of as balancing the gain in pairing energy as $\eta$ is increased
beyond $1$, allowing quarks on larger rings to pair, against the kinetic energy cost
of Cooper pairs with greater total momentum.   If the $\Delta/\delta\mu\rightarrow 0$
Ginzburg-Landau limit is not assumed, the pairing rings change from circular lines
on the Fermi surfaces into
ribbons of thickness $\sim\Delta$ and angular extent $\sim \Delta/\delta\mu$.
The condensate (\ref{singleplanewave}) carries a current, which is balanced by
a counter-propagating current carried by the unpaired quarks near
their Fermi surfaces that are not in the pairing ribbons. Hence, the state carries no net current.

After solving a gap equation for $\Delta$
and then evaluating the free energy of the phase with condensate (\ref{singleplanewave}),
one finds that this simplest ``crystalline'' phase is favored over two-flavor quark matter
with  either no pairing or BCS pairing only within  a narrow window
\be
0.707\, \Delta_{\rm 2SC} <  \delta\mu  < 0.754\,\Delta_{\rm 2SC}\ ,
\label{LOFFwindow}
\ee
where $\Delta_{\rm 2SC}$ is the gap parameter for the two-flavor phase 
with 2SC (2-flavor, 2-color)
BCS pairing  found at $\delta\mu=0$.
At the upper boundary of this window, $\Delta\rightarrow 0$ and one finds a second
order phase transition between the crystalline and unpaired phases.  At the lower boundary,
there is a first order transition between the crystalline and BCS paired phases.
The crystalline phase persists
in the weak coupling
limit only if $\delta\mu/\Delta_{\rm 2SC}$ is held fixed, within
the window (\ref{LOFFwindow}), while the standard weak-coupling limit
$\Delta_{\rm 2SC}/\mu\rightarrow 0$ is taken.  Looking  ahead to our context,
and recalling that in three-flavor quark matter $\delta\mu=M_s^2/(8\mu)$,
we see that at high densities one finds
the CFL phase (which is the three-flavor quark matter BCS phase) and
in some window of lower densities one finds a crystalline phase.
In the vicinity of the second order transition, where $\Delta\rightarrow 0$
and in particular where $\Delta/\delta\mu\rightarrow 0$ and, consequently given (\ref{LOFFwindow}),
$\Delta/\Delta_{\rm 2SC}\rightarrow 0$ a Ginzburg-Landau expansion
of the free energy order by order in powers of $\Delta$ is controlled.
Analysis within an NJL model shows that the results for $\Delta(\delta\mu)$  become
accurate in the limit $\delta\mu\rightarrow 0.754\, \Delta_{\rm 2SC}$ where $\Delta\rightarrow 0$,
as must be the case, and show that the Ginzburg-Landau approximation underestimates
$\Delta(\delta\mu)$ at all $\delta\mu$~\cite{Alford:2000ze,Bowers:2002xr}.

The Ginzburg-Landau
analysis can then be applied to more complicated crystal structures in which
Cooper pairs with several different $\bf q$'s, all with the same length but pointing
in different directions, arise~\cite{Bowers:2002xr}.  This analysis indicates that a face-centered cubic
structure constructed as the sum of eight plane waves with $\bf q$'s pointing at
the corners of a cube is favored, but it does not permit a quantitative evaluation of
$\Delta(\delta\mu)$. The Ginzburg-Landau expansion of the free energy has
terms that are quartic and sextic in $\Delta$ whose coefficients are both
large in magnitude and negative. To this order, $\Omega$ is not bounded from below.
This means that the Ginzburg-Landau analysis
predicts a strong first order phase transition
between the crystalline and unpaired phase, at some $\delta\mu$ significantly larger
than $0.754\, \Delta_{\rm 2SC}$, meaning that the crystalline
phase occurs over a range of $\delta\mu$ that is
much wider than (\ref{LOFFwindow}), but it precludes the 
quantitative evaluation of the $\delta\mu$ at 
which the transition occurs, of $\Delta$, or of $\Omega$.  

We shall find that in three-flavor quark matter, all the crystalline phases
that we analyze have Ginzburg-Landau free energies with positive sextic
coefficient, meaning that they can be used to evaluate $\Delta$, $\Omega$
and the location of the transition from unpaired quark matter to the crystalline
phase with a postulated crystal structure.  For the most favored crystal structures,
we find that the window in parameter space in which they occur is given
by (\ref{WideLOFFWindow}), which is in no sense narrow.

\subsection{Crystalline color superconductivity in  neutral three-flavor quark matter}

Our purpose in this paper is to analyze three-flavor crystalline color
superconductivity, with condensates as in Eq.~(\ref{udanduscondensates})
for a variety of choices of the sets of $\q{2}{a}$'s and $\q{3}{a}$'s, i.e.
for a variety of crystal structures.  We shall make weak coupling
(namely $\Delta_0,\delta\mu \ll \mu$) and Ginzburg-Landau (namely $\Delta\ll\Delta_0,\delta\mu$)
approximations througout.

The analysis of neutrality in three-flavor quark matter in a crystalline
color superconducting phase is very simple
in the Ginzburg-Landau limit in which $\Delta\ll \delta\mu$: because the construction of
this phase does {\it not} involve rearranging any  Fermi momenta prior to pairing,
and because the assumption $\Delta\ll \delta\mu$ implies that the pairing does not
significantly change any number densities, neutrality is achieved with the
same chemical potentials
$\mu_e=M_s^2/(4\mu)$ and $\mu_3=\mu_8=0$ as in unpaired quark matter, and
with Fermi momenta given in Eqs.~(\ref{pF1}), (\ref{pF2}), and
(\ref{pF3}) as in unpaired quark matter.
This result is correct only in the Ginzburg-Landau limit.

We consider a  condensate of the form
\begin{equation}
\langle\psi_{i\alpha}(x) C\gamma_5 \psi_{j\beta}(x)\rangle \propto 
 \sum_{I=1}^3\ \ 
\sum_{\q{I}{a}\in\setq{I}{}}
 \Delta_I  e^{2i\q{I}{a}\cdot\rr}\coleps\flaeps 
 \label{condensate}\;,
\end{equation}
where ${\bf q}_1^a$,  ${\bf q}_2^a$ and ${\bf q}_3^a$  and $\Delta_1$, $\Delta_2$  and
$\Delta_3$ are the wave vectors and gap parameters describing pairing between
the $(d,s)$, $(u,s)$ and $(u,d)$ quarks respectively, whose Fermi momenta are
split by $2\delta\mu_1$, $2\delta\mu_2$ and $2\delta\mu_3$ respectively.
{}From (\ref{pF3}), we see that $\delta\mu_2=\delta\mu_3=\delta\mu_1/2=M_s^2/(8\mu)$.
For each $I$, $\setq{I}{}$ is a set of momentum vectors that define the periodic spatial
modulation of the crystalline condensate describing pairing between the quarks
whose flavor is not $I$, and whose color is not $I$.  Our goal in this paper is to
compare condensates with different choices of $\setq{I}{}$'s, that is with different
crystal structures.   To shorten expressions, we will henceforth write
\be
\sum_{\q{I}{a}}\equiv
\sum_{\q{I}{a}\in\setq{I}{}}\ .
\label{SumOverqDefn}
\ee
The condensate (\ref{condensate}) has the color-flavor structure of the 
CFL condensate (obtained by setting all ${\bf q}$'s to zero) and is the
natural generalization to nontrivial crystal structures of the condensate
previously analyzed in Refs.~\cite{Casalbuoni:2005zp,Mannarelli:2006fy}, in
which each $\setq{I}{}$ contained only a single vector. 

In the derivation of the Ginzburg-Landau approximation in Section IV, we shall
make no further assumptions.  However, in Sections V and VI when
we evaluate the Ginzburg-Landau coefficients and give our results, 
we shall make the further simplifying assumption
that $\Delta_1=0$.  Given that $\delta\mu_1$ is twice $\delta\mu_2$ or $\delta\mu_3$,
it seems reasonable that $\Delta_1\ll \Delta_2,\Delta_3$.   We leave a quantitative
investigation of condensates with $\Delta_1\neq 0$ to future work.

\subsection{NJL Model, and Mean-Field Approximation}

As discussed in Section I, we shall work in an NJL model in which the quarks interact
via a point-like four-quark interaction, with the quantum numbers of single-gluon exchange,
analyzed in mean field theory.  By this we mean that the interaction term added to the
Lagrangian (\ref{lagr1}) is
\begin{equation}
{\cal L}_{\rm interaction} = 
-\frac{3}{8}\lambda(\bar{\psi}\Gamma^{A\nu}\psi)(\bar{\psi}\Gamma_{A\nu}\psi)
\label{interactionlagrangian}\ ,
\end{equation}
where we have suppressed the color and flavor indices that we showed
explicitly in (\ref{lagr1}), and have continued to suppress the Dirac indices.
The full expression for $\Gamma^{A\nu}$ is 
$(\Gamma^{A\nu})_{\alpha i,\beta j} = \gamma^\nu (T^A)_{\alpha \beta}\delta_{i j}$, 
where the $T^A$ are the 
color Gell-Mann matrices. The NJL coupling constant $\lambda$ has dimension -2,
meaning that an ultraviolet cutoff $\Lambda$ 
must be introduced as a second parameter in order
to fully specify the interaction.    Defining $\Lambda$ as the restriction that momentum
integrals be restricted to a shell around the Fermi surface, 
$\mu-\Lambda < |{\bf p}| < \mu + \Lambda$,  the CFL gap parameter
can then be evaluated:~\cite{reviews,Bowers:2002xr}
\be
\Delta_0=2^{\frac{2}{3}}\Lambda \exp\left[-\frac{\pi^2}{2\mu^2\lambda}\right]\ .
\end{equation}
We shall see in subsequent sections that in the limit in which
which $\Delta\ll \Delta_0,\delta\mu\ll\mu$, all our results can be expressed in terms of $\Delta_0$;
neither $\lambda$ nor $\Lambda$ shall appear.  This reflects the fact that in this
limit the physics of interest is dominated by quarks near the Fermi surfaces,  not near 
$\Lambda$, and so once $\Delta_0$ is used
as the parameter describing the strength of the attraction between quarks, $\Lambda$ 
is no longer visible; the cutoff $\Lambda$ only appears in the relation between
$\Delta_0$ and $\lambda$, not in any comparison among different possible
paired phases.
In our numerical evaluations in Section VI,
we shall take $\mu=500$~MeV, $\Lambda=100$~MeV,  and adjust $\lambda$ to be
such that $\Delta_0$ is $25$~MeV. 

In the mean-field approximation, the interaction Lagrangian (\ref{interactionlagrangian})
takes on the form
\begin{equation}
{\cal L}_{\rm interaction}=
\ha\bar{\psi}\Delta(x)\bar{\psi}^T + \ha\psi^T\bar{\Delta}(x)\psi,
\label{meanfieldapprox}
\end{equation}
where $\Delta(x)$ is related to the diquark condensate by the
relations
\begin{equation}
\begin{split}
\Delta(x) &= \frac{3}{4}\lambda\Gamma^{A\nu}\langle\psi\psi^T\rangle(\Gamma_{A\nu})^T \\
\bar{\Delta}(x) &= \frac{3}{4}\lambda
   (\Gamma^{A\nu})^T\langle\bp^T\bp\rangle\Gamma_{A\nu} \\
   &=\gamma^0\Delta^{\dagger}(x)\gamma^0\label{deltaislambdacondensate}\;.
\end{split}
\end{equation}
The ansatz (\ref{condensate}) can now be made precise: we take
\begin{equation}
\Delta(x)=\Delta_{CF}(x)\otimes C\gamma^5\label{spin structure}\;,
\end{equation}
with 
\begin{equation}
\Delta_{CF}(x)_{\cf} = \sum_{I=1}^3\sum_{\q{I}{a}}
 \Delta(\q{I}{a}) e^{2i\q{I}{a}\cdot \rr}\coleps\flaeps\label{precisecondensate}\ .
\end{equation}
We have introduced notation that 
allows for the possibility of gap parameters $\Delta(\q{I}{a})$
with different magnitudes for different $I$ {\it and for different} $a$.  In fact, we shall
only consider circumstances in which $\Delta(\q{I}{a})=\Delta_I$, as in (\ref{condensate}).
However, it will be very convenient in subsequent sections to keep track of 
which $\Delta_I$ in a complicated equation ``goes with'' which $\q{I}{a}$, making
this notation useful.

The full Lagrangian, given by the sum of (\ref{lagr1}) and (\ref{meanfieldapprox}),
is then quadratic and can be written very simply upon introducing the two component
Nambu-Gorkov spinor
\begin{equation}
\chi = \left( \begin{array}{c}
\psi    \\
\bp^T
\end{array} \right)\  {\rm and~hence}\ 
\bar{\chi} = \left( \begin{array}{cc} 
\bar{\psi}&\psi^T \end{array}
\right)\ ,
\end{equation}
in terms of which 
\begin{equation}
{\cal L} = \ha \bar{\chi} \left( \begin{array}{cc}
i\cross{\partial}+\cross{\mu} & \Delta(x)    \\
\bar{\Delta}(x) &   (i\cross{\partial}-\cross{\mu})^T
\end{array} \right) \chi \;.
\label{fulllagrangian}
\end{equation}
Here, $\cross\mu\equiv \mu\gamma_0$ and $\mu$ is the matrix (\ref{mu}), which
we have argued simplifies to 
\begin{equation}
\mu = \delta^{\alpha\beta}\otimes {\rm diag}\left(\mu_u,\mu_d,\mu_s\right)
\end{equation}
with the flavor chemical potentials given simply by (\ref{pF3}). In subsequent
sections, we shall also often use the notation $\cross{\mu}_i\equiv \mu_i \gamma_0$,
with $i=1,2,3$ corresponding to $u,d,s$ respectively.

The propagator corresponding to the Lagrangian (\ref{fulllagrangian}) is given by
\begin{equation}
\begin{split}
\langle\chi(x)\bar{\chi}(x')\rangle
 &=\Bigl( \begin{array}{cc}
   \langle\psi(x)\bp(x')\rangle & \langle\psi(x)\psi^T(x')\rangle    \\
   \langle\bp^T(x)\bp(x')\rangle & \langle\bp^T(x)\psi^T(x')\rangle
\end{array} \Bigr) \\
 &=\Bigl( \begin{array}{cc}
   iG(x,x') & iF(x,x')    \\
   i\bar{F}(x,x') & i\bar{G}(x,x')
\end{array} \Bigr) \;,
\label{propagator}
\end{split}
\end{equation}
where $G$ and $\bar{G}$ are the ``normal'' components of the propagator and $F$
and $\bar{F}$ are the ``anomalous'' components. They satisfy the coupled
differential equations
\begin{equation}
\begin{split}
\Bigl( \begin{array}{cc}
i\cross{\partial}+\cross{\mu} & \Delta(x)    \\
\bar{\Delta}(x) &   (i\cross{\partial}-\cross{\mu})^T
\end{array} \Bigr) &
\Bigl( \begin{array}{cc}
   G(x,x') & F(x,x')    \\
   \bar{F}(x,x') & \bar{G}(x,x')
\end{array} \Bigr)\\
&=
\Bigl( \begin{array}{cc}
  1  & 0     \\
  0  & 1
\end{array} \Bigr)\delta^{(4)}(x-x')\ .
\label{Green's equation}
\end{split}
\end{equation}
We can now rewrite (\ref{deltaislambdacondensate}) as
\begin{equation}
\begin{split}
\Delta(x) &= \frac{3i}{4}\lambda\Gamma^{A\nu} F(x,x) (\Gamma_{A\nu})^T \\
\bar{\Delta}(x) &= \frac{3i}{4}\lambda
   (\Gamma^{A\nu})^T \bar{F}(x,x) \Gamma_{A\nu}\ , 
\label{formalgapequation}
\end{split}
\end{equation}
either one of which is the self-consistency equation, or gap equation, that we
must solve.  Without further approximation, (\ref{formalgapequation}) is not tractable.
It yields an infinite set of coupled gap equations, one for each $\Delta(\q{I}{a})$, because
without further approximation it is not consistent to choose finite sets $\{{\bf q}_I\}$. 
When several plane waves are present in the condensate, they induce an infinite
tower of higher momentum condensates~\cite{Bowers:2002xr}.   The reason why
the Ginzburg-Landau approximation, to which we now turn, is such a simplification
is that it eliminates these higher harmonics.

\section{Ginzburg-Landau Approximation: Introduction}

The form of the Ginzburg-Landau expansion of the free energy can be derived using only
general arguments.  This, combined with results for two-flavor crystalline color
superconductivity from Ref.~\cite{Bowers:2002xr}, will allow us to draw some
partial conclusions in this Section.   

We shall only consider crystal structures in which
all the vectors $\q{I}{a}$ in the crystal structure $\{ {\bf q}_I\}$ are ``equivalent''.
By this we mean that a rigid rotation of the crystal structure can be found which
maps any $\q{I}{a}$ to any other $\q{I}{b}$ leaving the set $\{ {\bf q}_I \}$ invariant.
For such crystal structures, $\Delta(\q{I}{a})=\Delta_I$, meaning that the free energy
is a function only of $\Delta_1$, $\Delta_2$ and $\Delta_3$.  
As explained in Section II.D, the chemical potentials that maintain neutrality in
three-flavor crystalline color superconducting quark matter are the same as those
in neutral unpaired three-flavor quark matter.  Therefore, 
\be
\Omega_{\rm crystalline}=\Omega_{\rm unpaired} + \Omega(\Delta_1,\Delta_2,\Delta_3)\ ,
\label{CondensationEnergy}
\ee
with $\Omega_{\rm unpaired}$ given in (\ref{OmegaUnpaired}) with (\ref{mueneutral}), 
and 
with $\Omega(0,0,0)=0$. Our task is to evaluate the condensation energy 
$\Omega(\Delta_1,\Delta_2,\Delta_3)$.   Since our Lagrangian is baryon number
conserving and 
contains no
weak interactions, it is invariant under a global $U(1)$ symmetry for each flavor.
This means that $\Omega$ must be invariant under $\Delta_I \rightarrow e^{i\phi_I} \Delta_I$ 
for each $I$, meaning that each of the three $\Delta_I$'s can only appear in the combination
$\Delta_I^*\Delta_I$.  (Of course, the ground state can and does break these
$U(1)$ symmetries spontaneously; what we  need in the argument we are making here is only
that they are not explicitly broken in the Lagrangian.)  We conclude that if 
we expand $\Omega(\Delta_1,\Delta_2,\Delta_3)$ in powers of the $\Delta_I$'s up
to sextic order, it must take the form
\begin{widetext}
\begin{equation}
\begin{split}
\Omega(\{\Delta_I\})= &\frac{2\mu^2}{\pi^2}\Biggl[\sum_I P_I
\alpha_I \, \Delta_I^*\Delta_I \\
&+\ha\Biggl(\sum_I \beta_I(\Delta_I^*\Delta_I)^2
+\sum_{I>J} \beta_{IJ}\, \Delta_I^*\Delta_I\Delta_J^*\Delta_J\Biggr)\\
&+\frac{1}{3}\Biggl(\sum_I \gamma_I(\Delta_I^*\Delta_I)^3
+\sum_{I\neq J}\gamma_{IJJ}\,
\Delta_I^*\Delta_I\Delta_J^*\Delta_J\Delta_J^*\Delta_J
+\gamma_{123}\,\Delta_1^*\Delta_1\Delta_2^*\Delta_2\Delta_3^*\Delta_3\Biggr)\Biggr]
\label{GLexpansion}\;,
\end{split}
\end{equation}
\end{widetext}
where we have made various notational choices for later convenience.  The overall prefactor
of $2\mu^2/\pi^2$ is the density of states at the Fermi surface of unpaired quark 
matter with $M_s=0$; it will prove convenient that we have defined all the coefficients in
the Ginzburg-Landau expansion of the free energy relative to this.  We have defined
$P_I={\rm dim}\{{\bf q}_I\}$, the number of plane  waves in the crystal structure
for the condensate describing pairing between quarks whose flavor and color are
not $I$.  Writing the prefactor $P_I$ multiplying the quadratic term 
and writing the factors of $\ha$ and $\frac{1}{3}$ multiplying the 
quartic and sextic terms  ensures that the $\alpha_I$, $\beta_I$ and $\gamma_I$ coefficients
are defined the same way as in Ref.~\cite{Bowers:2002xr}.
The form of the
Ginzburg-Landau expansion (\ref{GLexpansion}) is model independent, whereas
the expressions for the coefficients $\alpha_I$, $\beta_I$, $\beta_{IJ}$, $\gamma_I$,
$\gamma_{IJJ}$ and $\gamma_{123}$ for a given ansatz
for the crystal structure  are model-dependent. In Section IV we shall
derive the Ginzburg-Landau approximation to our model, yielding expressions
for these coefficients which we then evaluate in Section V.

We see in  Eq.~(\ref{GLexpansion}) that there are some coefficients --- namely
$\alpha_I$, $\beta_I$ and $\gamma_I$ --- which multiply polynomials involving
only a single $\Delta_I$.   Suppose that we keep a single $\Delta_I$ nonzero,
setting the other two to zero.  This reduces the problem to one with two-flavor
pairing only, and the Ginzburg-Landau coefficients for this problem have been
calculated for many different crystal structures in Ref.~\cite{Bowers:2002xr}.
We can then immediately use these coefficients, called $\alpha$, $\beta$
and $\gamma$ in Ref.~\cite{Bowers:2002xr}, to determine our $\alpha_I$, 
$\beta_I$ and $\gamma_I$.  Using $\alpha_I$ as an example, we conclude
that
\begin{equation}
\begin{split}
\alpha_I&=
\alpha(q_I,\delta\mu_I)= -1
 +\frac{\delta\mu_I}{2 q_I}
 \log\left(\frac{q_I+\delta\mu_I}{q_I-\delta\mu_I}\right)\\
 &\qquad -\ha\log\left(\frac{\Delta_{\rm 2SC}^2}{4(q_I^2-\delta\mu_I^2)}\right)\ ,
\label{AlphaEqn}
\end{split}
\end{equation}
where $\delta\mu_I$ is the splitting between the Fermi surfaces of the quarks
with the two flavors other than $I$ and $q_I\equiv |\q{I}{a}|$ is the length of
the ${\bf q}$-vectors in the set $\{{\bf q}_I\}$.  (We shall see momentarily why
all have the same length.)
In (\ref{AlphaEqn}), $\Delta_{\rm 2SC}$ is the gap parameter
in the BCS state obtained with $\delta\mu_I=0$ and $\Delta_I$ nonzero with the
other two gap parameters set to zero.    Assuming that $\Delta_0\ll \mu$, this gap
parameter for 2SC (2-flavor, 2-color) BCS pairing is given by~\cite{Schafer:1999fe,reviews}
\be
\Delta_{\rm 2SC}= 2^{\frac{1}{3}}\Delta_0\ .
\ee
For a given $\delta\mu_I$ and $\Delta_0$, $\alpha_I$ given
in (\ref{AlphaEqn}) is minimized 
when~\cite{LOFF,Alford:2000ze,Bowers:2002xr}
\be
q_I=\eta\, \delta\mu_I {\rm ~with~} \eta=1.1997\ .
\label{EtaEqn}
\ee
In the Ginzburg-Landau approximation, in which the $\Delta_I$ are assumed
to be small, we must  first minimize the quadratic contribution to the
free energy, before proceeding to investigate the consequences of
the quartic and sextic contributions.  Minimizing $\alpha_I$ fixes the length
of all the ${\bf q}$-vectors in the set $\{{\bf q}_I\}$, thus eliminating the possibility
of higher harmonics.  
It is helpful to imagine the (three) sets $\setq{I}{}$ as
representing the vertices of (three) polyhedra in momentum space.
By minimizing $\alpha_I$, we have learned that each polyhedron $\setq{I}{}$ can be inscribed
in a sphere of radius $\eta \delta\mu_I$.   From the quadratic contribution to
the free energy, we do not learn anything about what shape polyhedra are
preferable.   In fact, the quadratic analysis in isolation would indicate
that if $\alpha_I<0$ (which happens for $\delta\mu_I<0.754\,\Delta_{\rm 2SC}$)
then modes with arbitarily many different $\hat{\bf q}_I$'s should condense.
It is the quartic and sextic coefficients that describe the interaction among the modes,
and hence control what shape polyhedra are in fact preferable.

The quartic and sextic coefficients $\beta_I$ and $\gamma_I$ can also be
taken directly from the two-flavor results of Ref.~\cite{Bowers:2002xr}.  They
are given by $\bar\beta/\delta\mu_I^2$ and $\bar\gamma/\delta\mu_I^4$ where
$\bar\beta$ and $\bar\gamma$ are dimensionless quantities depending only
on the directions of the vectors in the set $\setq{I}{}$.  They have been evaluated
for many crystal structures in Ref.~\cite{Bowers:2002xr}, resulting in two qualitative
conclusions.  Recall that, as reviewed in Section II.C,
the presence of a condensate
with some $\hat{\bf q}_I^a$ corresponds to pairing on a ring on each Fermi surface
with opening angle $67.1^\circ$.   The first qualitative conclusion is that any crystal
structure in which there are two $\hat{\bf q}_I^a$'s whose pairing rings intersect
has {\it very} large, positive, values of
both $\beta_I$ and $\gamma_I$, meaning that it is strongly disfavored.  
The second conclusion is that regular structures,
those in which there are many ways of adding four or six $\hat{\bf q}_I^a$'s  to form
closed figures in momentum space, are favored.  Consequently, according to 
Ref.~\cite{Bowers:2002xr} the favored crystal structure in the two-flavor
case has 8 $\hat{\bf q}_I^a$'s pointing towards the corners of a cube.  
Choosing the polyhedron in momentum space to be a cube yields a face-centered
cubic modulation of the condensate in position space.  

Because the $\beta_I$
and $\gamma_I$ coefficients in our problem can be taken over directly from
the two-flavor analysis, we can expect that it will be unfavorable for any of
the three sets $\setq{I}{}$ to have more than eight vectors, or to have any vectors
closer together than $67.1^\circ$.    At this point we cannot exclude the possibility
that the large positive $\beta_I$ and $\gamma_I$ indicating an unfavorable 
$\setq{I}{}$ could be offset by large negative values for the other coefficients
which we cannot read off from the two-flavor analysis.  However,
what we shall instead find in Section VI is that $\beta_{IJ}$ and $\gamma_{IIJ}$ are positive
in all cases that we have investigated.  This means that we know of no exceptions
to the rule that if a particular $\setq{I}{}$ is unfavorable as a two-flavor crystal structure, then
any three-flavor condensate in which this set of ${\bf q}$-vectors describes either
the $\De_1$, $\De_2$ or $\De_3$ crystal structure is also disfavored.

In Section IV we shall use our microscopic model to derive expressions
for all the coefficients in the Ginzburg-Landau expansion (\ref{GLexpansion}),
including rederiving those which we have taken above from
the two-flavor analysis of Ref.~\cite{Bowers:2002xr}.  The coefficients that
we cannot simply read off from a two-flavor analysis are those that multiply
terms involving more than one $\Delta_I$ and hence describe the interaction
between the three different $\Delta_I$'s.
Before evaluating the
expressions for the
coefficients in Section V, we shall make the further simpifying assumption that
$\Delta_1=0$, because the separation $\delta\mu_1$ between the $d$ and $s$
Fermi surfaces is twice as large as that between either and the intervening $u$ Fermi surface.
This simplifies (\ref{GLexpansion}) considerably, eliminating the $\gamma_{123}$ term
and all the $\beta_{IJ}$ and $\gamma_{IIJ}$ terms except $\beta_{32}$,
$\gamma_{223}$ and $\gamma_{332}$.  

\section{The Ginzburg-Landau approximation: Derivation}

We now derive the Ginburg-Landau approximation to the NJL model specified
in Section II.  We proceed by first making a  Ginzburg-Landau approximation to
the gap equation, and then formally integrate the gap equation in order to
obtain the free energy, since the gap equation is the variation of the free energy
with respect to the gap parameters. 

\begin{figure*}[t]
\includegraphics[width=4.5in]{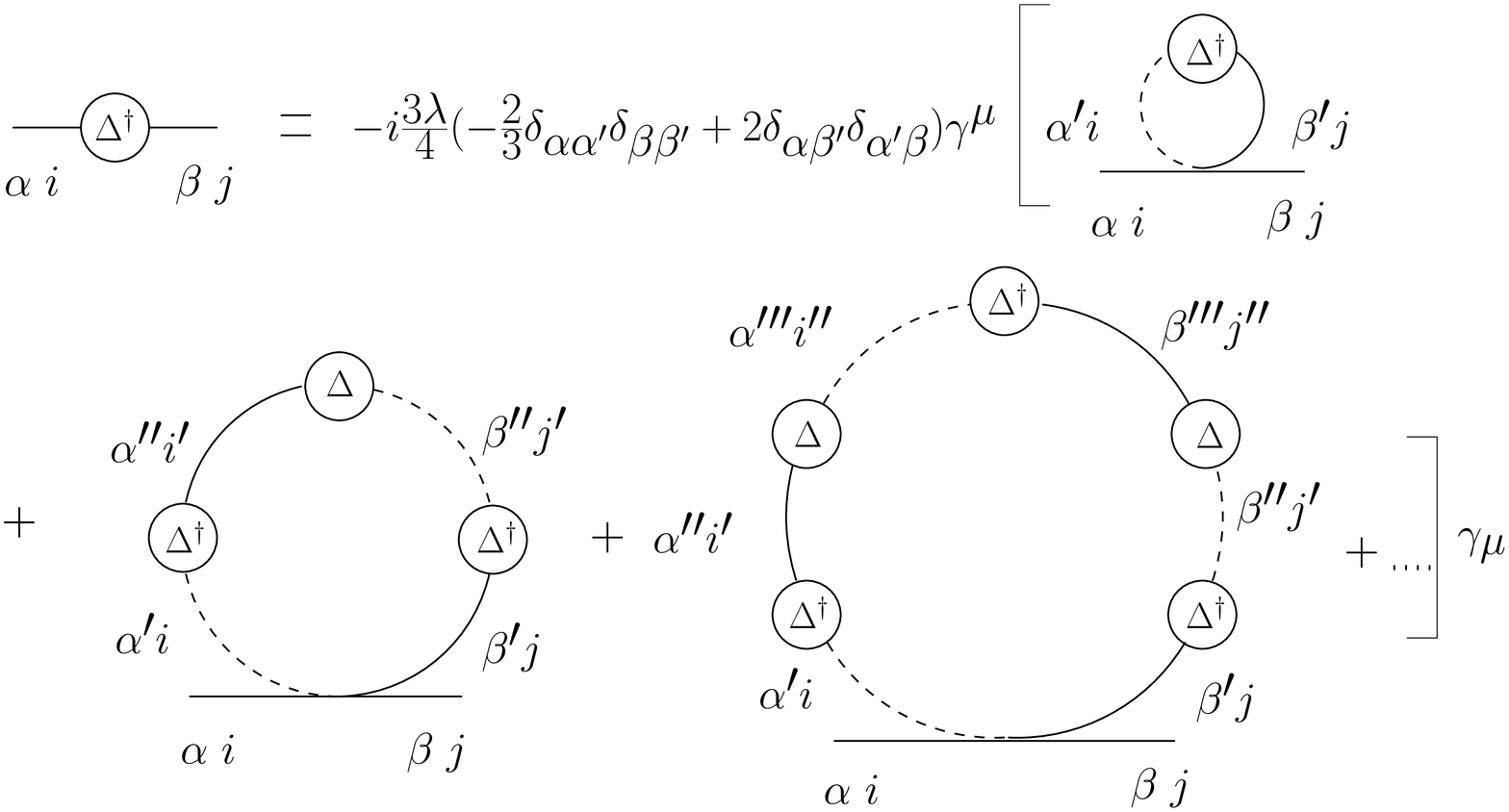}\hfill
\caption{ The gap equation. The labels $\alpha$, $\beta$ represent the
external color indices and $i$, $j$ represent the external flavor indices. All
the other color-flavor indices are contracted. $\Delta_{CF}$
(and $\Delta^\dagger_{CF}$) are matrices of the form (\ref{precisecondensate})
and carry the same color and flavor 
indices as the neighbouring propagators. The dashed lines 
represent
the propagator $(i\cross{\partial}-\cross{\mu})^{-1}$ and the solid lines
represent $(i\cross{\partial}+\cross{\mu})^{-1}$.   Evaluating the gap
equation involves substituting (\ref{precisecondensate}) for
$\Delta_{CF}$, doing the contraction over the internal color-flavor indices, and
evaluating the loop integrals in momentum space.}
\label{gl 3}
\end{figure*}

The gap equation (\ref{formalgapequation}) with which we closed
Section II is an infinite set of coupled equations, one for each $\Delta(\q{I}{a})$, 
with each equation containing arbitrarily high powers of the $\Delta$'s.  
In order to make a Ginzburg-Landau expansion, order by order in powers of the
$\Delta$'s, we first integrate (\ref{Green's equation}), obtaining
\begin{equation}
\begin{split}
G(x,x')&=G^{(0)}(x,x')-\intspace{z}G^{(0)}(x,z)\Delta(z)\bF(z,x') \\
\bF(x,x')&=-\intspace{z}\bG(x,z)\bD(z)G(z,x')\label{Green's
equation integrated} 
\end{split}
\end{equation}
with $G^{(0)}=(i\cross{\partial}+\cross{\mu})^{-1}$ and
$\bG=((i\cross{\partial}-\cross{\mu})^T)^{-1}$. We then expand these
equations order by order in $\Delta(x)$ by iterating them. To fifth
order, for $\bF$ we find
\begin{equation}
\begin{split}
\bF =& -\bG\bD G^{(0)} - \bG\bD G^{(0)}\Delta\bG\bD G^{(0)}\\
  \phantom{=}& -\bG\bD G^{(0)}\Delta\bG\bD G^{(0)}\Delta\bG\bD G^{(0)} +{\cal O}(\Delta^7)
\label{gl 1}\ ,
\end{split}
\end{equation}
where we have suppressed space-time coordinates and integrals for simplicity.  We then
substitute this expansion for $\bF$ into the right-hand side of the gap equation for
$\bar\Delta(x)$ in (\ref{formalgapequation}).  After using the $C\gamma_5$
Dirac structure of our ansatz (\ref{spin structure}) and the identity
$C(\gamma^\mu)^TC^{-1}=-\gamma^\mu$ to simplify the expression,
we obtain the gap equation satisfied by $\Delta_{CF}(x)$, the part of
our ansatz (\ref{spin structure},\ref{precisecondensate}) that describes the color, flavor and
spatial form of our condensate.  To order $\Delta^5$, we find
\begin{widetext}
\begin{equation}
\begin{split}
\Delta^{\dagger}_{CF} &= \frac{-3i\lambda}{4}(t^a)^T\gamma^\mu
 \Bigl[\frac{1}{i\cross{\partial}-\cross{\mu}}
       \Delta^{\dagger}_{CF} \frac{1}{i\cross{\partial}+\cross{\mu}}\\
       &+ \frac{1}{i\cross{\partial}-\cross{\mu}}
       \Delta^{\dagger}_{CF} \frac{1}{i\cross{\partial}+\cross{\mu}}
       \Delta_{CF} \frac{1}{i\cross{\partial}-\cross{\mu}}
       \Delta^{\dagger}_{CF} \frac{1}{i\cross{\partial}+\cross{\mu}}\\
       &+ \frac{1}{i\cross{\partial}-\cross{\mu}}
       \Delta^{\dagger}_{CF} \frac{1}{i\cross{\partial}+\cross{\mu}}
       \Delta_{CF} \frac{1}{i\cross{\partial}-\cross{\mu}}
       \Delta^{\dagger}_{CF} \frac{1}{i\cross{\partial}+\cross{\mu}}
       \Delta_{CF} \frac{1}{i\cross{\partial}-\cross{\mu}}
       \Delta^{\dagger}_{CF}
       \frac{1}{i\cross{\partial}+\cross{\mu}}\Bigr]_{(x,x)} \gamma_\mu t^a
       \label{gl 2}\ ,
\end{split}
\end{equation}
\end{widetext}
where the differential operators act on everything to their right and
where we have continued to simplify the notation by not writing the 
space-time, color and flavor arguments of the $\Delta_{CF}$'s 
and by not writing the integrals.
We then use the color Fierz identity
\begin{equation}
(t^a)_{\alpha'\alpha}(t^a)_{\beta'\beta} = \bigl(
-\frac{2}{3}\delta_{\alpha\alpha'}\delta_{\beta\beta'} +
2\delta_{\alpha'\beta}\delta_{\alpha\beta'} \bigr)
\end{equation}
to rewrite (\ref{gl 2}) as shown diagrammatically in Fig. \ref{gl 3}.

As written in (\ref{gl 2}) and shown in Fig.~\ref{gl 3}, what occurs on the left-hand
side of the gap equation is the space-dependent condensate from (\ref{precisecondensate}),
\begin{equation}
\Delta^*_{CF}(x)_{\cf}=\sum_I\coleps\flaeps\sum_{\q{I}{a}}\Delta^*(\q{I}{a})e^{-2i\q{I}{a}.\rr}\ ,
\end{equation}
whereas we now wish to turn the gap equation into a set of coupled equations
for the constants $\Delta(\q{I}{a})$. Doing so requires simplification of the
color-flavor structure of the right-hand side.  Our
ansatz for the color-flavor structure of the condensate, on the left-hand side,
is antisymmetric in both color and flavor.  However, direct evaluation of the
right-hand side yields terms that are symmetric in color and flavor, in addition
to the desired terms that are antisymmetric in both.  This circumstance is 
familiar from the analysis of the CFL phase~\cite{Alford:1998mk,reviews}, 
whose color-flavor structure we are after all employing.  In the presence of
a color and flavor antisymmetric condensate, a symmetric condensate must also
be generated because doing so does not change any symmetries.  The same
argument applies here also.  In the CFL phase, the symmetric condensate is
both quantitatively and parametrically suppressed relative to the antisymmetric
condensate, which is understandable based on the basic fact
that the QCD interaction is attractive
in the antisymmetric channel and repulsive in the symmetric channel.
We therefore expect that here too if we were to include color and flavor symmetric
condensates in our ansatz and solve for them, they would prove to be suppressed
relative to the antisymmetric condensates, and furthermore expect that, as in
the CFL phase, their inclusion would have negligible impact on the value of the
dominant antisymmetric condensate.  Hence, we drop the color and flavor
symmetric terms occurring on the right-hand side of the gap equation.
Upon so doing, the right-hand
side of the gap equation, which we denote $R_{\cf}$, has the structure
\begin{equation}
R_{\cf}(x) =\sum_I R_I({\bf r}) \coleps\flaeps\label{R decompose}
\end{equation}
Because $\coleps\flaeps$ are linearly independent tensors for each value of $I$,
in order for the gap equation to be satisfied for all values of $\alpha$,
$\beta$, $i$ and $j$ we must have
\begin{equation}
\sum_{\q{I}{a}}\Delta^*(\q{I}{a})e^{-2i\q{I}{a}.\rr}=R_I({\bf r}) \label{gl 3.5}\ 
\end{equation}
for all three values of $I$.   This is a set of $\sum_I P_I$ coupled equations
for the undetermined constants $\Delta(\q{I}{a})$.  (Recall that $P_I$ is the number
of vectors in the set $\setq{I}{}$.)  After transforming to momentum space, these
gap equations can be written as follows:
\begin{widetext}
\begin{equation}
\begin{split}
&\Delta^*(\q{I}{a})=-\frac{2{\mu}^2\lambda}{\pi^2}
  \Biggl[\sum_{\q{I}{b}}
 \Delta^*(\q{I}{b})\Pi_{jk}(\q{I}{b},\q{I}{a})\delta_{{\q{I}{b}-\q{I}{a}}}
\\
&\qquad +\sum_{{\q{I}{b}\q{I}{c}\q{I}{d}}}
 \Delta^*(\q{I}{b})\Delta(\q{I}{c})\Delta^*(\q{I}{d})
 \,{\cal{J}}_{jkjk}(\q{I}{b},\q{I}{c},\q{I}{d},\q{I}{a})
 \delta_{{\q{I}{b}-\q{I}{c}+\q{I}{d}-\q{I}{a}}}\\
 & \qquad + \ha\sum_{J}\sum_{{\q{J}{b}\q{J}{c}\q{I}{d}}}
 \Delta^*(\q{J}{b})\Delta(\q{J}{c})\Delta^*(\q{I}{d})
 \,{\cal{J}}_{kIkJ}(\q{J}{b},\q{J}{c},\q{I}{d},\q{I}{a})
 \delta_{{\q{J}{b}-\q{J}{c}+\q{I}{d}-\q{I}{a}}}\\
+&\sum_{\q{I}{b}\q{I}{c}\q{I}{d}\q{I}{e}\q{I}{f}}
 \Delta^*(\q{I}{b})\Delta(\q{I}{c})\Delta^*(\q{I}{d})\Delta(\q{I}{e})\Delta^*(\q{I}{f})
 \,{\cal{K}}_{jkjkjk} (\q{I}{b},\q{I}{c},\q{I}{d},\q{I}{e},\q{I}{f},\q{I}{a})
 \delta_{{\q{I}{b}-\q{I}{c}+\q{I}{d}-\q{I}{e}+\q{I}{f}-\q{I}{a}}}\\
+&\sum_{J}\,\sum_{\q{J}{b}\q{J}{c}\q{I}{d}\q{I}{e}\q{I}{f}}
 \Delta^*(\q{J}{b})\Delta(\q{J}{c})\Delta^*(\q{I}{d})\Delta(\q{I}{e})\Delta^*(\q{I}{f})
 \,{\cal{K}}_{kIkJkJ}(\q{J}{b},\q{J}{c},\q{I}{d},\q{I}{e},\q{I}{f},\q{I}{a})
 \delta_{{\q{J}{b}-\q{J}{c}+\q{I}{d}-\q{I}{e}+\q{I}{f}-\q{I}{a}}}\\
+&\ha\sum_{J}\sum_{\q{J}{b}\q{J}{c}\q{J}{d}\q{J}{e}\q{I}{f}}
 \Delta^*(\q{J}{b})\Delta(\q{J}{c})\Delta^*(\q{J}{d})\Delta(\q{J}{e})\Delta^*(\q{I}{f})
 \,{\cal{K}}_{kIkIkJ}(\q{J}{b},\q{J}{c},\q{J}{d},\q{J}{e},\q{I}{f},\q{I}{a})
 \delta_{{\q{J}{b}-\q{J}{c}+\q{J}{d}-\q{J}{e}+\q{I}{f}-\q{I}{a}}}\\
+&\ha\sum_{J,K}\sum_{{\q{I}{b}\q{J}{c}\q{K}{d}}\atop{\q{K}{e}\q{J}{f}}}
 \Delta^*(\q{I}{b})\Delta(\q{J}{c})\Delta^*(\q{K}{d})\Delta(\q{K}{e})\Delta^*(\q{J}{f})
 \,{\cal{K}}_{JKIJIK}(\q{I}{b},\q{J}{c},\q{K}{d},\q{K}{e},\q{J}{f},\q{I}{a})
 \delta_{{\q{I}{b}-\q{J}{c}+\q{K}{d}-\q{K}{e}+\q{J}{f}-\q{I}{a}}}\\
+&\frac{1}{4}\sum_{J,K}
 \sum_{{\q{J}{b}\q{J}{c}\q{I}{d}}\atop{\q{K}{e}\q{K}{f}}}
 \Delta^*(\q{J}{b})\Delta(\q{J}{c})\Delta^*(\q{I}{d})\Delta(\q{K}{e})\Delta^*(\q{K}{f})
 \,{\cal{K}}_{KIKJIJ}(\q{J}{b},\q{J}{c},\q{I}{d},\q{K}{e},\q{K}{f},\q{I}{a})
 \delta_{{\q{J}{b}-\q{J}{c}+\q{I}{d}-\q{K}{e}+\q{K}{f}-\q{I}{a}}}\\
+&\frac{1}{4}\sum_{J,K}
\sum_{{\q{J}{b}\q{K}{c}\q{I}{d}}\atop{\q{J}{e}\q{K}{f}}}
 \Delta^*(\q{J}{b})\Delta(\q{K}{c})\Delta^*(\q{I}{d})\Delta(\q{J}{e})\Delta^*(\q{K}{f})
 \,{\cal{K}}_{KIJKIJ}(\q{J}{b},\q{K}{c},\q{I}{d},\q{J}{e},\q{K}{f},\q{I}{a})
 \delta_{{\q{J}{b}-\q{K}{c}+\q{I}{d}-\q{J}{e}+\q{K}{f}-\q{I}{a}}}
\Biggr],\label{gl 4}
\end{split}
\end{equation}
\end{widetext}
where we have introduced a lot of notation that we now define and explain.
First, recall from (\ref{SumOverqDefn}) that $\sum_{\q{I}{b}}$ means a sum over all the
$\q{I}{b}$'s in the set $\setq{I}{}$.   The $\delta$'s are therefore Kronecker $\delta$'s,
indicating that only those ${\bf q}$-vectors that can be arranged to form a certain
closed two-,  four- or six-sided figure in momentum space are to be included
in the sum.  The sums over $J$ are always understood to be sums over $J\neq I$,
and the sums over $K$ are always understood to be sums over $K\neq J$ and $K\neq I$.
The remaining flavor subscripts in some terms which are not
summed, denoted $j$ or $k$, must always be chosen not equal to each other,
not equal to $I$, and not equal to $J$ if $J$ occurs.  (This appears to leave
an ambiguity related to the exchange of $j$ and $k$ in terms where both occur,
but we shall see that the functions $\Pi$, ${\cal J}$ and ${\cal K}$ each have
a cyclic symmetry that ensures that the two apparent choices of $j$ and $k$
are equivalent.)
The functions $\Pi$, ${\cal{J}}$ and ${\cal{K}}$ are proportional to the
various loop integrals that appear in the evaluation of the Feynman diagrams
in the gap equation of Fig.~\ref{gl 3}. They
are given by
\begin{widetext}
\begin{equation}
\begin{split}
\Pi_{i,j}({\bf{k}}_1,{\bf{k}}_2)&=-\frac{i\pi^2}{\bar{\mu}^2}\gamma^\mu
 \fourier{p}\frac{1}{(\cross{p}-\cross{\mu}_i)
 (\cross{p}+2\cross{{\bf{k}}}_1+\cross{\mu}_j)}\,\gamma_\mu\\
{\cal{J}}_{i,j,k,l}
({\bf{k}}_1,{\bf{k}}_2,{\bf{k}}_3,{\bf{k}}_4)&=
-\frac{i\pi^2}{{\mu}^2}\gamma^\mu\fourier{p}\frac{1}{
 (\cross{p}-\cross{\mu}_i)
 (\cross{p}+2\cross{{\bf{k}}}_1+\cross{\mu}_{j})
 (\cross{p}+2\cross{{\bf{k}}}_1-2\cross{{\bf{k}}}_2-\cross{\mu}_{k})}\\
 &\phantom{++}\phantom{++}\frac{1}{
 (\cross{p}+2\cross{{\bf{k}}}_1-2\cross{{\bf{k}}}_2+2\cross{{\bf{k}}}_3+\cross{\mu}_l)}
 \,\gamma_\mu\\
{\cal{K}}_{i,j,k,l,m,n}({\bf{k}}_1,{\bf{k}}_2,{\bf{k}}_3,{\bf{k}}_4,{\bf{k}}_5,{\bf{k}}_6)
&=-\frac{i\pi^2}{{\mu}^2}\gamma^\mu\fourier{p}\frac{1}{
 (\cross{p}-\cross{\mu}_i)
 (\cross{p}+2\cross{{\bf{k}}}_1+\cross{\mu}_{j})
 (\cross{p}+2\cross{{\bf{k}}}_1-2\cross{{\bf{k}}}_2+\cross{\mu}_{k})}\\
 &\phantom{++++}\frac{1}{
 (\cross{p}+2\cross{{\bf{k}}}_1-2\cross{{\bf{k}}}_2+2\cross{{\bf{k}}}_3+\cross{\mu}_{l})
 (\cross{p}+2\cross{{\bf{k}}}_1-2\cross{{\bf{k}}}_2+2\cross{{\bf{k}}}_3-2\cross{{\bf{k}}}_4-\cross{\mu}_{m})}\\
 &\phantom{++}\phantom{++}\frac{1}{
 (\cross{p}+2\cross{{\bf{k}}}_1-2\cross{{\bf{k}}}_2+2\cross{{\bf{k}}}_3-2\cross{{\bf{k}}}_4+2\cross{{\bf{k}}}_5+\cross{\mu}_n)}
 \,\gamma_\mu\label{piJK}\;,
\end{split}
\end{equation}
\end{widetext}
where $\cross{\mu}_i=\gamma^0\mu_i$ and $\cross{{\bf{k}}}=(0,{\bf
k})_\mu\gamma^\mu=-{\bf k}\cdot\gamma$.
The subscripts $i$, $j$ etc. on the functions $\Pi$, ${\cal J}$
and ${\cal K}$ are flavor indices that give the flavor of the quark lines in the propagators
going around the loops in Fig.~\ref{gl 3}.  In each term in (\ref{gl 4}) 
the choice of flavor indices in $\Pi$, ${\cal J}$ or ${\cal K}$ is determined
by the requirement that a given $\Delta(\q{I}{a})$ must connect two propagators
for quarks with flavors different from each other and $I$. For example, $\Delta_3$
always connects a $u$ and a $d$ quark.  The easiest way to see how
this provides the explanation for the (perhaps initially peculiar looking)
prescriptions for
the ${\cal J}$ and ${\cal K}$ functions in each term in the
gap equations (\ref{gl 4}) is to examine 
Fig.~\ref{FreeEnergyDiagrams} below, which depicts examples
of the contributions of $\Pi$, ${\cal J}$ and ${\cal K}$ to the free energy
which we shall discuss next.

The gap equations that we have derived must be equivalent to the
set of equations $\partial \Omega/\partial \Delta(\q{I}{a}) = 0$, because
solutions to the gap equation are stationary points of the free energy $\Omega$.
This means that integrating the gap equations determines $\Omega$ up
to an overall multiplicative constant, which we can fix by requiring that
we reproduce known results for the single-plane wave condensates,
and up to an additive constant which we fix by the requirement 
that $\Omega_{\rm crystalline}=\Omega_{\rm unpaired}$ when all 
$\Delta(\q{I}{a})$ are set to zero. We find
\begin{widetext}
\begin{equation}
\begin{split}
&\Omega \left( \left\{\Delta(\q{I}{a}) \right\} \right) = \frac{2{\mu}^2}{\pi^2}
 \sum_I
  \Biggl[
 \sum_{\q{I}{a}\q{I}{b}}\Delta^*(\q{I}{b})\Delta(\q{I}{a})\left(\Pi_{jk}(\q{I}{a},\q{I}{b})+\frac{\pi^2}{2\lambda\mu^2}\right)
 \delta_{{\q{I}{b}-\q{I}{a}}}
\\
&\qquad
 +\ha\sum_{{\q{I}{b}\q{I}{c}\q{I}{d}\q{I}{a}}}
 \Delta^*(\q{I}{b})\Delta(\q{I}{c})\Delta^*(\q{I}{d})\Delta(\q{I}{a})
 \,{\cal J}_{jkjk}(\q{I}{b},\q{I}{c},\q{I}{d},\q{I}{a})
 \delta_{{\q{I}{b}-\q{I}{c}+\q{I}{d}-\q{I}{a}}}\\
&\qquad
 +\ha\sum_{J>I}\sum_{{\q{J}{b}\q{J}{c}\q{I}{d}\q{I}{a}}}
 \Delta^*(\q{J}{b})\Delta(\q{J}{c})\Delta^*(\q{I}{d})\Delta(\q{I}{a})
 \,{\cal J}_{kIkJ}(\q{J}{b},\q{J}{c},\q{I}{d},\q{I}{a})
 \delta_{{\q{J}{b}-\q{J}{c}+\q{I}{d}-\q{I}{a}}}\\
&
 +\frac{1}{3}\sum_{\q{I}{b}\q{I}{c}\q{I}{d}\q{I}{e}\q{I}{f}\q{I}{a}}
 \Delta^*(\q{I}{b})\Delta(\q{I}{c})\Delta^*(\q{I}{d})\Delta(\q{I}{e})\Delta^*(\q{I}{f})\Delta(\q{I}{a})
 \,{\cal K}_{jkjkjk}(\q{I}{b},\q{I}{c},\q{I}{d},\q{I}{e},\q{I}{f},\q{I}{a})
 \delta_{{\q{I}{b}-\q{I}{c}+\q{I}{d}-\q{I}{e}+\q{I}{f}-\q{I}{a}}}\\
&
 +\ha\sum_{J\neq I}\sum_{\q{J}{b}\q{J}{c}\q{I}{d}\q{I}{e}\q{I}{f}\q{I}{a}}
 \Delta^*(\q{J}{b})\Delta(\q{J}{c})\Delta^*(\q{I}{d})\Delta(\q{I}{e})\Delta^*(\q{I}{f})\Delta(\q{I}{a})\\
 &
\qquad\qquad\qquad\qquad\qquad\qquad\qquad\qquad\qquad\qquad
{\cal K}_{kIkJkJ}(\q{J}{b},\q{J}{c},\q{I}{d},\q{I}{e},\q{I}{f},\q{I}{a})
 \delta_{{\q{J}{b}-\q{J}{c}+\q{I}{d}-\q{I}{e}+\q{I}{f}-\q{I}{a}}}\\
&
 +\frac{1}{4}\sum_{J\neq K\neq I\neq J}\sum_{\q{I}{b}\q{J}{c}\q{K}{d}\q{K}{e}\q{J}{f}\q{I}{a}}
 \Delta^*(\q{I}{b})\Delta(\q{J}{c})\Delta^*(\q{K}{d})\Delta(\q{K}{e})\Delta^*(\q{J}{f})\Delta(\q{I}{a})\\
 &
\qquad\qquad\qquad\qquad\qquad\qquad\qquad\qquad\qquad\qquad
 {\cal K}_{JKIJIK}(\q{I}{b},\q{J}{c},\q{K}{d},\q{K}{e},\q{J}{f},\q{I}{a})
 \delta_{{\q{I}{b}-\q{J}{c}+\q{K}{d}-\q{K}{e}+\q{J}{f}-\q{I}{a}}}\\
&
 +\frac{1}{12}\sum_{J\neq K\neq I\neq J}
 \sum_{\q{J}{b}\q{K}{c}\q{I}{d}\q{J}{e}\q{K}{f}\q{I}{a}}
 \Delta^*(\q{J}{b})\Delta(\q{K}{c})\Delta^*(\q{I}{d})\Delta(\q{J}{e})\Delta^*(\q{K}{f})\Delta(\q{I}{a})\\
 &
\qquad\qquad\qquad\qquad\qquad\qquad\qquad\qquad\qquad\qquad
 {\cal K}_{KIJKIJ}(\q{J}{b},\q{K}{c},\q{I}{d},\q{J}{e},\q{K}{f},\q{I}{a})
 \delta_{{\q{J}{b}-\q{K}{c}+\q{I}{d}-\q{J}{e}+\q{K}{f}-\q{I}{a}}}
\Biggr]\label{omega 2}\;.
\end{split}
\end{equation}
As in (\ref{gl 4}), 
in each term the flavor indices $j$ and $k$ (or just $k$) 
that are not summed over are understood
to differ from each other and from the summed indices $I$ (or $I$ and $J$).

\begin{figure*}[t]
\includegraphics[width=6.0in,angle=0]{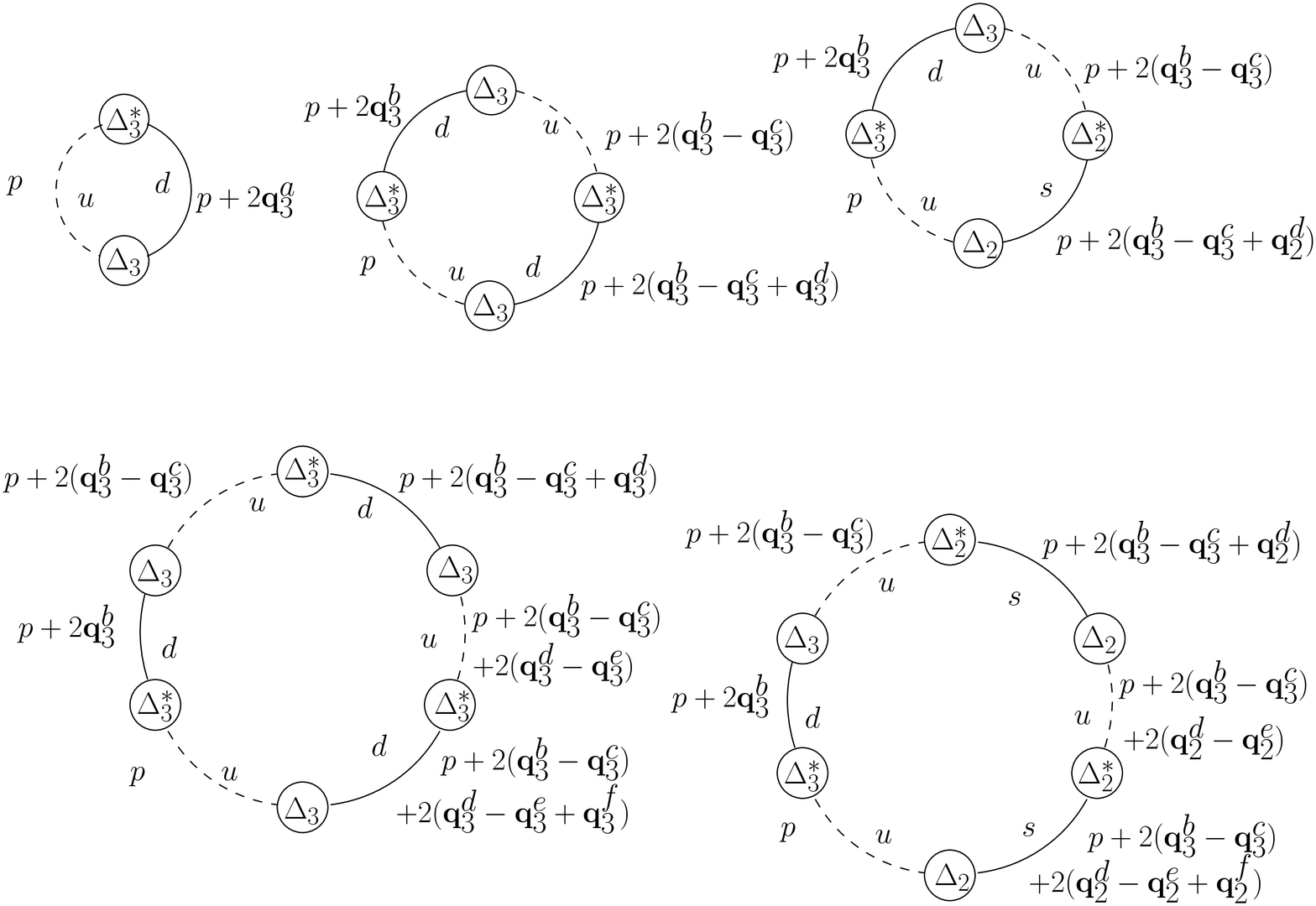}
\caption{Examples of contributions to the free energy.  The five diagrams depict a
$\Pi_{ud}$ contribution to $\alpha_3 \Delta^*_3\Delta_3$, a ${\cal J}_{udud}$
contribution to $\beta_3 (\Delta^*_3\Delta_3)^2$, a ${\cal J}_{udus}$
contribution to $\beta_{32}\Delta^*_3\Delta_3\Delta^*_2\Delta_2$, 
a ${\cal K}_{ududud}$ contribution to $\gamma_3(\Delta^*_3\Delta_3)^2$
and a ${\cal K}_{udusus}$  contribution to $\gamma_{322}\Delta^*_3\Delta_3(\Delta^*_2\Delta_2)^2$.
}
\vspace{-0.4 cm}
\label{FreeEnergyDiagrams}
\end{figure*}

As we discussed in Section III, we shall only consider crystal structures in which
each of the three sets $\setq{I}{}$ are regular, in the sense that all the $\q{I}{a}$
in one set $\setq{I}{}$ are equivalent.  
This means that $\Delta(\q{I}{a})=\Delta_I$,
which simplifies the free energy (\ref{omega 2}) to the form (\ref{GLexpansion})
which we derived on general grounds in Section III and  which
we reproduce here
\begin{equation}
\begin{split}
\Omega(\Delta_1,\Delta_2,\Delta_3)=&\frac{2\mu^2}{\pi^2}\Biggl[\sum_I P_I
\alpha_I \Delta_I^*\Delta_I+\ha\left(\sum_I \beta_I(\Delta_I^*\Delta_I)^2
+\sum_{I>J} \beta_{IJ}\Delta_I^*\Delta_I\Delta_J^*\Delta_J\right)\\
&+\frac{1}{3}\left(\gamma_I(\Delta_I^*\Delta_I)^3
+\sum_{I\neq J}\gamma_{IJJ}
\Delta_I^*\Delta_I\Delta_J^*\Delta_J\Delta_J^*\Delta_J
+\gamma_{123}\Delta_1^*\Delta_1\Delta_2^*\Delta_2\Delta_3^*\Delta_3\right)\Biggr]
\label{omega 3}
\end{split}
\end{equation}
for continuity. Now, however, we have obtained explicit expressions for 
all of the coefficients:
\begin{equation}
\begin{split}
\alpha_I &= \Pi_{jk}(\q{I}{a},\q{I}{a})+\frac{\pi^2}{2\lambda{\mu}^2}\\
\beta_I &= \sum_{\q{I}{b}\q{I}{c}\q{I}{d}\q{I}{a}}
{\cal J}_{jkjk}(\q{I}{b},\q{I}{c},\q{I}{d},\q{I}{a})
 \delta_{{\q{I}{b}-\q{I}{c}+\q{I}{d}-\q{I}{a}}} \\
\beta_{JI} &= \sum_{\q{J}{b}\q{J}{c}\q{I}{d}\q{I}{a}}
 \,{\cal J}_{kIkJ}(\q{J}{b},\q{J}{c},\q{I}{d},\q{I}{a})
 \delta_{{\q{J}{b}-\q{J}{c}+\q{I}{d}-\q{I}{a}}}\\
\gamma_I
&=\sum_{\q{I}{b}\q{I}{c}\q{I}{d}\q{I}{e}\q{I}{f}\q{I}{a}}
{\cal K}_{jkjkjk}(\q{I}{b},\q{I}{c},\q{I}{d},\q{I}{e},\q{I}{f}\q{I}{a})
 \delta_{{\q{I}{b}-\q{I}{c}+\q{I}{d}-\q{I}{e}+\q{I}{f}-\q{I}{a}}}\\
\gamma_{JII} &=\frac{3}{2}\sum_{\q{J}{b}\q{J}{c}\q{I}{d}\q{I}{e}\q{I}{f}\q{I}{a}}
{\cal K}_{kIkJkJ}(\q{J}{b},\q{J}{c},\q{I}{d},\q{I}{e},\q{I}{f},\q{I}{a})
 \delta_{{\q{J}{b}-\q{J}{c}+\q{I}{d}-\q{I}{e}+\q{I}{f}-\q{I}{a}}}\\
\gamma_{123} &=\frac{3}{4}\sum_{I\neq J\neq K\neq I}
 \sum_{\q{I}{b}\q{J}{c}\q{K}{d}\q{K}{e}\q{J}{f}\q{I}{a}}
{\cal K}_{JKIJIK}(\q{I}{b},\q{J}{c},\q{K}{d},\q{K}{e},\q{J}{f},\q{I}{a})
 \delta_{{\q{I}{b}-\q{J}{c}+\q{K}{d}-\q{K}{e}+\q{J}{f}-\q{I}{a}}}\\
 &\qquad+\frac{1}{4}\sum_{I\neq J\neq K\neq I}\sum_{\q{J}{b}\q{K}{c}\q{I}{d}\q{J}{e}\q{K}{f}\q{I}{a}}
{\cal K}_{KIJKIJ}(\q{J}{b},\q{K}{c},\q{I}{d},\q{J}{e},\q{K}{f},\q{I}{a})
 \delta_{{\q{J}{b}-\q{K}{c}+\q{I}{d}-\q{J}{e}+\q{K}{f}-\q{I}{a}}}\label{AlphaBetaGamma}\;.
\end{split}
\end{equation}
\end{widetext}
Here again, the unsummed indices $j$ and $k$ are chosen as described previously.
Since the free energy (\ref{omega 3}) is invariant under phase rotations of the 
$\Delta_I$ we can henceforth take all the $\Delta_I$ real and positive.
In Fig.~\ref{FreeEnergyDiagrams}, we give examples of contributions to the
free energy. These examples should make clear the choice of flavor subscripts
on the ${\cal J}$'s and ${\cal K}$'s in (\ref{AlphaBetaGamma})
and consequently in (\ref{gl 4}).  They also illustrate the origin of the Kronecker $\delta$'s
in so many of the expressions in this section:  each insertion of 
a $\Delta(\q{I}{a})$ (or $\Delta^*(\q{I}{a}$)) adds (or subtracts) momentum
$2 \q{I}{a}$ to (from) the loop, meaning that the Kronecker $\delta$'s arise due
to momentum conservation.   The diagrams also illustrate that $\Pi$,
${\cal J}$ and ${\cal K}$ are invariant under simultaneous cyclic permutation
of their flavor indices and momentum arguments, as this corresponds simply to
rotating the corresponding diagrams.

We have succeeded in deriving expressions for the Ginzburg-Landau coefficients
in our model; we shall turn to evaluating them in the next section.  Recall, however,
that upon setting $\Delta_1=0$ and keeping in mind that we can obtain results for $\alpha_I$,
$\beta_I$ and $\gamma_I$ from the two-flavor analyses in Ref.~\cite{Bowers:2002xr},
all that we need to do is evaluate $\beta_{32}$, $\gamma_{233}$ and $\gamma_{322}$
for the crystal structures we wish to investigate.  
We shall largely focus on 
crystal structures for which 
$\{\hat{\bf q}_2\}$ and $\{\hat{\bf q}_3\}$ are ``exchange symmetric'', meaning that
there is a sequence of rigid rotations
and reflections which when applied to
all the vectors in $\setq{2}{}$ and $\setq{3}{}$ together
has the effect of
exchanging $\{\hat{\bf q}_2\}$ and $\{\hat{\bf q}_3\}$.
If we choose an exchange symmetric crystal structure,
upon making the approximation
that $\delta\mu_2=\delta\mu_3$ and restricting our attention to solutions
with $\Delta_2=\Delta_3$ we have the further simplification that $\gamma_{322}=\gamma_{233}$.
Once we learn how to evaluate the loop integrals ${\cal J}$ and ${\cal K}$ in the
next Section, we will then in Section VI evaluate $\beta_{32}$ and $\gamma_{322}$
for various crystal structures, enabling us to evaluate the magnitudes of their
gaps and condensation energies.

\section{Calculating Ginzburg-Landau coefficients}

Calculating the Ginzburg-Landau coefficients (\ref{AlphaBetaGamma}) that
specify $\Omega(\Delta_1,\Delta_2,\Delta_3)$
for a given crystal structure involves
first evaluating the loop integrals $\Pi$, ${\cal J}$ and ${\cal K}$, defined 
in (\ref{piJK}), and then summing those that contribute to a given Ginzburg-Landau coefficient.
For example, we see from (\ref{AlphaBetaGamma}) that the
Ginzburg-Landau coefficient $\beta_{32}$ is given by 
summing ${\cal J}_{udus}(\q{3}{b},\q{3}{c},\q{2}{d},\q{2}{a})$ over all those
vectors $\q{3}{b}$ and $\q{3}{c}$ in the set $\setq{3}{}$ and all those 
vectors $\q{2}{d}$ and $\q{2}{a}$ in the set $\setq{2}{}$ which 
satisfy
$\q{3}{b}-\q{3}{c}+\q{2}{d}-\q{2}{a}=0$,
forming a 
closed four-sided figure in momentum space.  
Understanding how to evaluate 
the loop integrals $\Pi$, ${\cal J}$ and ${\cal K}$ requires
some explanation, which is our goal in this section.
Performing the sum required to evaluate a given
Ginzburg-Landau coefficient is then just bookkeeping, albeit nontrivial
bookkeeping for complicated crystal structures.

We are working in a weak-coupling limit in which $\delta\mu$, 
$|{\bf q}|=q=\eta\delta\mu$, and $\Delta_{\rm 2SC}$ are all much
smaller than $\mu$.  This means that we can choose our cutoff $\Lambda$
such that $\delta\mu,q,\Delta_{\rm 2SC}\ll \Lambda \ll \mu$.  Because
$\Lambda\ll\mu$, the integration measure in the expressions (\ref{piJK})
for $\Pi$, ${\cal J}$ and ${\cal K}$ simplifies as follows:
\begin{equation}
- \frac{i \pi^2}{{\mu}^2}\fourier{p} \approx
\int_{-\infty}^{+\infty}\frac{dp^0}{2\pi i}\int_{-\Lambda}^{\Lambda}
\frac{ds}{2}\int\frac{d\hat{{\bf p}}}{4\pi}\;,
\label{IntegrationMeasure}
\end{equation}
where $s\equiv |\vec{p}|-{\mu}$.
We now see by power counting that $\Pi$ is log-divergent as we
take $\Lambda\gg \delta\mu,q,\Delta_{\rm 2SC}$ whereas both
${\cal J}$ and ${\cal K}$ are $\Lambda$-independent in the large
$\Lambda$ limit.   Thus, in evaluating ${\cal J}$ and ${\cal K}$, we 
can safely take $\Lambda\rightarrow\infty$ whereas we must keep $\Lambda$
in the problem for a little longer in analyzing $\Pi$.  Explicit evaluation of 
$\Pi$ yields
\begin{equation}
\begin{split}
\Pi_{ud}({\bf q_3},{\bf q_3}) =& -1 + \frac{\delta\mu_3}{2q_3}
\log\left(\frac{q_3+\delta\mu_3}{q_3-\delta\mu_3}\right)\\
&-\ha\log\left(\frac{\Lambda^2}{q_3^2-\delta\mu_3^2}\right)\ .
\label{PiResult}
\end{split}
\end{equation}
We can now use
\be
\Delta_{\rm 2SC}=2\Lambda e^{-\frac{\pi^2}{2\lambda\mu^2}} 
\ee
and the relation between $\alpha_3$ and $\Pi_{ud}$ given in (\ref{AlphaBetaGamma})
to evaluate $\alpha_3$, obtaining the result (\ref{AlphaEqn}).   Notice that $\alpha_I$ 
depends on $\Lambda$ and $\lambda$ only through $\Delta_{\rm 2SC}$, and depends
only on the ratios $q_I/\Delta_{\rm 2SC}$ and $\delta\mu_I/\Delta_{\rm 2SC}$.
As discussed in Section III, $\alpha_I$ is negative for $\delta\mu_I/\Delta_{\rm 2SC}<0.754$,
and for a given value of this ratio for which $\alpha_I<0$, $\alpha_I$ is most negative
for $q_I/\Delta_{\rm 2SC}=\eta\,\delta\mu_I/\Delta_{\rm 2SC}$ with $\eta=1.1997$.
We therefore set $q_I=\eta\,\delta\mu_I$ henceforth and upon so doing obtain
\be
\begin{split}
\alpha(\delta\mu_I) =& -1 + \frac{1}{2\eta}\log\left(\frac{\eta+1}{\eta-1}\right) \\
&-\frac{1}{2}
\log\left(\frac{\Delta_{\rm 2SC}^2}{4 \delta\mu_I^2 (\eta^2-1)}\right)\\
=& -\frac{1}{2}
\log\left(\frac{\Delta_{\rm 2SC}^2}{4 \delta\mu_I^2 (\eta^2-1)}\right)\ ,
\label{AlphaEqn2}
\end{split}
\ee
where in the last line we have used the definition of $\eta$ derived from (\ref{AlphaEqn}).

The evaluation of $\beta_I$ and $\gamma_I$ is described in Ref.~\cite{Bowers:2002xr}. From 
the integration measure (\ref{IntegrationMeasure})
and the definitions of ${\cal J}$ and ${\cal K}$ (\ref{piJK}) we see that $\beta_I$ and $\gamma_I$
have dimension -2 and -4, respectively.  
Since they are independent of $\Lambda$ as long as $\Lambda\gg \delta\mu,q,\Delta_{\rm 2SC}$, 
and since $\lambda$ nowhere appears in their definition, there is no need to
introduce $\Delta_{\rm 2SC}$.  This means that the only dimensionful quantity on which
they can depend is $\delta\mu_I$ (since $q_I=\eta\delta\mu_I$ and since
the propagators are independent of $\mu$ in the weak-coupling limit) and so we
can write
\begin{equation}
\beta_I=\frac{\bar{\beta}_I}{\delta\mu_I^2}\,\, \mbox{and}\,\,
\gamma_I=\frac{\bar{\gamma_I}}{\delta\mu_I^4}\label{BetaIGammaI}\;,
\end{equation}
where $\bar{\beta}_I$ and $\bar{\gamma}_I$
are dimensionless quantities that depend only on the shape of the polyhedron
described by the set of vectors $\setq{I}{}$.  The evaluation of the ${\cal J}$ and ${\cal K}$\
loop integrals occurring in 
$\bar\beta$ and $\bar\gamma$ is described in Ref.~\cite{Bowers:2002xr}, and results
for many two-flavor crystal structures $\setq{3}{}$ are tabulated there.  The evaluation
is similar to but simpler than the evaluation of $\beta_{32}$ and $\gamma_{322}$,
to which we now turn.  

$\beta_{32}$ is the sum of ${\cal J}_{udus}(\q{3}{b},\q{3}{c},\q{2}{d},\q{2}{a})$, where
the momentum vectors satisfy
\begin{equation}
\q{3}{b}-\q{3}{c}+\q{2}{d}-\q{2}{a}=0\ .
\label{Jcondition}
\end{equation}
We now utilize the fact that $|\q{3}{b}|=|\q{3}{c}|=\eta\delta\mu_3$
and $|\q{2}{d}|=|\q{2}{a}|=\eta\delta\mu_2$ where $\delta\mu_3$ and $\delta\mu_2$
are similar in magnitude, but not precisely equal. (Recall from Section II.A that both are given
by $M_s^2/(8\mu)$ to this order, but that they differ at order $M_s^4/\mu^3$.)
Because $\delta\mu_2\neq\delta\mu_3$, the condition (\ref{Jcondition}) can only
be satisfied if
$\q{3}{b}=\q{3}{c}$, and $\q{2}{d}=\q{2}{a}$.  We must therefore evaluate
\begin{widetext}
\begin{equation}
{\cal J}_{udus}(\q{3}{b},\q{3}{b},\q{2}{a},\q{2}{a}) =-\frac{i\pi^2}{\mu^2}\gamma^\mu\left[
\fourier{p}\frac{1}{(\cross{p}-\cross{\mu}_u)
(\cross{p}+2\cross{{\bf q}}_{3}^{b}+\cross{\mu}_d)(\cross{p}-\cross{\mu}_u)
(\cross{p}+2\cross{{\bf q}}_{2}^{a}+\cross{\mu}_s)}
\right] \gamma_\mu \label{J23}
\end{equation}
We now expand the propagators in the weak-coupling limit, in
which $p^0$, $s$,  $|{\bf q}|$, $(\mu_d-\mu_u)$ and $(\mu_u-\mu_s)$
are all small compared to $\mu_u$, as follows:
\begin{equation}
\begin{split}
\frac{1}{\cross{p}+2\cross{q}+\cross{\mu}_i}&=
 \frac{(p^0+\mu_i)\gamma^0-({\bf  p}+2{\bf q})\cdot{\bf \gamma}}
 {(p^0+\mu_i-|{\bf p}+2{\bf q}|)(p^0+\mu_i+|{\bf p}+2{\bf q}|)}\\
 &\approx \frac{\mu_u\gamma^0 - {\bf p}\cdot{\bf \gamma}}
 {(p^0+\mu_u-(\mu_u-\mu_i)-|{\bf p}|-2{\bf q}\cdot\hat{{\bf p}})(2\mu_u)}\\
 &\approx
 \frac{1}{2}\left(\frac{\gamma^0-\hat{{\bf p}}\cdot\gamma}
 {p^0-s+(\mu_i-\mu_u)-2{\bf q}\cdot\hat{{\bf p}}}\right)\label{largemu}\;.
\end{split}
\end{equation}
Similarly,
\begin{equation}
\frac{1}{\cross{p}+2\cross{q}-\cross{\mu}_i}\approx
 \frac{1}{2}\left(\frac{\gamma^0+\hat{{\bf p}}\cdot\gamma}{p^0+s-(\mu_i-\mu_u)+2{\bf q}\cdot\hat{{\bf p}}}
 \right)\ .
\end{equation}
Eq.~(\ref{J23}) then simplifies to
\begin{equation}
{\cal J}_{udus}(\q{3}{b},\q{3}{b},\q{2}{a},\q{2}{a}) =
\int\frac{d\hat{{\bf p}}}{4\pi}\int_{-\infty}^{+\infty}\frac{dp^0}{2\pi i}
\int_{-\infty}^{+\infty} ds\left[\frac{1}{(p^0+s)^2
(p^0-s-\hat{{\bf p}}\cdot2\q{3}{b}+2\delta\mu_{3})
(p^0-s-\hat{{\bf p}}\cdot2\q{2}{a}-2\delta\mu_{2})}\right]\label{J23simplified}
\end{equation}
\end{widetext}
where we have used $\delta\mu_3=\ha(\mu_d-\mu_u)$ and $\delta\mu_2=\ha(\mu_u-\mu_s)$.

 To integrate (\ref{J23simplified}), we Wick rotate $p^0$ to $ip^4$ and then 
do the $s$ integral by contour integration. This gives two contibutions with 
different sign factors,
sign$(p^4)$ and sign$(-p^4)$, which are complex conjugates of each other.
Combining the two, the integration over $p^4$ is of form $2\Re
e\int_{\epsilon}^{+\infty}dp^4(...)$ where we have started the $p^4$ integration
from the infinitesimal positive number $\epsilon$ instead of zero, thus
definining the principal value of the integral. The integration 
over $p^4$ can now be carried out safely to obtain
\begin{widetext}
\begin{equation}
\begin{split}
{\cal J}_{udus}(\q{3}{b},\q{3}{b},\q{2}{a},\q{2}{a})&=
-\frac{1}{4}\int\frac{d\hat{{\bf p}}}{4\pi}
\Re e\left[\frac{1}{(i\epsilon-\hat{{\bf p}}\cdot\q{3}{b}+\delta\mu_{3})
(i\epsilon-\hat{{\bf p}}\cdot\q{2}{a}-\delta\mu_{2})}\right] \\
 &=-\frac{1}{4\delta\mu_{2}\delta\mu_{3}}
 \int\frac{d\hat{{\bf p}}}{4\pi}
\Re e \left[\frac{1}{(i\epsilon-\eta\hat{{\bf p}}\cdot\hat{{\bf q}}_3^b+1)
(i\epsilon-\eta\hat{{\bf p}}\cdot\hat{{\bf q}}_{2}^{a}-1)}\right]\label{J23 unevaluated}\;,
\end{split}
\end{equation}
\end{widetext}
where $\eta=\frac{|\q{3}{}|}{\delta\mu_{3}}=\frac{|\q{2}{}|}{\delta\mu_{2}}$.
{}From rotational symmetry it follows that the value of (\ref{J23 unevaluated}) depends only
on the angle between the momentum vectors 
$\hat{{\bf q}}_{3}^{b}$ and $\hat{{\bf q}}_{2}^{a}$, which we denote by $\phi$. 
We therefore define the dimensionless quantities
\begin{equation}
\bar{J}_{32}(\phi)=
\delta\mu_{2}\delta\mu_{3}\,{\cal J}_{udus}(\q{3}{b},\q{3}{b},\q{2}{a},\q{2}{a})
\end{equation}
and, correspondingly,
\be
\bar{\beta}_{32}=\delta\mu_2\delta\mu_3\beta_{32}\ .
\label{Beta32BarDefn}
\ee
$\bar{J}_{32}$ can be evaluated analytically by using Feynman parameters to
simplify the integrand in (\ref{J23 unevaluated}). The result is
\begin{widetext}
\begin{equation}
\bar{J}_{32}(\phi)=\frac{1}{4\,\eta\cos(\phi/2)}\left[\frac{1}{\sqrt{\eta^2\sin^2(\phi/2)-1}}
 \arctan\left(\frac{\sqrt{\eta^2\sin^2(\phi/2)-1}}{\eta\cos(\phi/2)}
 \right)\right]\label{J23 evaluated}\ .
\end{equation}
\end{widetext}
This completes the  evaluation of the loop integral ${\cal J}$ needed to
calculate $\beta_{32}$ for any crystal structure.  We summarize the 
calculation by noting that for a given crystal structure, $\beta_{32}$ depends
only on the shape of the polyhedra defined by $\setq{2}{}$ and $\setq{3}{}$ 
and on their relative orientation, depends
on the Fermi surface splittings $\delta\mu_3$ and $\delta\mu_2$, and is 
obtained using (\ref{Beta32BarDefn}) with
\begin{equation}
\bar\beta_{32}=
\sum_{{\bf q}_{3}^{b},{\bf q}_{2}^{a}}
\bar{J}_{32}(\angle\hat{{\bf q}}_{3}^{b}\hat{{\bf q}}_{2}^{a})\;,
\label{BetaBarResult}
\end{equation}
where $\bar{J}_{32}(\phi)$ is given by (\ref{J23 evaluated}).

We turn now to the evaluation of $\gamma_{322}$.  
From (\ref{AlphaBetaGamma}), 
\begin{equation}
\gamma_{322}
=\frac{3}{2}\sum_{{\q{3}{b},\q{3}{c},}\atop{\q{2}{d},\q{2}{e},\q{2}{f},\q{2}{a}}}
{\cal K}_{udusus}(\q{3}{b},\q{3}{c},\q{2}{d},\q{2}{e},\q{2}{f},\q{2}{a})\;,
\label{gamma322 form}
\end{equation}
and we again use the fact that the ${\bf q}_3$'s and ${\bf q}_2$'s do not
have precisely the same length to conclude that the momentum vectors must satisfy
both
\be
\q{3}{b}=\q{3}{c}
\ee
and
\be
\q{2}{d}-\q{2}{e}+\q{2}{f}-\q{2}{a}=0
\label{condition for K}\;.
\end{equation}
In the following expressions, it is always understood that  (\ref{condition for K})
is satisfied although we will not complicate equations by eliminating one 
of the $\q{2}{}$'s in favor of the other three.
We can see without calculation that, unlike ${\cal J}$, ${\cal K}$ will not
reduce to depending only on a single angle between two momentum vectors.
It will depend on the shape made by the four $\q{2}{}$ vectors satisfying 
(\ref{condition for K}), which can in fact be specified by two
angles, as well as on the angles that specify the direction of
$\q{3}{b}$ relative to the shape made by the four $\q{2}{}$'s.

The expression for ${\cal K}$ is given in (\ref{piJK})
and can also be read off from the bottom right Feynman diagram 
in Fig. \ref{FreeEnergyDiagrams}. It is given by
\begin{widetext}
\begin{equation}
\begin{split}
{\cal K}_{udusus}&(\q{3}{b},\q{3}{b},\q{2}{d},\q{2}{e},\q{2}{f},\q{2}{a})
=-\frac{i\pi^2}{\mu^2}\times
\gamma^\mu\fourier{p}
\Biggl[\frac{1}{(\cross{p}-\cross{\mu}_u)
(\cross{p}+2\cross{{\bf q}}_{3}^{b}+\cross{\mu}_d)(\cross{p}-\cross{\mu}_u)}\\
&\phantom{+++}\frac{1}{
(\cross{p}+2\cross{{\bf q}}_{2}^{d}+\cross{\mu}_s)(\cross{p}+2(\cross{{\bf q}}_{2}^{d}-\cross{{\bf q}}_{2}^{e})-\cross{\mu}_u)
(\cross{p}+2(\cross{{\bf q}}_{2}^{d}-\cross{{\bf q}}_{2}^{e}+\cross{{\bf
q}}_{2}^{f})+\cross{\mu}_s)}\Biggr] \gamma_\mu\;. \label{K322}
\end{split}
\end{equation}
After simplifying the propagators using (\ref{largemu}), we can rewrite
equation (\ref{K322}) as
\begin{equation}
\begin{split}
{\cal K}_{udusus}&(\q{3}{b},\q{3}{b},\q{2}{d},\q{2}{e},\q{2}{f},\q{2}{a}) =
\int\frac{d\hat{{\bf p}}}{4\pi}\int_{-\infty}^{+\infty}\frac{dp^0}{2\pi i}
\int_{-\infty}^{+\infty} ds\Biggl[\frac{1}{(p^0+s)^2
(p^0+s+\hat{{\bf p}}\cdot2(\q{2}{d}-\q{2}{e}))}\\
&\phantom{+++}\frac{1}{
(p^0-s-\hat{{\bf p}}\cdot2\q{3}{b}+2\delta\mu_{3})
(p^0-s-\hat{{\bf p}}\cdot2\q{2}{d}-2\delta\mu_{2})
(p^0-s-\hat{{\bf p}}\cdot2(\q{2}{d}-\q{2}{e}+\q{2}{f})-2\delta\mu_{2})}\Biggr]
\label{K322simplified}
\end{split}
\end{equation}
Unlike in the evaluation of ${\cal J}_{udus}$, we are not able to
do the $s$ and $p^0$ integrals analytically without introducing
Feynman parameters to simplify the integrand at this stage, before
doing any of the integrals.
We introduce one set of Feynman parameters, $x_1,\,x_2$, to
collect denominators of the form $p^0+s+..$ and another set, $y_1,\,y_2,\,y_3$, to 
collect the denominators of form $p^0-s+..$. 
This reduces the integral to
\begin{equation}
\begin{split}
&{\cal K}_{udusus}(\q{3}{b},\q{3}{b},\q{2}{d},\q{2}{e},\q{2}{f},\q{2}{a}) =
 \int_0^1 {\prod_{n=1}^2}dx_n\,\delta\left(1-\sum_{n=1}^2x_n\right)
 \int_0^1 {\prod_{m=1}^3}dy_m\,\delta\left(1-\sum_{m=1}^3y_m\right)\\
&\times\int\frac{d\hat{{\bf p}}}{4\pi}\int_{-\infty}^{+\infty}\frac{dp^0}{2\pi i}
\int_{-\infty}^{+\infty} ds\left[\frac{4(1-x_2)}{(p^0+s+2x_2\hat{{\bf
p}}\cdot[\q{2}{d}-\q{2}{e}])^3}\right]\\
& \qquad\qquad\qquad\left[
\frac{1}{
p^0-s-2\hat{{\bf p}}\cdot[y_1\q{3}{b}+y_2\q{2}{d}+
   y_3(\q{2}{d}-\q{2}{e}+\q{2}{f})]
   +y_1 2\delta\mu_{3}-y_2 2\delta\mu_{2}-y_3 2\delta\mu_{2}}\right]^3\ .
\label{K322simplifiedfeyn}
\end{split}
\end{equation}
 We now perform the $p^0$ and $s$ integrations in (\ref{K322simplifiedfeyn}), 
 following steps analogous to the
integration arising in the expression for ${\cal J}_{udus}$. i.e. Wick rotate 
$p^0$ to $ip^4$, do the $s$ integral by contour integration, add the two complex 
conjugate contributions thus obtained to write the integration over $p^4$
as $2\Re e\int_{\epsilon}^{+\infty}dp^4(...)$ and then perform the integration 
over $p^4$. This gives us
\begin{equation}
\begin{split}
&{\cal K}_{udusus}(\q{3}{b},\q{3}{b},\q{2}{d},\q{2}{e},\q{2}{f},\q{2}{a}) =-\frac{3}{8}
 \Re e \int_0^1 {\prod_{n=1}^2}dx_n\,\delta\left(1-\sum_{n=1}^2x_n\right)
 \int_0^1 {\prod_{m=1}^3}dy_m\,\delta\left(1-\sum_{m=1}^3y_m\right)\\
&\times\int\frac{d\hat{{\bf p}}}{4\pi}
\left[\frac{1-x_2}{i\epsilon+\hat{{\bf p}}\cdot[x_2(\q{2}{d}-\q{2}{e})
-(y_1\q{3}{b}+y_2\q{2}{d}+
   y_3(\q{2}{d}-\q{2}{e}+\q{2}{f}))]
   +y_1 \delta\mu_{3}-y_2 \delta\mu_{2}-y_3 \delta\mu_{2}}\right]^4\ .
\label{K322simplified ds}
\end{split}
\end{equation}
Finally, we do the $d\hat{{\bf p}}$ integral and obtain
\begin{equation}
\begin{split}
&{\cal K}_{udusus}(\q{3}{b},\q{3}{b},\q{2}{d},\q{2}{e},\q{2}{f},\q{2}{a}) =\frac{1}{8}
 \Re e \int_0^1 {\prod_{n=1}^2}dx_n\,\delta\left(1-\sum_{n=1}^2x_n\right)
 \int_0^1 {\prod_{m=1}^3}dy_m\,\delta\left(1-\sum_{m=1}^3y_m\right)(1-x_2)\\
&\phantom{+}\times
\frac{|x_2(\q{2}{d}-\q{2}{e})
      -(y_1\q{3}{b}+y_2\q{2}{d}+ y_3(\q{2}{d}-\q{2}{e}+\q{2}{f}))|^2
      +3[y_1 \delta\mu_{3}-y_2 \delta\mu_{2}-y_3 \delta\mu_{2}]^2}
   {\left[|x_2(\q{2}{d}-\q{2}{e}) - (y_1\q{3}{b}+y_2\q{2}{d}+
      y_3(\q{2}{d}-\q{2}{e}+\q{2}{f}))|^2
      -[y_1 \delta\mu_{3}-y_2 \delta\mu_{2}-y_3 \delta\mu_{2}+i\epsilon]^2\right]^3}\ .
\label{K322simplified ds dn}
\end{split}
\end{equation}
Noting that we can replace $\q{2}{}$ by $\eta\hat{{\bf q}}_2^{}$ and  $\q{3}{}$ 
by $\eta\hat{{\bf q}}_3^{}$, we conclude that, as expected, ${\cal K}_{udusus}$
depends only upon 
the shape of the polyhedra defined by $\setq{2}{}$ and $\setq{3}{}$ and
on the Fermi surface splittings $\delta\mu_3$ and $\delta\mu_2$.
 We cannot simplify (\ref{K322simplified ds dn}) further for general $\delta\mu_2$,
$\delta\mu_3$. However, if we now set $\delta\mu_2=\delta\mu_3=\delta\mu$, 
which is corrected only at order $M_s^4/\mu^3$, we
can then factor out the dependence on the Fermi surface splitting, since the only
dimensionful quantity in the integrand is then $\delta\mu$. Defining, for
$\delta\mu_2=\delta\mu_3=\delta\mu$,
\begin{equation}
{\cal{K}}_{udusus}(\q{3}{b},\q{3}{b},\q{2}{d},\q{2}{e},\q{2}{f},\q{2}{a}) 
=\frac{1}{\delta\mu^4}\bar{K}_{322}(\q{3}{b},\q{3}{b},\q{2}{d},\q{2}{e},\q{2}{f},\q{2}{a}) 
\;,
\end{equation}
and using $|\q{I}{}|=\eta\delta\mu_I$, for all the momentum vectors, we find 
that $\bar{K}_{322}$ is given by 
\begin{equation}
\bar{K}_{322}(\q{3}{b},\q{3}{b},\q{2}{d},\q{2}{e},\q{2}{f},\q{2}{a}) =
\frac{1}{8}\int_0^1 dx_2(1-x_2)\int_0^1 dy_1
\int_0^{1-y_1} dy_2 \\
\Re
 e \frac{\eta^2|{\bf a}(x_2,y_1,y_2)|^2+3(1-2y_1)^2}{\left[\eta^2|{\bf a}(x_2,y_1,y_2)|^2-(1-2y_1)^2+i\epsilon\right]^3}
 \label{Kbar322}\;,
\end{equation}
where 
\be
{\bf a}=x_2\left(\hat{{\bf q}}_2^d-\hat{{\bf q}}_2^e\right)-\left(y_1\hat{{\bf q}}_3^b + y_2\hat{{\bf q}}_2^d +
(1-y_1-y_2)(\hat{\bf q}_2^d-\hat{\bf q}_2^e+\hat{{\bf q}}_2^f)\right)\ .
\ee 
\end{widetext}
For general arguments we were not able to do the integrals
that remain in  (\ref{Kbar322})
analytically and therefore evaluated it numerically. Since
$\bar{K}_{322}(\q{3}{b},\q{3}{b},\q{2}{d},\q{2}{e},\q{2}{f},\q{2}{a})$ is
the limit of the function
$\bar{K}_{322}(\q{3}{b},\q{3}{b},\q{2}{d},\q{2}{e},\q{2}{f},\q{2}{a},\epsilon)$ as
$\epsilon\rightarrow 0$, we numerically evaluated
the integral appearing in (\ref{Kbar322}) at four values of $\epsilon$ and
extrapolated (using a cubic polynomial to fit the values) to $\epsilon = 0$.
Finally 
\be
\bgma_{322}=\gamma_{322}\delta\mu^4
\label{GammaBarDefn}
\ee
is found by 
summing 
$\bar{K}_{322}$ evaluated with all possible choices of momentum vectors  
$(\q{3}{b},\q{3}{b},\q{2}{d},\q{2}{e},\q{2}{f},\q{2}{a})$ 
satisfying (\ref{condition for K}) and multiplying this sum by $3/2$.

\section{Results \label{results}}

\subsection{Generalities}

We shall assume that $\Delta_1=0$ throughout this section. As described previously,
this simplification is motivated by the fact that $\Delta_1$ describes the pairing
of $d$ and $s$ quarks, whose Fermi surfaces are twice as far apart from each
other as either is from that of the $u$ quarks.
We shall focus most of our attention on exchange symmetric
crystal structures, as defined at the end of Section IV, 
in which the polyhedra
defined by 
$\{\hat{\bf q}_2\}$ and
$\{\hat{\bf q}_3\}$ are related by a rigid rotation.  In Section VI.D we will discuss one example
in which  $\{\hat{\bf q}_2\}$ and
$\{\hat{\bf q}_3\}$ are not exchange symmetric, and we have evaluated others.  However,
as none that we have investigated prove to be favorable, we shall 
make the notational simplifications that come with assuming 
that $\{\hat{\bf q}_2\}$ and
$\{\hat{\bf q}_3\}$ are exchange symmetric, as this implies $\alpha_2=\alpha_3\equiv\alpha$,
$P_2={\rm dim}\setq{2}{}=P_3={\rm dim}\setq{3}{}\equiv P$,
$\beta_2=\beta_3\equiv\beta$ and $\gamma_{322}=\gamma _{233}$.  The final
simplification we employ is to make the approximation that 
$\delta\mu_2=\delta\mu_3\equiv \delta\mu = M_s^2/(8\mu)$. As described in Section II.A,
this approximation is corrected by terms of order $M_s^4/\mu^3$.  Upon making
all these simplifying assumptions and approximations, the free energy (\ref{omega 3})
reduces to
\begin{equation}
\begin{split}
\Omega(\Delta_2,&\Delta_3)=\frac{2{\mu}^2}{\pi^2}\Biggl[
P\alpha(\delta\mu)\bigl(\Delta_2^2+\Delta_3^2\bigr)\\
&\phantom{}+\ha\frac{1}{\delta\mu^2}\bigl(\bar{\beta}(\Delta_2^4+\Delta_3^4)
 +\bar{\beta}_{32}\Delta_2^2\Delta_3^2\bigr)\\
 &\!\!\!\!\!+\frac{1}{3}\frac{1}{\delta\mu^4}\bigl(\bar{\gamma}(\Delta_2^6+\Delta_3^6)
 +\bar{\gamma}_{322}(\Delta_2^2\Delta_3^4+\Delta_2^4\Delta_3^2)\bigr)\Biggr]
 \label{congruent equal free energy}\;,
\end{split}
\end{equation}
where $\bar\beta$, $\bar\gamma$, $\bar\beta_{32}$ and $\bar\gamma_{322}$ 
are the dimensionless constants that we must calculate for each crystal structure
as described in Section V, and
where  the $\delta\mu$-dependence of $\alpha$ is given by Eq.~(\ref{AlphaEqn2}).
 
 In order to find the extrema of $\Omega(\Delta_2,\Delta_3)$ in $(\Delta_2,\Delta_3)$-space,
 it is convenient to write
$(\Delta_2,\Delta_3)$ as $\sqrt{2}(\Delta_r\cos \theta,\Delta_r \sin\theta)$
in terms of which the free energy (\ref{congruent equal free energy}) is given by 
\begin{equation}
\begin{split}
\Omega &(\Delta_r,\theta)=\\ 
 &\phantom{}\frac{2{\mu}^2}{\pi^2}
 \Biggl[2 P\alpha(\delta\mu)\Delta_r^2
     +\frac{2}{\delta\mu^2}\bar{\beta}\Delta_r^4
	 +\frac{8}{3\delta\mu^4}\bar{\gamma}\Delta_r^6\\
     &\!\!\!\!\!\!
	 +\left(\frac{\Delta_r^4}{2\delta\mu^2}
	  (\bar{\beta}_{32}-2\bar{\beta})
     +\frac{2\Delta_r^6}{3\delta\mu^4}(\bar{\gamma}_{322}-3\bar{\gamma})\right)\sin^22\theta\Biggr]
 \label{radial free energy}.
\end{split}
\end{equation}
Because ${\sin}^2(2\theta)$ has extrema only at $\theta={\pi}/{4}$ and $\theta = 0,\pi/2$,
we see that extrema of $\Omega(\Delta_2,\Delta_3)$ either have $\Delta_2=\Delta_3=\Delta$,
or have one of $\Delta_2$ and $\Delta_3$ vanishing.  The latter class of extrema are
two-flavor crystalline phases.  We are interested in the solutions with 
$\Delta_2=\Delta_3=\Delta$.   The stability of these solutions relative
to those with only one of the $\Delta$'s nonzero appears to be controlled by the sign of 
the factor that multiplies $\sin^22\theta$ in 
(\ref{radial free energy}). However, we shall show in Appendix \ref{stability}
that the three-flavor crystalline phases that we construct, with 
$\Delta_2=\Delta_3=\Delta$, are electrically neutral whereas the two-flavor solutions
in which only one of the $\Delta$'s is nonzero are not.  
Setting $\Delta_2=\Delta_3=\De$, the free energy becomes
\begin{equation}
\Omega(\Delta) =  \frac{2{\mu}^2}{\pi^2}
 \left[2P\alpha(\delta\mu)\Delta^2
+\frac{\Delta^4}{2\delta\mu^2}\bar{\beta}_{\rm eff}
+\frac{\Delta^6}{3\delta\mu^4}\bar{\gamma}_{\rm eff}\right]
 \label{effective free energy}\;, 
\end{equation}
where we have defined
\be
\begin{split}
\bar{\beta}_{\rm eff}&=2\bar{\beta} + \bar{\beta}_{32}\\
\bar{\gamma}_{\rm eff}&=2\bar{\gamma}+2\bar{\gamma}_{322}\ .
\end{split}
\label{BetaGammaEff}
\ee
We have arrived at a familiar-looking sextic order Ginzburg-Landau free energy
function, whose coefficients we will evaluate for specific crystal structures
in VI.B and VI.D.  First, however, we review the physics described by
this free energy depending on whether $\bar\beta_{\rm eff}$ and
$\bar\gamma_{\rm eff}$ are positive or negative.

If $\bar{\beta}_{\rm eff}$ and $\bar{\gamma}_{\rm eff}$ are both positive,
the free energy (\ref{effective free energy}) describes a second order
phase transition between the crystalline color superconducting phase
and the normal phase at the $\delta\mu$ at which $\alpha(\delta\mu)$
changes sign.  From (\ref{AlphaEqn2}), this critical point occurs
where $\delta\mu=0.754\, \Delta_{\rm 2SC}$.  In plotting our results,
we will take the CFL gap to be $\Delta_0=25$~MeV,
making $\Delta_{\rm 2SC}=2^{1/3}\Delta_0=31.5$~MeV.
Recalling that $\delta\mu=M_s^2/(8\mu)$,
this puts the second order phase transition at 
\be
\frac{M_s^2}{\mu}\Biggr|_{\rm \alpha=0}=6.03\, \Delta_{\rm 2SC} = 7.60\,\Delta_0 = 190.0~{\rm MeV}\ .
\label{AlphaZero}
\ee
(The authors of Refs.~\cite{Casalbuoni:2005zp,Mannarelli:2006fy} 
neglected to notice that it is $\Delta_{\rm 2SC}$, rather
than the CFL gap $\Delta_0$, that occurs in Eqs.~(\ref{AlphaEqn}) 
and (\ref{AlphaEqn2}) and therefore controls 
the $\delta\mu$ at 
which $\alpha=0$. 
In analyzing the crystalline phase in isolation, this
is immaterial since either $\Delta_0$ or $\Delta_{\rm 2SC}$ could be taken as
the parameter defining the strength of the interaction between quarks.  However,
in Section VI.E we shall compare the free energies of the CFL, gCFL and crystalline
phases, and in making this comparison it is important to take into account that
$\Delta_{\rm 2SC}=2^{1/3}\Delta_0$.)  For values of $M_s^2/\mu$ that
are smaller than (\ref{AlphaZero}) (that is, lower densities), $\alpha<0$ and
the free energy is minimized by a nonzero $\Delta=\Delta_{\rm min}$ given by
\begin{equation}
\Delta_{\rm min} = \delta\mu \sqrt{\frac{1}{2\bar\gamma_{\rm eff}}\Bigl(-\bar\beta_{\rm eff}
          +\sqrt{\bar\beta_{\rm eff}^2-8P\alpha(\delta\mu)\bar\gamma_{\rm eff}}
          \Bigr)}\label{Deltamin2nd}\ ,
\end{equation}
and thus describes a crystalline color superconducting phase.

If $\bar\beta_{\rm eff}<0$ and $\bar\gamma_{\rm eff}>0$, then the free energy
(\ref{effective free energy}) describes a first order
phase transition between unpaired and crystalline quark matter occurring
at 
\be
\alpha = \alpha_* =  \frac{3\,\bar\beta^2_{\rm eff}}{32\, P \bar\gamma_{\rm eff}}\ .
\label{AlphaStar}
\ee
At this positive value of $\alpha$, the function $\Omega(\Delta)$ has a minimum at $\Delta=0$
with $\Omega=0$, 
initially rises quadratically with increasing $\Delta$,
and is then turned back downward by
the negative quartic term before being turned back upwards again
by the positive sextic term, yielding a second
minimum at
\be
\Delta=\delta\mu \sqrt{ \frac{3\, |\bar\beta_{\rm eff }|}{4\,\bar\gamma_{\rm eff}}}\ ,
\label{DeltaAlphaStar}
\ee
also with $\Omega=0$, which describes a crystalline color superconducting
phase.
For $\alpha<\alpha_*$, the crystalline phase is favored over unpaired quark matter.
Eq.~(\ref{AlphaEqn2}) must be used to determine the value of $\delta\mu$, and
hence $M_s^2/\mu$, at which 
$\alpha=\alpha_*$ and the first order phase transition occurs.  If $\alpha_*\ll 1$,
the transition occurs at a value of $M_s^2/\mu$ that is greater than (\ref{AlphaZero}) by
a factor $(1+\alpha_*)$.  See Fig.~\ref{OmegaVsDeltaFig} 
for an explicit example of plots of $\Omega$ versus $\Delta$
for various values of $\alpha$ for one of the crystal structures that we analyze 
in Section VI.D which turns out to have a first order phase transition.

A necessary condition for the Ginzburg-Landau approximation to be quantitatively
reliable is that the sextic term in the free energy is small
in magnitude compared to the quartic, meaning
that $\De^2\ll \dmu^2 |\bar\beta_{\rm eff}/\bar\gamma_{\rm eff}|$.  If the transition
between the unpaired and crystalline phases
is second order, then this condition is satisfied close enough to the transition
where $\De\rightarrow 0$.  However, if $\bar\beta_{\rm eff}<0$ and $\bar\gamma_{\rm eff}>0$,
making the transition first order, we see from (\ref{DeltaAlphaStar}) that at the
first order transition itself $\Delta$ is large enough to make the quantitative
application of the Ginzburg-Landau approximation marginal.  This is a familiar 
result, coming about whenever a Ginzburg-Landau approximation predicts
a first order phase transition
because at the first order phase transition the quartic
and sextic terms are balanced against each other.   Even though
it is quite a different problem, it is worth recalling the Ginzburg-Landau analysis
of  the crystallization of a solid from a liquid~\cite{Chaikin}. 
There too, a Ginzburg-Landau analysis predicts a first-order
phase transition and thus predicts its own quantitative downfall.
However, it remains important as a qualitative guide: it predicts
a body-centered cubic crystal structure, and most elementary
solids are body-centered cubic near their melting point.
We shall find that our Ginzburg-Landau analysis predicts a first
order phase transition; knowing that it is therefore at the edge
of its quantitative reliability, we shall focus in Sections VI.E and VII on qualitative
conclusions.

If $\bar\gamma_{\rm eff}<0$, then the Ginzburg-Landau expansion
of the free energy to sextic order in (\ref{effective free energy}) is
not bounded from below.  The transition must be first order, with 
higher-than-sextic order terms making the free energy bounded.
In this circumstance, all we learn from (\ref{effective free energy}) 
is that the transition is first order; we cannot obtain an estimate
of the transition point or of $\Delta$ at the first order transition.
Even though $\bar\gamma$ is negative for many crystal
structures~\cite{Bowers:2002xr}, in all the three-flavor crystalline
phases that we present 
in Section VI.D  
we find that $\bar\gamma_{322}$ is positive and sufficiently large
that $\bar\gamma_{\rm eff}=2\bar\gamma+2\bar\gamma_{322}$ is
positive.  We therefore need not discuss the $\bar\gamma_{\rm eff}<0$ 
case any further.

\subsection{Two plane wave structure}

We begin with the simplest three-flavor  ``crystal'' structure in which
$\setq{2}{}$ and $\setq{3}{}$ each contain only a single vector, yielding
a condensate 
\begin{equation}
\Delta_{\cf} = e^{2i\q{2}\cdot\rr}\Delta_2\epsilon_{2\alpha\beta}\epsilon_{2ij}
+e^{2i\q{3}\cdot\rr}\Delta_{3}\epsilon_{3\alpha\beta}\epsilon_{3ij}\;,
\label{TwoPlaneWave}
\end{equation}
in which the $\langle us\rangle$ and $\langle ud\rangle$ condensates are
each plane waves.  As explained in the previous subsection, we shall
seek solutions with $\Delta_2=\Delta_3=\Delta$.
We begin with such a simple ansatz both because it has been analyzed
previously in Refs.~\cite{Casalbuoni:2005zp,Mannarelli:2006fy} and because
it will yield a qualitative lesson which will prove extremely helpful in winnowing the space
of multiple plane wave crystal structures.

\begin{figure}[t]
\includegraphics[width=3in,angle=0]{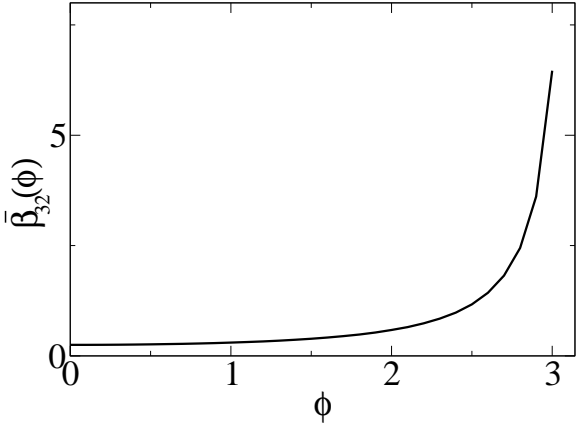}
\caption{$\bar\beta_{32}(\phi)=\bar J_{32}(\phi)$ 
for the two plane wave ``crystal'' structure with condensate
(\ref{TwoPlaneWave}). $\phi$ is the angle between $\q{2}{}$ and $\q{3}{}$.
For more complicated crystal structures, $\bar\beta_{32}$ is given by 
the sum in (\ref{BetaBarResult}), meaning that it is a sum of $\bar J_{32}(\phi)$
evaluated at various values of $\phi$ corresponding to the various angles between
a vector in $\setq{2}{}$ and a vector in $\setq{3}{}$.
}
\label{Jbar23vsphi}
\end{figure}

Let us now walk through the evaluation of all the coefficients in the free energy
(\ref{effective free energy}) for this two-plane wave structure.  First, $P=1$ (one vector
in each of $\setq{2}{}$ and $\setq{3}{}$) and as always $\alpha(\dmu)$ is given by (\ref{AlphaEqn2}).
Next, we obtain the results for $\bar\beta_2=\bar\beta_3$
and $\bar\gamma_2=\bar\gamma_3$ from the analysis of the single plane wave condensate
in the two flavor model of Ref.~\cite{Bowers:2002xr}:
\begin{equation}
\begin{split}
\bar{\beta}_2&=\frac{1}{4}\frac{1}{\eta^2-1}=0.569\\
\bar{\gamma}_2&=\frac{1}{32}\frac{\eta^2+3}{(\eta^2-1)^3}=1.637\;.
\end{split}
\end{equation}
We now turn to $\bar{\beta}_{32}$ and $\bar{\gamma}_{322}$ 
which describe the interaction between
the $\langle us\rangle$ and $\langle ud\rangle$ condensates and which we have
calculated in Section V.  In general, $\bar\beta_{32}$ is given by (\ref{BetaBarResult})
but in this instance since $\setq{2}{}$ and $\setq{3}{}$ each contain only a single
vector the sum in this equation reduces simply to
\be
\bar\beta_{32}=\bar J_{32}(\phi)
\ee
where $\phi$ is the angle between $\q{2}{}$ and $\q{3}{}$ and
where $\bar J_{32}(\phi)$ is given in Eq.~(\ref{J23 evaluated}).
$\bbta_{32}$  is plotted as a function of $\phi$ in Fig. \ref{Jbar23vsphi}.
For this simple crystal structure, $\bbta_{32}$ was calculated previously
in Refs.~\cite{Casalbuoni:2005zp,Mannarelli:2006fy}. 

\begin{figure}[t]
\includegraphics[width=3in,angle=0]{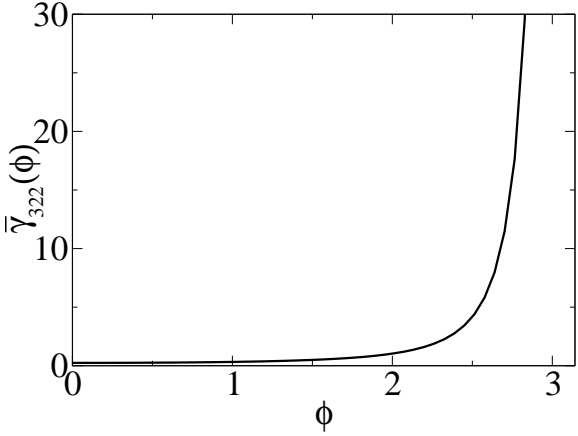}
\caption{$\bar\gamma_{322}(\phi)$ for the two plane wave ``crystal''
structure with condensate (\ref{TwoPlaneWave}). $\phi$ is the angle between
$\q{2}{}$ and $\q{3}{}$.  $\bar\gamma_{322}(0)=0.243$ and $\bar\gamma_{322}(\phi)$
increases monotonically with $\phi$.
}
\label{3by2Kbar23vsphi}
\end{figure}

Turning to $\bgma_{322}$, this is given by 
\be
\bgma_{322}=\frac{3}{2}\bar K_{322}(\q{3}{},\q{3}{},\q{2}{},\q{2}{},\q{2}{},\q{2}{})
\label{Kbar322special}
\ee
where $\bar K_{322}$ is given by Eq.~(\ref{Kbar322}).  As occurred
in the evaluation of $\beta_{32}$,
the sum over ${\bf q}$-vectors in the general expression (\ref{AlphaBetaGamma})
has reduced to evaluating $\bar K_{322}$
just once, because $\setq{2}{}$ and $\setq{3}{}$ each contain only a single
vector.   For the special case where the last four arguments
of $\bar K_{322}$ are the same, as in (\ref{Kbar322special}), $\bar K_{322}$ depends
only on $\phi$, the angle between $\q{2}{}$ and $\q{3}{}$, and the 
integrals in (\ref{Kbar322}) can all be evaluated analytically, yielding
\begin{widetext}
\begin{equation}
\begin{split}
\bar{K}_{322}(\phi) =& 
  \frac{1}{64\left(\eta\cos\frac{\phi}{2}\right)^{3}\left(\eta^2\ssq-1\right)^{3/2}}\\
  &\times\left[
  \eta^2 \arctan(b(\phi))\ssq
  +\frac{\left(\eta^2\ssq-1\right)b(\phi)}{1+b(\phi)^2}
  +\frac{b(\phi)\left(b(\phi)^2-1\right)}{\left(b(\phi)^2+1\right)^2}\right]\ ,
\label{K23 evaluated}
\end{split}
\end{equation}
\end{widetext}
where 
\be
b(\phi)=\frac{\sqrt{\eta^2\ssq-1}}{\eta\cos\frac{\phi}{2}}\ .
\ee
$\bgma_{322}$
is plotted as a function of $\phi$ in Fig. \ref{3by2Kbar23vsphi}.

We note that for any angle $\phi$, both $\bar{\beta}_{32}$ and
$2\bar{\gamma}_{322}$ are
positive quantities which when added to the positive $2\bar{\beta}$ and
$2\bar{\gamma}$ give positive $\bar\beta_{\rm eff}$ and $\bar\gamma_{\rm eff}$, 
respectively.  Hence, we see that upon making 
this two plane wave ``crystal'' structure ansatz we find a second order
phase transition between the crystalline and unpaired phases, for all
choices of the angle $\phi$.   We also note that both $\bar\beta_{32}(\phi)$
and $\bar\gamma_{322}(\phi)$ increase monotonically with $\phi$, and
diverge as $\phi\rightarrow\pi$.  This tells us that within this two plane
wave ansatz, the most favorable orientation is $\phi=0$, namely
$\q{2}{}\parallel\q{3}{}$.  Making this choice yields the smallest possible
$\bar\beta_{\rm eff}$ and $\bar\gamma_{\rm eff}$ within this ansatz, and
hence the largest possible $\Delta$ and condensation energy, again within
this ansatz.
The divergence at $\phi\rightarrow\pi$
tells us that choosing $\q{2}{}$ and
$\q{3}{}$ precisely antiparallel exacts an infinite free energy price
in the combined Ginzburg-Landau and weak-coupling
limit in which $\Delta\ll \delta\mu,\Delta_0\ll \mu$,
meaning that in this limit if we chose $\phi=\pi$ we find $\Delta=0$.
Away from the Ginzburg-Landau limit, when the pairing rings
on the Fermi surfaces widen into bands, choosing $\phi=\pi$ exacts a finite
price meaning that $\Delta$  is nonzero but smaller than that for any other
choice of $\phi$.  All these results
confirm conclusions drawn in Refs.~\cite{Casalbuoni:2005zp,Mannarelli:2006fy}
based only upon the result for
$\bar\beta_{32}(\phi)$.

The high cost of choosing $\q{2}{}$ and
$\q{3}{}$ precisely antiparallel can be understood 
qualitatively as arising from the fact that in this case the
ring of states on the $u$-quark Fermi surface that ``want to'' pair
with $d$-quarks coincides precisely with the ring that ``wants to''
pair with $s$-quarks~\cite{Mannarelli:2006fy}.  
(For example, if $\q{2}{}$ and $\q{3}{}$ point in the $-z$ and $+z$ directions,
$\Delta_2$ ($\Delta_3$) describes pairing between $s$-quarks ($d$-quarks) 
within a ring in the northern hemisphere
of the $s$- ($d$-)Fermi surface and $u$-quarks within a ring in the southern hemisphere
of the $u$-Fermi surface.
The rings on the $u$-Fermi surface coincide, as illustrated in 
Fig.~2 of Ref.~\cite{Mannarelli:2006fy}.)
In the most favorable case
within the two-plane wave ansatz,
where $\q{2}{}\parallel\q{3}{}$, the two pairing rings on the $u$-quark Fermi
surface are centered on antipodal points~\cite{Mannarelli:2006fy}.
(For example, if $\q{2}{}$ and $\q{3}{}$ both point in the $+z$ direction,
$\Delta_2$ ($\Delta_3$) describes pairing of $s$-quarks ($d$-quarks) 
within a ring in the southern (northern) hemisphere
of the $s$- ($d$-)Fermi surface and $u$-quarks within rings in the (northern) southern hemisphere
of the $u$-Fermi surface.)

The simple two plane wave ansatz (\ref{TwoPlaneWave}) has been
analyzed in the same NJL model that we employ upon making
the weak-coupling approximation but without
making a Ginzburg-Landau approximation in Ref.~\cite{Mannarelli:2006fy}.
All the qualitative lessons that we have learned from the Ginzburg-Landau approximation
remain valid, including the favorability of the choice $\phi=0$,
but we learn further that in the two plane wave case
the Ginzburg-Landau approximation always underestimates $\Delta$~\cite{Mannarelli:2006fy}.
We also see from Ref.~\cite{Mannarelli:2006fy} that the $\Delta$ at which the
Ginzburg-Landau approximation breaks down shrinks as $\phi\rightarrow\pi$.
We can understand this result as follows. The sextic term in the free
energy (\ref{effective free energy}) is small compared to the quartic term 
only if $\Delta^2\ll \dmu^2\bar\beta_{\rm eff}/\bar\gamma_{\rm eff}$,
making this a necessary condition for the quantitative validity of the 
Ginzburg-Landau approximation.  
As $\phi\rightarrow\pi$, $\bgma_{\rm eff}$ diverges more strongly than $\bbta_{\rm eff}$:
from (\ref{J23 evaluated}) and (\ref{Kbar322}) we find that as $\phi\rightarrow\pi$,
\begin{equation}
\begin{split}
\bbta_{\rm eff}\sim\bar{J}_{32} &\sim
\frac{\pi}{8\eta\sqrt{\eta^2-1}}\left(\frac{1}{\cos(\frac{\phi}{2})}\right) \\
\bgma_{\rm eff}\sim \frac{3}{2}\bar{K}_{322} &\sim
\frac{3\pi}{256\eta(\sqrt{\eta^2-1})^3}\left(\frac{1}{\cos(\frac{\phi}{2})}\right)^3\;.
\end{split}
\end{equation}
Therefore the Ginzburg-Landau calculation predicts that its own breakdown
will occur at a $\Delta$ that decreases with increasing $\phi$, as found 
in Ref.~\cite{Mannarelli:2006fy} by explicit comparsion with a calculation that does
not employ the Ginzburg-Landau approximation.

\subsection{Implications for more plane waves: qualitative principles for favorable crystal structures}

In this subsection we ask what lessons we can learn from the
evaluation of the Ginzburg-Landau coefficients for the two plane wave
``crystal'' structure in Section VI.B 
for crystal
structures with more than one vector in $\setq{2}{}$ and $\setq{3}{}$.

First,
we can conclude that $\bar\beta_{32}$ is positive for {\it any} choice of 
$\setq{2}{}$ and $\setq{3}{}$.  The argument is simple: $\bar\beta_{32}$ is
given in general by (\ref{BetaBarResult}), a sum over $\bar J_{32}$ evaluated at a host of angles
corresponding to all angles between a vector in $\setq{2}{}$ and $\setq{3}{}$.
But, we see from Fig.~\ref{Jbar23vsphi} that $\bar J_{32}$ is positive at any angle. 

Second, we
{\it cannot} draw such a conclusion about $\gamma_{322}$. 
This coefficient is a sum over contributions of the form 
$\bar K_{322}(\q{3}{b},\q{3}{b},\q{2}{d},\q{2}{e},\q{2}{f},\q{2}{a})$ where the last four momentum
vector arguments, selected from $\setq{2}{}$, must satisfy (\ref{condition for K}).  
The calculation in Section VI.B whose result is plotted in Fig.~\ref{3by2Kbar23vsphi}
only demonstrates that those contributions in which the four $\q{3}{}$ arguments
are selected to all be the same vector are positive.
For any crystal structure in which $\setq{2}{}$ contains two or more vectors, there
are other contributions to $\bar\gamma_{322}$ that we have not evaluated
in this section
which depend on one $\q{2}{}$ vector and several $\q{3}{}$ vectors, and thus on
more than one angle.  We know of instances where individual contributions
$\bar K_{322}(\q{3}{b},\q{3}{b},\q{2}{d},\q{2}{e},\q{2}{f},\q{2}{a})$ 
in crystal structures that we describe below {\it are} negative. However, we
have found no crystal structure for which $\bar\gamma_{322}$ is negative.  

The final lesson we learn is that crystal structures in which any of the
vectors in $\setq{2}{}$ are close  to antiparallel to any of the vectors
in $\setq{3}{}$ are strongly disfavored.  (The closer to antiparallel, the
worse, with the free energy penalty for $\Delta\neq 0$ 
diverging for the precisely antiparallel case, driving $\Delta$ to zero.)
If a vector 
in $\setq{2}{}$ is antiparallel (or close to antiparallel) to one in $\setq{3}{}$,
this yields infinite (or merely large) positive contributions to $\bar\beta_{32}$ and
to $\bar\gamma_{322}$ and hence to $\bar\beta_{\rm eff}$
and $\bar\gamma_{\rm eff}$.   In the case of $\bar\beta_{32}$, these large 
positive contributions cannot be cancelled since all contributions are positive.
In the case of $\bar\gamma_{322}$, negative contributions are possible but we
know of no instances of divergent negative contributions to $\bar\gamma_{322}$
or indeed to any other coefficient in the Ginzburg-Landau expansion.
The divergent positive contributions are associated with the tangential intersection (in
the case of $\bar\beta$ and $\bar\gamma$~\cite{Bowers:2002xr})
or coincidence (in the case of of $\bar\beta_{32}$ and $\bar\gamma_{322}$)
of pairing rings on Fermi surfaces.  We know of no configuration of rings that
leads to an infinitely favorable (as opposed to unfavorable) free energy in
the combined Ginzburg-Landau and weak-coupling limits.  So, although
we do not have a proof that the divergent positive contributions to $\bar\gamma_{322}$
arising as vectors in $\setq{2}{}$ and $\setq{3}{}$ approach one another's
antipodes are uncancelled, we also see no physical argument for
how this could conceivably arise.   Certainly in all example crystal structures
that we have considered, 
$\bar\beta_{32}$ and $\bar\gamma_{322}$ and hence $\bar\beta_{\rm eff}$
and $\bar\gamma_{\rm eff}$ diverge as vectors in $\setq{2}{}$ and $\setq{3}{}$ approach one 
another's antipodes.

We can now summarize the qualitative principles that we have arrived at 
for constructing favorable crystal structures for three-flavor crystalline
color superconductivity.  First, as described in
Section III the sets $\setq{2}{}$ and $\setq{3}{}$ should each
separately be chosen to yield crystal structures which, seen
as separate two-flavor crystalline phases, are as favorable as possible.
In Section III we have reviewed the results of Ref.~\cite{Bowers:2002xr}
for how this should be done, and the conclusion that the most
favored 
$\setq{2}{}$ or $\setq{3}{}$ in isolation consists of eight
vectors pointing at the corners of a cube.  Second, the new addition
in the three-flavor case is the qualitative principle that 
$\setq{2}{}$ and $\setq{3}{}$ should be rotated with respect to each other
in such a way as to best keep vectors in one set away from the antipodes
of vectors in the other set.

\begin{table*}[h]
\begin{tabular}{||c||l||c||}
\hline\hline
Structure & \qquad\qquad\qquad\qquad\qquad\qquad\qquad\qquad Description & Largest Angle\\
\hline\hline
2PW & $\setq{2}{}$ and $\setq{3}{}$ coincide; each contains one vector. (So, 2 plane waves
with $\q{2}{}\parallel\q{3}{}$.)& $0^\circ$\\
\hline\hline
SqX & $\setq{2}{}$ and $\setq{3}{}$ each contain two antiparallel vectors. The four vectors together form& $90^\circ$ \\
& a square; those from $\setq{2}{}$ and those from $\setq{3}{}$ each form
one stroke of an ``X''.&\\ 
\hline
Tetrahedron & 
$\setq{2}{}$ and $\setq{3}{}$ each contain two vectors. The four together form
a tetrahedron. & $109.5^\circ$\\ 
\hline\hline
2Triangles & $\setq{2}{}$ and $\setq{3}{}$ coincide; each contains three vectors forming a triangle.
&$120^\circ$ \\
 \hline\hline
Cube X & $\setq{2}{}$ and $\setq{3}{}$ each contain 4 vectors forming a rectangle. The 8 
vectors together  &
$109.5^\circ$\\
See Eq.~(\ref{CubeXStructure}) & form a cube. The 2 rectangles intersect to look like an ``X'' 
if viewed end-on. &\\
\hline
 2Tet & $\setq{2}{}$ and $\setq{3}{}$ coincide; each contains four vectors forming a tetrahedron.& 
 $109.5^\circ$\\
\hline
Twisted & $\setq{2}{}$ and $\setq{3}{}$ each contain four vectors forming a square  which
could be one face of &$143.6^\circ$\\
Cube & a cube. Instead, the eight vectors together form the polyhedron obtained by twisting&\\ 
& the top face of a cube by $45^\circ$ relative to its bottom face.&\\ 
\hline\hline
2Octa90xy & $\setq{2}{}$ and $\setq{3}{}$ each contain 6 vectors forming an octahedron.
The $\setq{2}{}$ vectors point & $135^\circ$ \\
& along the positive and negative axes. The $\setq{3}{}$-octahedron is rotated relative to the & \\
&  $\setq{2}{}$-octahedron 
by $90^\circ$ about the $(1,1,0)$-axis. & \\
\hline
2Octa45xyz & $\setq{2}{}$ and $\setq{3}{}$ each contain 6 vectors forming an octahedron.
The $\setq{2}{}$ vectors point & $143.6^\circ$ \\
& along the positive and negative axes. The $\setq{3}{}$-octahedron is rotated relative to the & \\
& $\setq{2}{}$-octahedron 
by $45^\circ$ about the $(1,1,1)$-axis. & \\
\hline\hline
2Cube45z & $\setq{2}{}$ and $\setq{3}{}$ each contain 8 vectors forming a cube. The
$\setq{2}{}$ vectors point along    & $143.6^\circ$\\
See Eq.~(\ref{2Cube45zStructure}) & $(\pm 1,\pm 1,\pm 1)$. The $\setq{3}{}$-cube is rotated relative to that
by $45^\circ$ about the $z$-axis.&\\
\hline
2Cube45xy & $\setq{2}{}$ and $\setq{3}{}$ each contain 8 vectors forming a cube. The
$\setq{2}{}$ vectors point along  & $154.5^\circ$\\ 
& $(\pm 1,\pm 1,\pm 1)$. The $\setq{3}{}$-cube is rotated
relative to that
by $45^\circ$ about the $(1,1,0)$-axis.&\\
\hline\hline
\end{tabular}
\caption{Descriptions of the crystal structures whose Ginzburg-Landau coefficients are given in 
Table~II.
The third column is the largest angle between any vector in $\setq{2}{}$ and any vector
in $\setq{3}{}$.  Other things being equal, we expect that the larger the largest angle, meaning
the closer vector(s) in $\setq{2}{}$ get to vector(s) in $\setq{3}{}$, the bigger the $\bar\beta_{32}$
and $\bar\gamma_{322}$ and hence the bigger the $\bar\beta_{\rm eff}$ and
$\bar\gamma_{\rm eff}$, and hence the less favorable the structure.}
\end{table*}

\begin{table*}[h]
\begin{tabular}{||c||c|c||c||c|c||c||c|c||}
\hline\hline
Structure & $\bar{\beta}$ & $\bar{\beta}_{32}$ & $\bar{\beta}_{\rm eff}$ &
$\bar{\gamma}$ & $\bar{\gamma}_{322}$ & $\bar{\gamma}_{\rm eff}$ & $\alpha_*$ & 
$\Delta(\alpha_*)/\dmu$ \\
\hline\hline
2PW      &0.569   &0.250  &1.388   &1.637   &0.243   &3.760 & 0& 0 \\
\hline
SqX          &0.138   &1.629  &1.906   &1.952   &2.66   &9.22 & 0 & 0\\
Tetrahedron  &-0.196  &2.146  &1.755   &1.450    &7.21   &17.29 & 0 & 0\\
\hline
2Triangles & -1.976& 4.647 & 0.696 &1.687 & 13.21 & 29.80 &0 &0 \\
\hline
CubeX        &-10.981 &6.961  &-15.001 &-1.018  &19.90  &37.76 & 0.140 & 0.548\\
2Tet      &-5.727  &7.439  &-4.015  &4.350   &30.35  &69.40 & 0.0054 & 0.208\\
Twisted Cube &-16.271 &12.445 &-20.096 &-37.085 &315.5 &556.8& 0.0170 &0.165 \\
\hline
2Octa90xy & -31.466 & 18.665 & -44.269 & 19.711 & 276.9 & 593.2 & 0.0516 & 0.237 \\
2Octa45xyz & -31.466 & 19.651 & -43.282 & 19.711 & 297.7 & 634.9 & 0.0461 & 0.226 \\
\hline
2Cube45z &-110.757&36.413 &-185.101&-459.24 &1106.  &1294.& 0.310 & 0.328 \\
2Cube45xy&-110.757&35.904 &-185.609&-459.24 &11358. &21798.& 0.0185 &0.0799\\
\hline\hline
\end{tabular}
\caption{Ginzburg-Landau coefficients for three-flavor crystalline color superconducting
phases with various crystal structures, described in Table~I. $\alpha_*$ 
is the $\alpha$ at which the transition from unpaired quark matter to a given crystalline
phase occurs: $\alpha_*=0$ if $\beta_{\rm eff}>0$ and the transition is second order;
$\alpha_*$ is given by (\ref{AlphaStar}) if  $\beta_{\rm eff}<0$ and the transition is first order.
For a first order transition, $\Delta(\alpha_*)$, 
given in (\ref{DeltaAlphaStar}), is the magnitude of the gap at the transition.
}
\end{table*}

\subsection{Multiple plane waves \label{multiple}}

In Table I we describe 11 different crystal structures that we have
analyzed, and in Table II we give the coefficients
that specify each Ginzburg-Landau free energy (\ref{congruent equal free energy}).
The $\bar\beta$'s and $\bar\gamma$'s were calculated as described in 
Ref.~\cite{Bowers:2002xr}; $\bar\beta_{32}$'s and $\bar\gamma_{322}$'s
were calculated as described in Section V.  We also give the
combinations $\bar\beta_{\rm eff}$ and $\bar\gamma_{\rm eff}$ defined 
in (\ref{effective free energy}) that specify the free energy as in (\ref{BetaGammaEff}).
In those cases in which $\bar\beta_{\rm eff}<0$, the phase transition
between the crystalline phase and the unpaired phase is first order,
occurring where $\alpha=\alpha_*$ with $\alpha_*$ given by (\ref{AlphaStar}).
At the first order phase transition, unpaired quark matter with $\Delta=0$
and crystalline quark matter with $\Delta(\alpha_*)$ given in (\ref{DeltaAlphaStar})
have the same free energy. We give both $\alpha_*$ and $\Delta(\alpha_*)$ 
in Table II.  

The first row of the Tables describes the simple ``crystal structure'' analyzed
in detail in Section VI.B, in which both $\setq{2}{}$ and $\setq{3}{}$\
contain just a single vector, with $\q{2}{}\parallel\q{3}{}$ as we have seen
that this is the most favorable choice for the angle between $\q{2}{}$
and $\q{3}{}$.  This condensate carries a baryon number 
current which means that
the unpaired gapless fermions (in ``blocking regions'' in momentum
space~\cite{Alford:2000ze,Bowers:2001ip}) must carry a current that is equal in magnitude but opposite
in direction~\cite{Alford:2000ze}.  
The analysis of this ``crystal structure'' in Sections VI.B and VI.C
has proved instructive, giving us qualitative insight that we
shall use to understand all the other crystal structures. However, in
all rows in the Tables other than the first we have chosen
crystal structures with condensates that carry no net current, meaning
that the gapless fermions need carry no current. There is nothing in
our mean-field analysis that precludes condensates carrying a net
current, but we do not analyze them here primarily because it simplifies our task
but also because we expect that, beyond mean-field theory, a phase containing
gapless fermions carrying a net current is unlikely to be the favored ground state.

Let us next examine the last two rows of the Tables. Here, we consider two
crystal structures in which $\setq{2}{}$ and $\setq{3}{}$ each contain 
eight vectors forming cubes.  Since the cube is the most favorable
two-flavor crystal structure according to the analysis of Ref.~\cite{Bowers:2002xr},
evident in the large negative $\bar\beta$ and $\bar\gamma$ for both
these crystal structures in Table II,
this should be a good starting point.  We cannot have
the two cubes coincident, as in that case there are vectors from $\setq{2}{}$ and
vectors from $\setq{3}{}$ separated by a 180$^\circ$ angle, yielding infinite
positive contributions to both $\bar\beta_{32}$ and $\bar\gamma_{322}$. 
So, we rotate the $\setq{3}{}$-cube relative to the $\setq{2}{}$ cube, in
two different ways in the 2Cube45z and 2Cube45xy crystal structures described
in Table I.  

We explain explicitly in Appendix B why translating one cube relative
to the other in position space by a fraction of a lattice spacing does not
alleviate the problem: a relative rotation of the $\langle us \rangle$
and $\langle ud \rangle$ condensates is required.
Qualitatively, this reflects the nature of the difficulty that  occurs when a
$\setq{2}{}$ vector is opposite to a $\setq{3}{}$ vector. It can be thought
of as arising because the $\langle us \rangle$ and $\langle ud \rangle$
condensates both want to ``use'' those up quarks lying on the same ring on
the up Fermi surface. It therefore makes sense that a relative rotation is required.
Quantitatively, what we show in Appendix B is that $\Omega$ does not
change if we translate the $\langle us \rangle$ condensate relative
to the $\langle ud \rangle$ condensate.

We see in Table I that in the 2Cube45z structure, the largest angle between
vectors in $\setq{2}{}$ and $\setq{3}{}$ is 143.6$^\circ$ whereas in the
2Cube45xy structure, that largest angle
is $154.5^\circ$ meaning that the rotation we have chosen does a less good job
of keeping $\setq{2}{}$-vectors away from the antipodes of $\setq{3}{}$
vectors.  Correspondingly, we see in Table II that 2Cube45xy has a much larger
$\bar\gamma_{322}$ and hence $\bar\gamma_{\rm eff}$, and hence has
a first order phase transition occurring at a smaller $\alpha_*$ and with
a smaller $\Delta(\alpha_*)$.  This is an example confirming our 
general principle that, other things being equal, crystal structures in
which  $\setq{3}{}$ vectors come closer to $\setq{2}{}$ vectors will be
disfavored.  According to this principle, the 2Cube45z crystal structure
should be particularly favorable as it employs the relative rotation
between the two cubes that does the best possible job of keeping
them apart. 

We now turn to crystal structures with fewer than 16 plane waves.  By having
fewer than 8 plane waves in $\setq{2}{}$ and $\setq{3}{}$, we are no longer
optimizing the two-flavor $\bar\beta$ and $\bar\gamma$.  However, with
fewer vectors it is possible to keep the $\setq{2}{}$- and $\setq{3}{}$-vectors farther 
away from each other's antipodes.  We list two crystal structures in which
$\setq{2}{}$ and $\setq{3}{}$ have 6 waves forming octahedra. These
are not particularly favorable two-flavor structures --- $\bar\gamma$ is positive
rather than being large and negative for the cube. 2Octa45xyz has the
same largest angle between $\setq{2}{}$- and $\setq{3}{}$-vectors
as 2Cube45z, but its significantly more positive $\bar\beta$ and $\bar\gamma$
make it significantly less favorable.  Choosing the 2Octa90xy structure
instead reduces the largest angle between $\setq{2}{}$- and $\setq{3}{}$-vectors
from 143.6$^\circ$ to 135$^\circ$, which improves $\bar\beta_{\rm eff}$
and $\bar\gamma_{\rm eff}$, but only slightly.

We investigate three crystal structures in which $\setq{2}{}$ and $\setq{3}{}$
each contain 4 vectors.  Among these, the Twisted Cube is strongly disfavored
by its significantly larger largest angle between $\setq{2}{}$- and $\setq{3}{}$-vectors.
CubeX and 2Tet are both constructed by choosing $\setq{2}{}$ and $\setq{3}{}$
as subsets containing half the vectors from a cube. In the 2Tet structure, 
we choose the tetrahedra coincident since this does the best job of
keeping vectors in $\setq{2}{}$ and $\setq{3}{}$ away from each other's
antipodes. (Choosing the two tetrahedra so that their union forms a cube
is the worst possible choice, as vectors in $\setq{2}{}$ and $\setq{3}{}$
are then antipodal.)  In the CubeX structure, we choose the two rectangles
such that their union forms a cube, as this does the best job of reducing
the largest angle between vectors in $\setq{2}{}$ and $\setq{3}{}$; making
the rectangles coincident would have been the worst possible choice.
CubeX and 2Tet have the same largest angle, but they differ considerably
in that the $\setq{2}{}$ and $\setq{3}{}$ rectangles that make up CubeX are
more favorable two-flavor structures (lower $\bar\beta$ and $\bar\gamma$)
than the tetrahedra that make up 2Tet.  We see from Table II that the CubeX structure,
with only 8 vectors in total, is particularly favorable: it is not possible to tell
from Table II whether it is more or less favorable than 2Cube45z, since one
has the larger $\alpha_*$ while the other has the larger $\Delta(\alpha^*)$. We shall
evaluate their free energies below, and confirm that they are indeed comparable,
and that these two structures have the lowest free energy of any in the Tables.

In the remaining rows of the Tables, we 
investigate one crystal structure in which $\setq{2}{}$ and $\setq{3}{}$
each contain 3 vectors, and two in which each contain 2 vectors. These structures
all have positive $\bar\beta_{\rm eff}$, and hence second order phase transitions, and
so are certainly not favored.

Inspecting the results in Table II shows that in all cases where we have investigated
different three-flavor crystal structures built from the same $\setq{2}{}$ and $\setq{3}{}$, 
the one with the relative rotation between the two polyhedra that yields the smaller
largest angle between vectors in $\setq{2}{}$ and $\setq{3}{}$ is favored.  And, in
all cases where we have investigated two crystal structures with same largest
angle between vectors in $\setq{2}{}$ and $\setq{3}{}$, the one built from the
more favorable two-flavor crystal structure is favored.  We thus find no exceptions
to the qualitative principles we described in Section VI.C.  However, these 
qualitative principles certainly do not explain all the features of the results
in Table II. For example, we have no qualitative understanding of why
2Cube45z and 2Cube45xy have such similar $\bar\beta_{32}$, whereas
2Cube45xy has a much larger $\bar\gamma_{322}$ as expected. For 
example, we have no qualitative understanding of why $\bar\gamma_{322}$
increases much more in going from 2Cube45xy to 2Cube45z than it
does in going from 2Octa90xy to 2Octa45xyz.  The calculations must
be done; the qualitative principles are a good guide, but not a substitute.

\begin{figure*}[t]
\includegraphics[width=5in,angle=0]{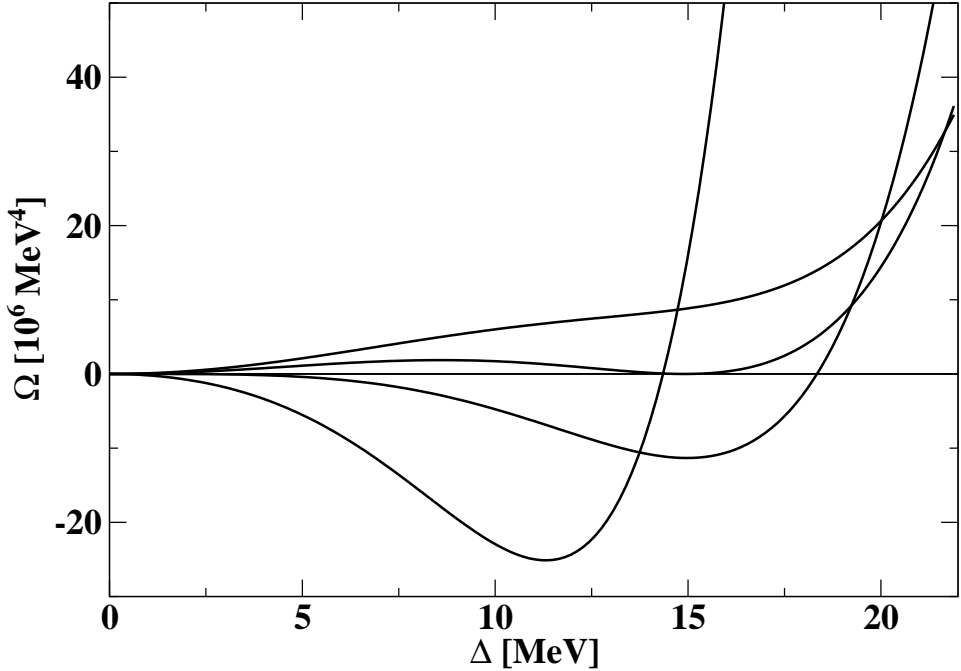}
\caption{Free energy $\Omega$ vs. $\Delta$ for the CubeX crystal structure, 
described in Table II, at four values of $M_s^2/\mu$. From top curve to bottom curve,
as judged from the left half of the figure, the curves are
$M_s^2/\mu=240$,  $218.61$, $190$, and $120$~MeV, corresponding to $\alpha=0.233$,
$0.140$, 0, $-0.460$.
The first order phase transition occurs at $M_s^2/\mu=218.61$~MeV.  The values
of $\Delta$ and $\Omega$ at the minima of curves like these are what we
plot in Figs.~\ref{deltavsx} and \ref{omegavsx}.
}
\label{OmegaVsDeltaFig}
\end{figure*}

\begin{figure*}[t]
\includegraphics[width=5in,angle=0]{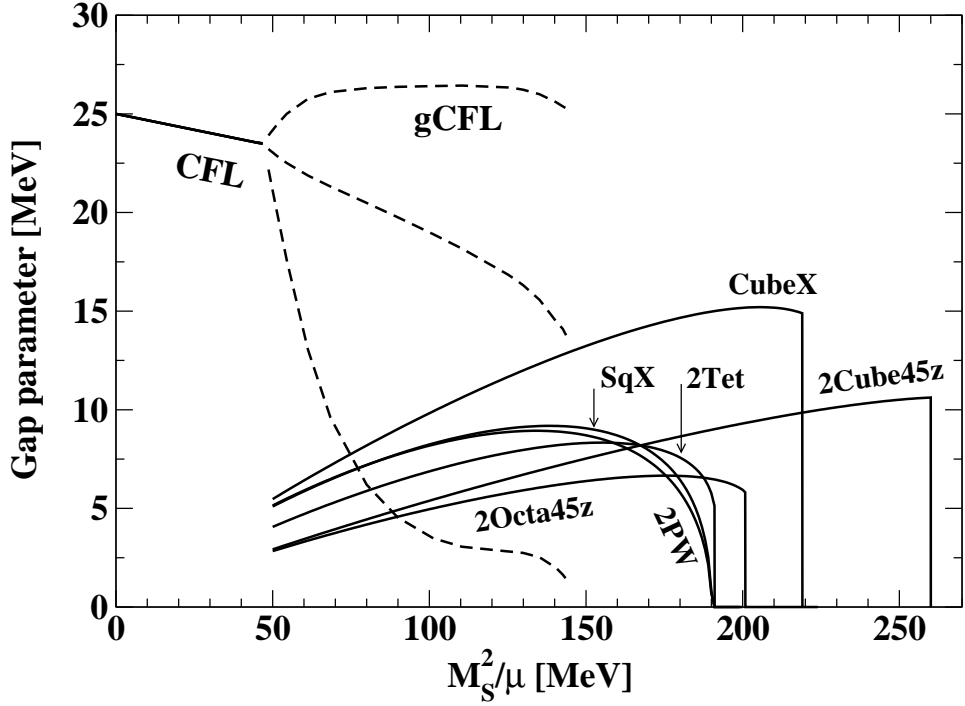}
\caption{Gap parameter $\Delta$ versus $M_s^2/\mu$ for three-flavor crystalline color superconducting
phases with various crystal structures.  The crystal structures are described in Table II.
For comparison, we also show the CFL gap parameter and the gCFL gap
parameters $\Delta_{1}$, $\Delta_2$ and $\Delta_3$~\cite{Alford:2003fq,Alford:2004hz}.  Recall that the
splitting between Fermi surfaces is proportional to $M_s^2/\mu$, and that small (large)
$M_s^2/\mu$ corresponds to high (low) density.
}
\label{deltavsx}
\end{figure*}

\begin{figure*}[t]
\includegraphics[width=5in,angle=0]{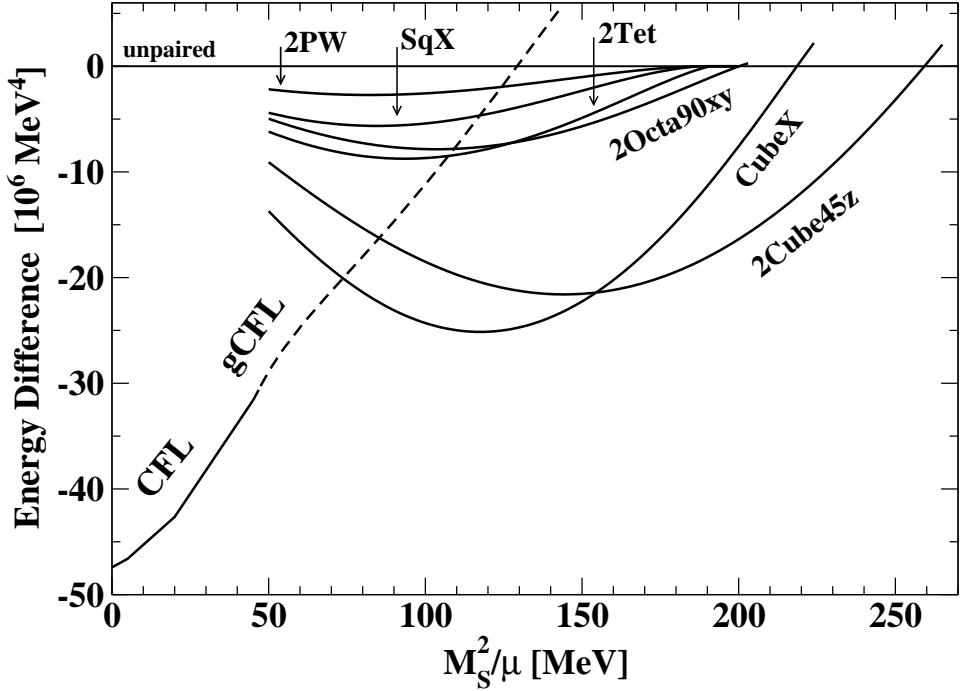}
\caption{Free energy $\Omega$ versus $M_s^2/\mu$ for
the three-flavor crystalline color superconducting phases
with various crystal structures whose gap parameters are
plotted in Fig.~\ref{deltavsx}. The crystal structures are
described in Table II.  Recall that the gCFL phase
is known to be unstable, meaning that in
the regime where the gCFL phase free energy is plotted, the true ground state of three-flavor quark
matter must be some phase whose free energy lies below the dashed line. We see that the
three-flavor crystalline color superconducting quark matter phases with the most favorable 
crystal structures that we have found, namely 2Cube45z and CubeX 
described in (\ref{2Cube45zStructure}) and (\ref{CubeXStructure}), have sufficiently 
robust condensation energy (sufficiently negative $\Omega$) that they are candidates
to be the ground state of three-flavor quark matter over a wide swath of $M_s^2/\mu$,
meaning over a wide range of densities.
}
\label{omegavsx}
\end{figure*}

The final crystal structure that we describe is one in which $\setq{2}{}$
is a cube while $\setq{3}{}$ is an octahedron, with the six $\setq{3}{}$-vectors
pointing at the centers of the faces of the $\setq{2}{}$-cube.  So, if
the $\setq{2}{}$-vectors are taken to point along the $(\pm 1,\pm 1,\pm 1)$
directions then the $\setq{3}{}$ vectors point along the positive and negative
axes. We chose to investigate this structure because it seems 
particularly symmetric and because it has an unusually small largest
angle between vectors in $\setq{2}{}$ and $\setq{3}{}$ given the
large number of vectors in total: $125.3^\circ$.  Because
$\setq{2}{}$ and $\setq{3}{}$ are not congruent, $\bar\beta_2\neq \bar\beta_3$
and $\bar\gamma_2\neq\bar\gamma_3$.  All these coefficients can be
found in Table II.  We find $\bar\beta_{32}=24.510$, 
$\bar\gamma_{322}=419.9$ and
$\bar\gamma_{233}=4943.$  Because $\{\hat{\bf q}_2\}$ and $\{\hat{q}_3\}$ are not
exchange symmetric, the general argument that we gave in Section VI.A for why
extrema of $\Omega(\Delta_2,\Delta_3)$ --- i.e. solutions
to the gap equations --- occur at $\Delta_2=\Delta_3$ does not apply. However, we find
that at the solution $\Delta_2$ and $\Delta_3$ differ by
less than 20\%. The large values of $\bar\gamma_{233}$ and $\bar\gamma_{322}$
make this crystal structure quite unfavorable --- even though it has
a (weak) first order phase transition, its free energy turns out
to be comparable only to that of the 2PW structure, far above the free energy
of the favored CubeX and 2Cube45z structures. Furthermore, the
arguments of Appendix A do not apply to a crystal structure like this, meaning
that we do not expect this solution with $\Delta_2\neq\Delta_3$ to be 
neutral.  For this reason, and because it appears to be free-energetically
unfavorable anyway,  we will not investigate
it further.  We cannot say whether choosing $\setq{2}{}$ and $\setq{3}{}$
to not be exchange symmetric generically yields an unfavorable crystal structure, as
we have not investigated many possibilities.

We have certainly not done an exhaustive search of three-flavor crystal structures.
For example, we  have only scratched the surface in investigating
structures in which $\setq{2}{}$ and $\setq{3}{}$ are not exchange symmetric.
We have investigated the structures that are the best that we
can think of according to the qualitative principles described in Section VI.C.
Readers should feel free to try others.  (We are confident that in 2Cube45z
we have found the most favorable structure obtained by rotating one 
cube relative to another. We are not as confident that CubeX is the best possible
structure with fewer than 8+8 vectors.)  As we shall see in Section VI.E, however, the
two most favorable structures that we have found, 2Cube45z and CubeX,
are impressively robust and do a very good job of making the case
that three-flavor crystalline color superconducting phases are the ground
state of cold quark matter over a wide range of  densities.  If 
even better crystal structures can be found, this will only further strengthen this case.

\subsection{Free energy comparisons}

We can now evaluate and plot the gap parameter $\Delta$ and free energy
$\Omega(\Delta)$ for all the crystal structures described in Table I, whose
Ginzburg-Landau coefficients are given in Table II.  
For a given
crystal structure, $\Omega(\Delta)$ is given by Eq.~(\ref{effective free energy}), 
with $\bar\beta_{\rm eff}$ and $\bar\gamma_{\rm eff}$ taken from Table II.
The quadratic coefficient $\alpha$ is related to $\delta\mu$ by Eq.~(\ref{AlphaEqn2}).
Recall that we have made the approximation that
$\delta\mu_2=\delta\mu_3=\delta\mu=M_s^2/(8\mu)$, valid up to corrections
of order $M_s^3/\mu^4$.   At any value of $M_s^2/\mu$, we can
evaluate $\alpha(\delta\mu)$ and hence $\Omega(\Delta)$, 
determine $\Delta$ by minimizing $\Omega$,
and finally evaluate the free energy $\Omega$ at the minimum.  
In Fig.~\ref{OmegaVsDeltaFig},
we give an example of $\Omega(\Delta)$ for various $M_s^2/\mu$
for one crystal structure with a first
order phase transition (CubeX), illustrating how the first order phase
transition is found, and how the $\Delta$ solving the gap
equations --- i.e. minimizing $\Omega$ --- is found .
We plot $\Delta$
and $\Omega$ at the minimum versus $M_s^2/\mu$ in Figs.~\ref{deltavsx} and \ref{omegavsx}
for some of the crystal structures in Tables I and II. 

In Figs.~\ref{deltavsx} and \ref{omegavsx}, we show two examples of 
crystal structures for which the phase transition to the unpaired state
is second order: 2PW and SqX. (See Table I for descriptions of these
structures.)  The second order phase transition occurs at 
$M_s^2/\mu = 7.60 \Delta_0 = 190.0$~MeV, where $\alpha=0$. (See Eq.~(\ref{AlphaZero}).)
We show four examples of crystal structures with first
order phase transitions, occurring where $\alpha=\alpha_*>0$ meaning
at some $M_s^2/\mu > 190.0$~MeV.
We show the two most favorable structures that we
have found: CubeX and 2Cube45z. And, we show two examples (2Tet
and 2Octa90xy) of
structures with first order phase transitions that are more favorable than
the structures with a second order transition, but less favorable than
CubeX and 2Cube45z.

In Figs.~\ref{deltavsx} and \ref{omegavsx}, we have chosen
the interaction strength between quarks such that the CFL gap parameter at $M_s=0$
is $\Delta_0=25$~MeV.  However, our results for both the
gap parameters and the condensation energy for any
of the crystalline phases can easily be scaled to any value of
$\Delta_0$.  We saw in Section V that the quartic and sextic coefficients in
the Ginzburg-Landau free energy do not depend on $\Delta_0$.  
And, recall from Eq.~(\ref{AlphaEqn2}) that
$\Delta_0$ enters $\alpha$ only through the combination $\Delta_{\rm 2SC}/\delta\mu$,
where $\Delta_{\rm 2SC}=2^{\frac{1}{3}}\Delta_0$ and $\delta\mu= M_s^2/(8\mu)$.
This means that if we pick a $\Delta_0\neq 25$~MeV, the 
curves describing the gap parameters for the
crystalline phases in Fig.~\ref{deltavsx} are precisely unchanged
if we rescale both the vertical and horizontal axes proportional to
$\Delta_0/25$~MeV.  In the case of Fig.~\ref{omegavsx}, the vertical
axis must be rescaled by $(\Delta_0/25~{\rm MeV})^2$.
Of course, the weak-coupling
approximation $\Delta_0\ll\mu$, which we have used for example in simplifying
the propagators in (\ref{largemu}), will break down if we scale $\Delta_0$ to be too large.
We cannot evaluate up to what $\Delta_0$ we can scale our results
reliably without doing a calculation that goes beyond the weak-coupling
limit. However, such calculations have been done for the gCFL phase 
in Ref.~\cite{Fukushima:2004zq},
where it turns out that the gaps and
condensation energies plotted Figs.~\ref{deltavsx} and \ref{omegavsx} 
scale with $\Delta_0$ and $\Delta_0^2$ to good accuracy 
for $\Delta_0\leq 40$~MeV with $\mu=500$~MeV, but the scaling is
significantly less accurate for $\Delta_0=100$~MeV.
Of course, for $\Delta_0$ as large as $100$~MeV, any quark matter in
a compact star is likely to be in the CFL phase.  Less symmetrically
paired quark matter, which our results suggest is in a crystalline
color superconducting phase, will occur in compact stars only
if $\Delta_0$ is smaller, in the range where our results can be 
expected to scale well.

The qualitative behavior of $\Delta$ at smaller $M_s^2/\mu$, 
well to the left of the unpaired/crystalline phase transitions in Fig.~\ref{deltavsx}, can
easily be understood.  
The quadratic, quartic and sextic coefficients in
the free energy (\ref{effective free energy}) are $\alpha(\delta\mu)$, 
$\beta_{\rm eff}=\bar\beta_{\rm eff}/\delta\mu^2$
and $\gamma_{\rm eff}=\bar\gamma_{\rm eff}/\delta\mu^4$.  If $\alpha$ tended to
a constant at small $\delta\mu$, then the solution 
$\Delta_{\rm min}$ that minimizes $\Omega$ would be proportional to
$\delta\mu$.
(See Eq.~(\ref{Deltamin2nd}).) In fact, from (\ref{AlphaEqn2}) 
we see that $\alpha\propto\log\delta\mu$ at small $\delta\mu$,
meaning that, according to (\ref{Deltamin2nd}), $\Delta_{\rm min}$ should vanish
slightly more slowly than linear as  
 $M_s^2/\mu\propto \delta\mu \rightarrow 0$, as in Fig.~\ref{deltavsx}.  
And, since the $\Delta$'s vanish for $\delta\mu\rightarrow 0$, so do
the condensation energies of Fig.~\ref{omegavsx}.

Fig.~\ref{deltavsx} can be used to evaluate the validity of the Ginzburg-Landau 
approximation.  The simplest criterion is to compare the $\Delta$'s for the
crystalline phases to the CFL gap parameter $\Delta_0$.
This is the correct criterion in the vicinity of the 2nd order phase  transition point,
where $\delta\mu = M_s^2/(8\mu) \approx \Delta_0$.  Well to the left, it is more
appropriate to compare the $\Delta$'s for the crystalline phase to $\delta\mu=M_s^2/(8\mu)$.
By either criterion, we see that all the crystal structures with first order
phase transitions (including the two that are most favored) have $\Delta$'s
that are large enough that the Ginzburg-Landau approximation is at the edge
of its domain of validity, a result which we expected based on the general
arguments in Section VI.A.  Note that the Ginzburg-Landau
approximation is controlled for those structures with second order phase
transitions only near the second order phase transition, again a result that
can be argued for on general grounds.

Fig.~\ref{omegavsx} makes manifest one of the central conclusions of our
work.  The three-flavor crystalline color superconducting
phases with the two most favored crystal structures that we have
found are robust by any measure.  Their condensation energies 
reach about half that of  the CFL phase at $M_s=0$, remarkable given
that in the CFL phase pairing occurs over the whole of all three Fermi surfaces.
Correspondingly, these two crystal structures are favored over the wide
range of $M_s^2/\mu$ seen in Fig.~\ref{omegavsx} and
given in Eq.~(\ref{WideLOFFWindow}).   

Taken literally, Fig.~\ref{omegavsx} indicates that within the regime (\ref{WideLOFFWindow})
of the phase diagram occupied by crystalline  color superconducting quark matter,
the 2Cube45z phase is favored at lower densities and the CubeX phase is favored at 
higher densities.   Although, as detailed in Sections VI.C and VI.D,  we do have
qualitative arguments why 2Cube45z and CubeX are favored over other phases,
we have no qualitative argument why one should be favored over the other.
And, we do not trust that the Ginzburg-Landau approximation is sufficiently 
quantitatively reliable to trust the conclusion that one phase is favored at higher
densities while the other is favored at lower ones.  We would rather leave the
reader with the conclusion that these are the two most favorable phases we have
found, that both are robust, that the crystalline color superconducting phase
of three-flavor quark matter with one crystal structure or the other occupies
a wide swath of the QCD phase diagram,
and that their free energies are similar enough to each other that
it will take a beyond-Ginzburg-Landau calculation to compare them reliably.

\section{Conclusions, Implications, and Future Work}

We have evaluated the gap parameter and free energy for three-flavor
quark matter in crystalline color superconducting phases with varied
crystal structures, within a Ginzburg-Landau approximation.  Our 
central results are shown in Figs.~\ref{deltavsx} and \ref{omegavsx}. 
Descriptions of the crystal structures that we have investigated, together with the
coefficients for the Ginzburg-Landau free energy (\ref{effective free energy})
for each structure, are given in Tables I and II.    

We have found two qualitative rules that guide our understanding of
what crystal structures are favored in three-flavor crystalline quark matter.
First, the $\langle ud \rangle$ and $\langle us \rangle$ condensates separately
should be chosen to have favorable free energies, as evaluated in the
two-flavor model of Ref.~\cite{Bowers:2002xr}.  Second, the $\langle ud \rangle$
and $\langle us \rangle$ condensates should be rotated relative to each other
in such a way as to maximize the angles between the wave vectors
describing the crystal structure of the $\langle ud \rangle$ condensate and the antipodes of
the wave vectors describing the $\langle us \rangle$ condensate.  This
second qualitative rule can be understood as minimizing the ``competition''
between the two condensates for up quarks on the up Fermi surface,
as first elucidated in a simpler setting in Ref.~\cite{Mannarelli:2006fy}.

Fig.~\ref{omegavsx} shows that over most of the range of $M_s^2/\mu$ where it was once considered
a possibility, the gCFL phase can be replaced by a {\it much} more favorable three-flavor
crystalline color superconducting phase.  However, Fig.~\ref{omegavsx} also
indicates that it is hard to find a crystal structure which yields a crystalline
phase that has lower free energy than the gCFL phase at the lowest values
of $M_s^2/\mu$ (highest densities) in the ``gCFL window'', closest to the
CFL$\rightarrow$gCFL transition.  This narrow window where the gCFL curve
remains the lowest curve in Fig.~\ref{omegavsx} is therefore the most likely place
in the QCD phase diagram in which to find the gCFL phase augmented
by current-carrying meson condensates described in 
Refs.~\cite{Kryjevski:2005qq,Gerhold:2006dt}.
Except within this window, the crystalline color superconducting
phases with either the CubeX or the 2Cube45z crystal structure
provide an attractive resolution to the instability of the gCFL phase.

The three-flavor crystalline color superconducting
phases with the CubeX and 2Cube45z crystal structures 
have condensation energies that can be as large as half
that of the CFL phase.  This robustness  makes them the
lowest free energy phase that we know of, and hence
a candidate for the ground state of QCD, over a wide range
of densities.  To give a sense of the implications of the range of 
$M_s^2/\mu$ 
over which crystalline color superconductivity is favored,
given by Eq.~(\ref{WideLOFFWindow}) and shown in Fig.~\ref{omegavsx},
if we suppose that $\Delta_0=25$~MeV and $M_s=250$~MeV,
the window (\ref{WideLOFFWindow}) translates to $240 {\rm MeV} < \mu < 847 {\rm MeV}$.
With these choices of parameters, then, the 
lower part of this range of $\mu$ (higher part of the range of $M_s^2/\mu$) is certainly
superseded by nuclear matter.  And, the high end of this range extends far
beyond the $\mu\sim 500$~MeV characteristic of the quark matter at the densities
expected 
at the very center of compact stars. Our result therefore suggests that if compact stars
have quark matter cores, it is entirely reasonable to suppose that the {\it entire}
quark matter core could be in a crystalline color superconducting phase.
Of course, if $\Delta_0$ is larger, say $\sim 100$~MeV, the entire quark matter
core could be in the CFL phase.  And, there are reasonable values
of $\Delta_0$ and $M_s$ for which the outer layer of a possible quark matter
core would be in a crystalline phase while the inner core would not.  We do not know
$\Delta_0$ and $M_s$ well enough to answer the question of what phases of
quark matter occur in compact stars. However, our results add 
the possibility that as much as all of the 
quark matter in a compact star could be in a crystalline color superconducting phase
to the menu of options that must ultimately be winnowed by confrontation
with astrophysical observations.

We have identified two particularly favorable crystal structures, using
the qualitative rules described above and by direct calculation.
We do not believe that our Ginzburg-Landau approximation is sufficiently
accurate to trust its determination of which of these two structures is
more favorable. For this reason, we wish to leave the reader with a picture
of both
the 2Cube45z and CubeX 
crystal structures in position space. In the 2Cube45z phase, the color-flavor
and position space  dependence of the condensate, defined in (\ref{spin structure})
and (\ref{precisecondensate}),
is given by
\begin{widetext}
\begin{equation}
\begin{split}
\Delta_{CF}(x)_{\cf} =& \epsilon_{2\alpha\beta}\epsilon_{2ij} \, 
2\Delta \Biggl[ 
\cos \frac{2\pi}{a} \left( x+y+z\right) + \cos \frac{2\pi}{a} \left(-x+y+z\right)\\
&\qquad\qquad\qquad\qquad\qquad+\cos \frac{2\pi}{a} \left(x-y+z\right) 
+ \cos \frac{2\pi}{a}\left(-x-y+z\right) \Biggr]\\
&+ \epsilon_{3\alpha\beta}\epsilon_{3ij} \, 2\Delta \Biggl[
\cos \frac{2\pi}{a} \left( \sqrt{2} x +z\right) + \cos \frac{2\pi}{a} \left(\sqrt{2} y+z\right)\\
&\qquad\qquad\qquad\qquad\qquad+\cos \frac{2\pi}{a} \left(-\sqrt{2}y+z\right) 
+ \cos \frac{2\pi}{a}\left(-\sqrt{2}x+z\right) \Biggr]\ ,
\label{2Cube45zStructure}
\end{split}
\end{equation}
\end{widetext}
where $\alpha$ and $\beta$ ($i$ and $j$) are color (flavor) indices and 
where
\be
a = \frac{\sqrt{3}\pi}{q} = \frac{4.536}{\delta\mu} = \frac{\mu}{1.764 M_s^2}
\label{LatticeSpacing}
\ee
is the lattice spacing of the face-centered cubic crystal structure.  For
example, with $M_s^2/\mu=100, 150, 200$~MeV the lattice
spacing is  $a=72, 48, 36$~fm.  Eq.~(\ref{2Cube45zStructure}) can equivalently
be written as
\begin{equation}
\Delta_{CF}(x)_{\cf} = \epsilon_{2\alpha\beta}\epsilon_{2ij} \Delta_2(\rr)
+ \epsilon_{3\alpha\beta}\epsilon_{3ij} \Delta_3(\rr)\ ,
\end{equation}
with (\ref{2Cube45zStructure}) providing the expressions for $\Delta_2(\rr)$
and $\Delta_3(\rr)$.
A three-dimensional contour
plot that can be seen as depicting 
either $\Delta_2(\rr)$ or $\Delta_3(\rr)$ separately
can be found in Ref.~\cite{Bowers:2002xr}.
We have not found an informative way of depicting the entire condensate
in a single contour plot.  Note also that in (\ref{2Cube45zStructure})
and below in our description of the CubeX phase, we make an arbitrary
choice for the relative position of $\Delta_3(\rr)$ and $\Delta_2(\rr)$.  We show
in Appendix B that one can be translated relative to the other at no cost
in free energy.  Of course, as we have investigated in detail in Section VI,
rotating one relative to the other changes the
Ginzburg-Landau coefficients $\bar\beta_{32}$ and $\bar\gamma_{322}$ 
and hence the free energy.

\begin{figure*}[t]
\vspace{-0.35in}
\includegraphics[width=3.7in,angle=0]{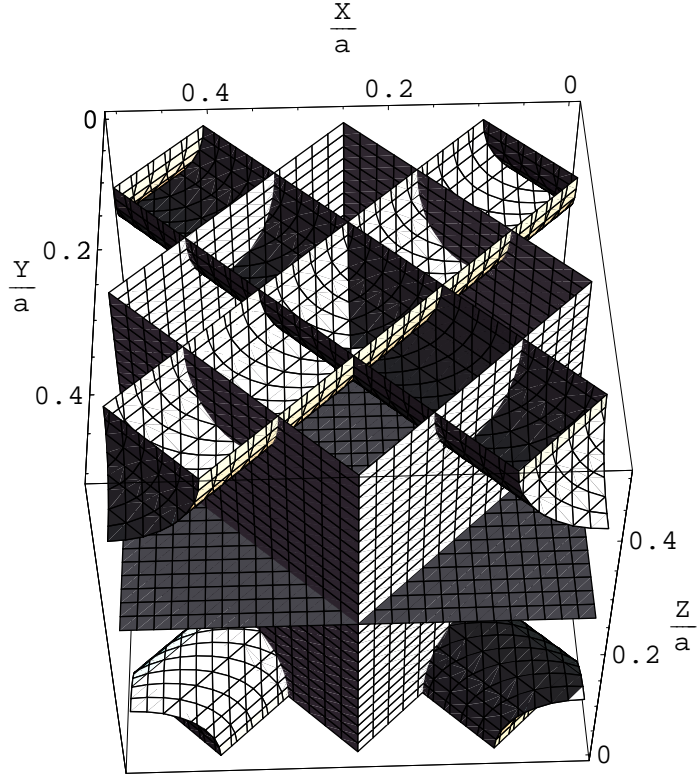}
\vspace{-0.15in}
\caption{The CubeX crystal structure of Eq.~(\ref{CubeXStructure}). 
The figure extends from 0 to $a/2$ in the $x$, $y$ and $z$ directions.
Both $\Delta_2(\rr)$ and $\Delta_3(\rr)$
vanish at the horizontal plane.  $\Delta_2(\rr)$ vanishes on the darker vertical planes,
and $\Delta_3(\rr)$ vanishes on the lighter vertical planes.   
On the upper
(lower) dark cylinders and the lower (upper) two small corners of dark cylinders,
$\Delta_2(\rr)=  +3.3 \Delta$ ($\Delta_2(\rr)=  -3.3 \Delta$).
On the upper
(lower) lighter cylinders and the lower (upper) two small corners of lighter cylinders,
$\Delta_3(\rr)=  -3.3 \Delta$ ($\Delta_3(\rr)=  +3.3 \Delta$).  
Note that the largest value of $|\Delta_I(\rr)|$ is $4\Delta$, occurring along lines 
at the centers of the cylinders. The lattice spacing is $a$ when one takes into
account the signs of the condensates; if one looks only at $|\Delta_I(\rr)|$, 
the lattice spacing is $a/2$. $a$ is given in (\ref{LatticeSpacing}).
In (\ref{CubeXStructure}) and
hence in this figure, we have made a particular choice for the relative position of
$\Delta_3(\rr)$ versus $\Delta_2(\rr)$.  We show in Appendix B that one can be
translated relative to the other with no cost in free energy. 
}
\label{CubeXContours}
\end{figure*}

In the CubeX phase, the color-flavor and position space dependence of 
the condensate is given by
\begin{widetext}
\begin{equation}
\begin{split}
\Delta_{CF}(x)_{\cf} =& \epsilon_{2\alpha\beta}\epsilon_{2ij} \, 
2\Delta \Biggl[ 
\cos \frac{2\pi}{a} \left( x+y+z\right) + \cos \frac{2\pi}{a}\left(-x-y+z\right) \Biggr]\\
&+ \epsilon_{3\alpha\beta}\epsilon_{3ij} \, 2\Delta \Biggl[
\cos \frac{2\pi}{a} \left( -x+y+z\right) + \cos \frac{2\pi}{a}\left(x-y+z\right) \Biggr]\ .
\label{CubeXStructure}
\end{split}
\end{equation}
\end{widetext}
We provide a depiction of this condensate in Fig.~\ref{CubeXContours}.

The gap parameter $\Delta$ is large enough in both the 2Cube45z and CubeX
phases that the Ginzburg-Landau approximation that we have used to obtain
our results is being pushed to the limits of its validity. Therefore, although
we expect that the qualitative lessons that we have learned about 
the favorability of crystalline phases in three-flavor quark matter
are valid, and expect that the relative favorability of the 2Cube45z and CubeX
structures and the qualitative size of their $\Delta$ and condensation energy
are trustworthy, we do not expect quantitative reliability of our results.
There is therefore strong motivation to analyze crystalline
color superconducting quark matter with these two crystal structures
without making a Ginzburg-Landau approximation.  It will be very interesting to
see whether the Ginzburg-Landau approximation underestimates
$\Delta$ and the condensation energy for the crystalline
phases with CubeX and 2Cube45z
crystal structures, as it does for the much simpler 2PW structure (in which
$\Delta_2(\rr)$ and $\Delta_3(\rr)$ are each single plane waves)~\cite{Mannarelli:2006fy}.

Even prior to having a beyond-Ginzburg-Landau analysis available, having
an ansatz (actually, two ans\"atze) for the crystal structure and a good qualitative
guide to the scale of $\Delta$ and $\Omega(\Delta)$ should allow significant progress
toward the calculation of astrophysically relevant observables. For example, it
would be interesting to evaluate the effects of a crystalline color superconducting
core on the rate at which a neutron star  cools by neutrino emission. The 
specific heat of crystalline color superconducting quark matter is linear
with $T$ because of the presence of
gapless quark excitations at the boundaries of the regions in momentum
space where there are unpaired quarks~\cite{Casalbuoni:2003sa}. Calculating
the heat capacity of the CubeX and 2Cube45z structures should therefore
yield only quantitative changes relative to that for unpaired 
quark matter, unlike in the gCFL case where the heat capacity is parametrically
enhanced~\cite{Alford:2004zr}.  The neutrino emissivity should turn out to be significantly
suppressed relative to that in unpaired quark matter. The evaluation of the phase
space for direct URCA neutrino emission from the CubeX and 2Cube45z phases
will be a nontrivial calculation, given that 
thermally excited gapless quarks 
occur only on patches of the Fermi surfaces, separated by the (many)
pairing rings. (The direct URCA processes $u + e \rightarrow s + \nu$
and $s\rightarrow  u + e + \bar\nu$ require $s$, $u$ and $e$ to all be within $T$
of a place in momentum space where they are  gapless and at the same
time to have $\bf p_u + \bf p_e = \bf p_s$ to within $T$.  Here, $T\sim$~ keV is
very small compared to all the scales relevant to the description of
the crystalline phase itself.)

Beginning with
Ref.~\cite{Alford:2000ze}, one of the motivations for the study of crystalline
color superconducting quark matter has been the possibility that, if present
within the core of a compact star, it could provide a region within which
rotational vortices are pinned and hence a locus for the origin
of (some) pulsar glitches.  Or, the presence
of crystalline quark matter within neutron stars could be ruled out
if it predicts glitch phenomenology in qualitative disagreement with
that observed.  

There are two key microphysical properties of crystalline
quark matter that must be estimated before glitch phenomenology can
be addressed. The first is the pinning force.  Estimating this will require
analyzing how the CubeX and 2Cube45z respond when rotated. We expect
vortices to form, and expect the vortices to be pinned at the intersections
of the nodal planes at which condensates vanish.   Analyzing the vortices
in three-flavor crystalline phases will be nontrivial.  One complication
is that because
baryon number current can be carried by gradients
in the phase  of either the $\langle us \rangle$
crystalline condensate or the $\langle ud \rangle$ condensate or both,
and the most favorable vortex or vortices that form upon rotating the
CubeX and 2Cube45z phases will have to be determined.
Another complication arises because the vortex core size, $1/\Delta$, is only a factor
of three to four smaller than the lattice spacing $a$.  This means that
the vortices cannot be thought of as pinned by an unchanged crystal;
the vortices themselves will qualitatively deform the crystalline condensate in
their vicinity.  

The second microphysical quantity that is required is
the shear modulus of the crystal.  After all, if vortices are well-pinned but
the crystalline condensate can easily deform under shear stress, the vortices
will be able to move regardless of the pinning force.  Glitches occur if
vortices are pinned and immobile while the spinning pulsar's angular velocity 
slows over years,  with the glitch being triggered by the catastrophic unpinning
and motion of long-immobile vortices.  In order to immobilize vortices, and hence
make glitches a possibility, both the pinning force and the shear modulus must 
be sufficient.  The shear modulus can
be related to the coefficients in the low energy effective theory that
describes the phonon modes of the 
crystal~\cite{Casalbuoni:2001gt,Casalbuoni:2002pa,Casalbuoni:2002my}.  
This effective theory
has been analyzed, with its coefficients calculated, 
for the two-flavor crystalline color superconductor
with face-centered cubic symmetry~\cite{Casalbuoni:2002my}.
Extending this analysis to three-flavor crystalline color superconducting phases
with the 2Cube45z and CubeX crystal structures is a priority
for future work.  

Now that we have two well-motivated candidates
for the favored crystal structure of the  three-flavor crystalline color superconducting
phase of cold quark matter, favorable over a very wide range of intermediate
densities, the challenge becomes calculating the shear modulus and
the pinning force exerted on rotational vortices in these structures.
These are the prerequisites to determining whether observations of
pulsar glitches can be used to rule out (or in) the presence of quark matter
in the crystalline color superconducting phase within compact stars.

\begin{acknowledgments}

We thank K.~Fukushima, M.~Mannarelli and A.~Schmitt for useful discussions. 
We acknowledge the
hospitality of the Nuclear Theory Group at LBNL. This research was
supported in part by the Office of Nuclear Physics of the Office of
Science of the U.S.~Department of Energy under contract
\#DE-AC02-05CH11231 and cooperative research agreement
\#DF-FC02-94ER40818.

\end{acknowledgments}

\appendix

\section{Neutrality of solutions with $\Delta_2=\Delta_3$}
\label{stability}

In Section VI.A, we gave a general analysis of the free energy
$\Omega(\Delta_2,\Delta_3)$. We showed that if
we write  $(\Delta_2,\Delta_3)$ as $\sqrt{2}(\Delta_r \cos \theta,\Delta_r \sin \theta)$
the free energy takes the form (\ref{radial free energy}), and therefore 
has extrema only at $\theta=\pi/4$ (namely $\Delta_2=\Delta_3=\Delta_r$) or
$\theta=0,\pi/2$ (namely a two flavor crystalline
phase with only one $\Delta_I$ nonzero).  
As we have explained in Section II.D, in the strict Ginzburg-Landau limit
in which $\Delta_I/\delta\mu \rightarrow 0$ any solution $(\Delta_2,\Delta_3)$ is
neutral.  (The argument is that choosing $\mu_e=M_s^2/(4\mu)$ as in
neutral unpaired quark matter suffices since, unlike BCS
superconductivity, crystalline color superconductivity does not require
any modification of the unpaired Fermi momenta prior to pairing and since
in the Ginzburg-Landau limit the modifications to number densities due
to the pairing itself vanishes.)  In this Appendix, we take a small step
away from the strict Ginzburg-Landau limit.  We assume that $\Delta_r$
is small, but do not work in the limit in which it vanishes.  We then show
that the only solutions with 
$\mu_e=M_s^2/(4\mu)$ and, consequently,
$\delta\mu_2=\delta\mu_3=\delta\mu=M_s^2/(8\mu)$ which are electrically
neutral are those with $\Delta_2=\Delta_3=\Delta_r$.  The two-flavor
crystalline phases with only one $\Delta_I$ nonzero
are not neutral in three-flavor quark matter.

The result of this Appendix allows us to neglect solutions which have
only one $\Delta_I$ nonzero. This is fortunate, because there
are many two-flavor crystal structures for which
the sextic coefficient $\bar\gamma$ is negative, meaning
that to sextic order the Ginzburg-Landau potential $\Omega(\Delta_2,\Delta_3)$
often has runaway directions along the $\Delta_2$ and $\Delta_3$ axes~\cite{Bowers:2002xr}. 
Furthermore, if the coefficient multiplying $\sin^2\theta$ in (\ref{radial free energy})
is negative, for example if $\bar\beta$ and $\bar\gamma$ are both negative
while $\bar\beta_{32}$ and $\bar\gamma_{322}$ are both positive as is the 
case for both the CubeX and the 2Cube45z crystal structures on which
we focus, then the extremum of $\Omega(\Delta_2,\Delta_3)$
that we find with $\Delta_2=\Delta_3$ appears to be a local maximum with
respect to variation of $\theta$ away from $\pi/4$ while keeping $\mu_e$ fixed. 
We show in this Appendix that upon fixing $\mu_e=M_s^2/(4\mu)$ any solution with
$\Delta_2\neq \Delta_3$ is not neutral. For this reason, all these complications
can be neglected, and we are correct to focus only on solutions with $\Delta_2=\Delta_3$. 

The more formal way to proceed would be to define
an $\Omega_{\rm neutral}(\Delta_2,\Delta_3)$, obtained by varying $\mu_e$
(and $\mu_3$ and $\mu_8$ too) at a given value of the $\Delta$'s in order to 
obtain neutrality, and then finding $\Delta_2$ and $\Delta_3$ that minimize
$\Omega_{\rm neutral}(\Delta_2,\Delta_3)$.  We have done a partial version of this
investigation in a few 
cases and have found that, as expected, $\Omega_{\rm neutral}$ does have a minimum
with $\mu_e$ very close to $M_s^2/(4\mu)$ and $\Delta_2$ very close to $\Delta_3$.
A full exploration in this vein requires evaluating the Ginzburg-Landau coefficients
without assuming $\delta\mu_2\approx \delta\mu_3$ and, more challenging, requires
reformulating our analysis to include nonzero $\mu_3$ and $\mu_8$.  We have
not attempted the latter, and it is in this sense that our preliminary investigation
referred to above was ``partial''.  We leave this to future work, and turn now
to the promised derivation of the neutrality of solutions with $\Delta_2=\Delta_3$
and $\mu_e=M_s^2/(4\mu)$.

We shall only consider crystal structures for which 
$\{\hat{\bf q}_2\}$ and $\{\hat{\bf q}_3\}$ are exchange symmetric, as this is the symmetry
that
allows the free energy to have extrema along the
line $\Delta_2=\Delta_3$. 
(Recall that by exchange symmetric we mean that there is a sequence of rigid rotations
and reflections which when applied to
all the vectors in $\setq{2}{}$ and $\setq{3}{}$ together
has the effect of
exchanging $\{\hat{\bf q}_2\}$ and $\{\hat{\bf q}_2\}$.) 
Because we wish to evaluate $\partial\Omega/\partial\mu_e$ at
$\mu_e=M_s^2/(4\mu)$, we must restore
$\mu_e$ to our expression for the free energy $\Omega$, rather than
setting it to $M_s^2/(4\mu)$ from the beginning.
Recall from (\ref{CondensationEnergy}) that $\Omega_{\rm crystalline}$ is the sum of
the free energy for unpaired quark matter, which we know satisfies
$\partial\Omega_{\rm unpaired}/\partial\mu_e=0$ at $\mu_e=M_s^2/(4\mu)$,
and $\Omega(\Delta_2,\Delta_3)$. Upon restoring the $\mu_e$-dependence,
the latter is given by
\begin{equation}
\begin{split}
\Omega(&\mu_e,\Delta_2,\Delta_3) =\\
 &\phantom{}\frac{2{\mu}^2}{\pi^2}\Biggl[
    P \alpha(\delta\mu_{2}) \Delta_2^2
    +P \alpha(\delta\mu_{3})\Delta_3^2 \\
   &\phantom{+}+\ha\left(\frac{1}{\delta\mu_{2}^2}\bar{\beta}_2\Delta_2^4
   +\frac{1}{\delta\mu_{3}^2}\bar{\beta}_3\Delta_3^4
    +\frac{1}{\delta\mu_{3}\delta\mu_{2}}\bar{\beta}_{32}\Delta_2^2\Delta_3^2\right)\\
    &+\frac{1}{3}\Biggl(\frac{1}{\delta\mu_{2}^4}\bar{\gamma}_2\Delta_2^6
    +\frac{1}{\delta\mu_{3}^4}\bar{\gamma}_3\Delta_3^6
    +{\gamma}_{233}(\delta\mu_2,\delta\mu_3)\Delta_2^2\Delta_3^4\\
    &\phantom{+}+{\gamma}_{322}(\delta\mu_2,\delta\mu_3)\Delta_3^2\Delta_2^4
    \Biggr)\Biggr]\label{congruent free energy}\;,
\end{split}
\end{equation}
where $\delta\mu_2$ and $\delta\mu_3$ can no longer be taken to be equal,
as they are given by
\begin{eqnarray}
\dmu_{3}&=&\frac{\mu_e}{2}\nonumber\\
\dmu_{2}&=&\frac{M_s^2}{4\mu}-\frac{\mu_e}{2}\ ,
\end{eqnarray}
which in particular means that
\be
\frac{\partial\delta\mu_3}{\partial\mu_e}=-
\frac{\partial\delta\mu_2}{\partial\mu_e}= \frac{1}{2}\ .
\label{ddeltamudmue}
\ee
Because $\{\hat{\bf q}_2\}$ and $\{\hat{\bf q}_3\}$ are exchange symmetric, $\bar\beta_2=\bar\beta_3=\bar\beta$
and $\bar\gamma_2=\bar\gamma_3=\bar\gamma$.
Because $\delta\mu_2\neq\delta\mu_3$, however, the coefficients $\gamma_{322}$
and $\gamma_{233}$ are not equal and, furthermore, their $(\delta\mu_2,\delta\mu_3)$-dependence
cannot be factored out as in (\ref{Beta32BarDefn}) or (\ref{GammaBarDefn}).
The coefficient $\gamma_{322}$ depends on $\delta\mu_2$ and $\delta\mu_3$
through its dependence
on ${\cal K}_{udusus}$: $\gamma_{322}=(3/2)\sum 
{\cal K}_{udusus}(\q{3}{b},\q{3}{b},\q{2}{d},\q{2}{e},\q{2}{f},\q{2}{a})$.
${\cal K}_{udusus}$ is given in (\ref{K322simplified ds dn}).
Note that its dependence on $\delta\mu_2$ and $\delta\mu_3$
comes via  $\q{2}{}=\eta\,\delta\mu_2\,\hat{\bf q}_2$ and
$\q{3}{}=\eta\,\delta\mu_3\,\hat{\bf q}_3$ in addition to the
explicit dependence visible in (\ref{K322simplified ds dn}). 
Similarly $\gamma_{233}=(3/2)\sum 
{\cal K}_{usudud}(\q{2}{b},\q{2}{b},\q{3}{d},\q{3}{e},\q{3}{f},\q{3}{a})$ where
${\cal K}_{usudud}$ has the same form as (\ref{K322simplified ds dn})
except that $\delta\mu_2$ and $\delta\mu_3$ are interchanged.
Using the definitions (\ref{Beta32BarDefn}) and (\ref{GammaBarDefn}), one can confirm that 
(\ref{congruent free energy}) reduces to 
(\ref{congruent equal free energy}) if we take $\delta\mu_2=\delta\mu_3$
and hence $\gamma_{322}=\gamma_{233}$.  

We now differentiate $\Omega$ given in (\ref{congruent free energy}) with respect to 
$\mu_e$, noting (\ref{ddeltamudmue}), obtaining
\begin{widetext}
\begin{equation}
\begin{split}
\frac{\partial\Omega}{\partial\mu_e}
&=
\frac{\partial\Omega}{\partial\Delta_2}\frac{d\Delta_2}{d\mu_e}
 +\frac{\partial\Omega}{\partial\Delta_3}\frac{d\Delta_3}{d\mu_e}
 +\frac{{\mu}^2}{\pi^2}\Biggl[ P\frac{d\alpha(\delta\mu_{3})}{d\mu_{3}}
    \Delta_3^2
    -P\frac{d\alpha(\delta\mu_{2})}{d\mu_{2}}\Delta_2^2
    +
    \frac{\bar\beta}{\delta\mu_{2}^3}\Delta_2^4
    -\frac{\bar\beta}{\delta\mu_{3}^3}\Delta_3^4\\
&    +\left(
	 \frac{1}{2\delta\mu_{2}^2\delta\mu_{3}}
	 -\frac{1}{2\delta\mu_{2}\delta\mu_{3}^2}\right)\bar\beta_{32}
     \Delta_2^2\Delta_3^2  
     +
	\frac{4\bar\gamma}{3\delta\mu_{2}^5}\Delta_2^6
    -\frac{4\bar\gamma}{3\delta\mu_{3}^5}\Delta_3^6 \\
&    +\frac{1}{3}\left(\frac{\partial{\gamma}_{233}(\delta\mu_2,\delta\mu_3)}{\partial\delta\mu_3}
    -\frac{\partial{\gamma}_{233}(\delta\mu_2,\delta\mu_3)}{\partial\delta\mu_2}
\right)\Delta_2^2\Delta_3^4
    +\frac{1}{3}\left(\frac{\partial{\gamma}_{322}(\delta\mu_2,\delta\mu_3)}{\partial\delta\mu_3}
    -\frac{\partial{\gamma}_{322}(\delta\mu_2,\delta\mu_3)}{\partial\delta\mu_2}
\right)\Delta_2^4\Delta_3^2
	\Biggr]\ .
\label{symmetric neutrality}
\end{split}
\end{equation}
We shall only evaluate $\partial\Omega/\partial\mu_e$ at values of $\Delta_2$
and $\Delta_3$ which are solutions to the gap equations $\partial\Omega/\partial\Delta_2=0$
and $\partial\Omega/\partial\Delta_3=0$, meaning that the first two terms in
(\ref{symmetric neutrality}) vanish. Furthermore, we shall only
evaluate $\partial\Omega/\partial\mu_e$ at 
$\mu_e=M_s^2/(4\mu)$, where
$\dmu_{2}=\dmu_{3}=\dmu$, and at solutions for which
$\Delta_2=\Delta_3$. Under these circumstances, the
terms involving $\alpha$, $\bar{\beta}_2$, $\bar{\beta}_{32}$ and
$\bar{\gamma}_2$ vanish and (\ref{symmetric neutrality}) 
becomes
\begin{equation}
\frac{\partial\Omega}{\partial\mu_e}
\Biggr|_{\mu_e=\x{4},\ \Delta_2=\Delta_3=\Delta_{\rm min}}
= \frac{\mu^2}{3\pi^2}\left[
	  \frac{\partial{\gamma}_{233}
	  }{\partial\delta\mu_3}
	  -\frac{\partial{\gamma}_{233}
	  }{\partial\delta\mu_2}
	  +\frac{\partial{\gamma}_{322}
	  }{\partial\delta\mu_3}
    -\frac{\partial{\gamma}_{322}
    }{\partial\delta\mu_2}
	\right]\Delta_{\rm min}^6\Biggr|_{\delta\mu_2=\delta\mu_3=\delta\mu}
\label{symmetric neutrality2}
\end{equation}
\end{widetext}
We argue that this vanishes as follows. 
Consider a particular term that contributes to
$\partial{\gamma_{322}}/{\partial\delta\mu_2}$, $\partial{{\cal K}_{udusus}
(\q{3}{b},\q{3}{b},\q{2}{d},\q{2}{e},\q{2}{f},\q{2}{a})}/{\partial\delta\mu_2}$. 
This
is a complicated integral of a function which depends on the unit momentum
vectors 
$(\hat{\bf q}_{3}^{b},\hat{\bf q}_{2}^{d},\hat{\bf q}_{2}^{e},\hat{\bf q}_{2}^{f},
\hat{\bf q}_{2}^{a})$ 
and on $\delta\mu_2$ and $\delta\mu_3$. 
From rotational invariance,  we know that
the
value of the integral can depend on the relative orientation of the unit 
momentum vectors  and on $\delta\mu_2$ and $\delta\mu_3$ but
must be independent of common rotations of all the unit vectors. 
Now, all
the crystal structures that we consider are exchange symmetric, meaning
that for every
quintuple of
unit momentum vectors,
$(\hat{\bf q}_{3}^{b},\hat{\bf q}_{2}^{d},\hat{\bf q}_{2}^{e},\hat{\bf q}_{2}^{f},
\hat{\bf q}_{2}^{a})$ with the 
first chosen from $\setq{3}{}$ and the last four
chosen from $\setq{2}{}$
there exists a quintuple 
$(\hat{\bf q}_{2}^{b},\hat{\bf q}_{3}^{d},\hat{\bf q}_{3}^{e},\hat{\bf q}_{3}^{f},
\hat{\bf q}_{3}^{a})$ with the 
first chosen from $\setq{2}{}$ and the last four
chosen from $\setq{3}{}$ such that the unit vectors in each of these two quintuples
have the same relative orientation among themselves.
This means that for every term
$\partial{{\cal K}_{udusus}
(\q{3}{b},\q{3}{b},\q{2}{d},\q{2}{e},\q{2}{f},\q{2}{a})}/{\partial\delta\mu_2}$. 
occurring in $\partial{\gamma_{322}}/{\partial\delta\mu_2}$, 
there is a corresponding term
$\partial{{\cal K}_{usudud} 
(\q{2}{b},\q{2}{b},\q{3}{d},\q{3}{e},\q{3}{f},\q{3}{a})}/{\partial\delta\mu_3}$
occurring in
$\partial{\gamma_{233}}/{\partial\delta\mu_3}$ such that
$\partial{{\cal K}_{usudud} 
(\q{2}{b},\q{2}{b},\q{3}{d},\q{3}{e},\q{3}{f},\q{3}{a})}/{\partial\delta\mu_3}$ is
related to $\partial{{\cal K}_{udusus}
(\q{3}{b},\q{3}{b},\q{2}{d},\q{2}{e},\q{2}{f},\q{2}{a})}/{\partial\delta\mu_2}$ by
the interchange of $\delta\mu_2$ and $\delta\mu_3$. Consequently,
for $\delta\mu_2=\delta\mu_3$ the two contributions cancel pair by pair when we 
evaluate
$\partial{\gamma_{322}}/{\partial\delta\mu_2}
-\partial{\gamma_{233}}/{\partial\delta\mu_3}$ or  
$\partial{\gamma_{322}}/{\partial\delta\mu_3}
-\partial{\gamma_{233}}/{\partial\delta\mu_2}$. 
In this way, the right hand side of
(\ref{symmetric neutrality2}) vanishes, as we set out to show. 
We conclude that solutions to the gap equations with $\Delta_2=\Delta_3$
and $\mu_e=M_s^2/(4\mu)$ meaning $\delta\mu_2=\delta\mu_3$ 
are neutral.  

It is easy to see that the cancellations required
in the proof of neutrality do not occur for solutions with $\Delta_2\neq\Delta_3$.
For example,
following a derivation analogous to that above,
we find that 
a solution with
$\Delta_2=0$ and
only $\Delta_3$ nonzero is neutral with $\mu_e=M_s^2/(4\mu)$
only if
\begin{equation}
    P\frac{\partial\alpha(\delta\mu_{3})}{\partial\delta\mu_3}\Delta_3^2
    -\frac{\bar\beta_3}{\delta\mu_{3}^3}\Delta_3^4
    -\frac{4\bar\gamma_3}{3\delta\mu_{3}^5}\Delta_3^6 = 0 \ ,
\end{equation}
a condition which has no reason to be satisfied.  The study of solutions
with $\Delta_2\neq\Delta_3$ therefore requires that they be constructed
from the beginning with $\delta\mu_2\neq\delta\mu_3$ and with
$\mu_e$ fixed by the neutrality condition.  We leave this to future work,
focussing in Section VI on solutions with $\Delta_2=\Delta_3$ which,
we have proved here, are neutral.

\section{Translating $\langle us\rangle$ relative to $\langle ud \rangle$ does not avoid repulsion}

We have seen in Section VI  that crystal structures in which a vector
from $\setq{2}{}$ and a vector from $\setq{3}{}$ make a 180$^\circ$ angle
are strongly disfavored, with infinite quartic
and sextic Ginzburg-Landau coefficients $\bar\beta_{32}$ and $\bar\gamma_{322}$.
Suppose we consider a structure like that 
in which $\setq{2}{}$ and $\setq{3}{}$ are coincident
cubes, a disastrous choice.  The way that we have improved upon this
disastrous choice in Section VI.D is to rotate one cube relative to the other.
Indeed, if we choose a 45$^\circ$ rotation about an axis perpendicular to
a face of the cube, we obtain the 2Cube45z structure which is 
one of the two crystal structures that
we find to be most favorable.  In this Appendix, we ask whether
we can instead avoid the infinite free energy cost of antipodal pairs
by translating the $\langle ud \rangle$ condensate relative to the $\langle us \rangle$
condensate in position space, rather than  rotating it.  We find that the
answer is no, and furthermore show that the Ginzburg-Landau free energy
$\Omega$ that we have evaluated does not change if 
the $\langle ud \rangle$ condensate is translated relative to the $\langle us \rangle$
condensate.

Corresponding to each $\setq{I}{}$ in momentum space we get a function 
$\Delta_I(\rr)$ in position space which varies as $\Delta_I(\rr)\sim
\sum_{\q{I}{a}}e^{2i\q{I}{a}\cdot\rr}$.
To analyze the effects of translating $\Delta_2(\rr)$ relative to $\Delta_3(\rr)$,
it is helpful to restore the notation of (\ref{gl 4})
with $\Delta(\q{I}{a})$ representing the gap
parameter corresponding to the momentum component $\q{I}{a}$. 
$\Delta_2(\rr)$ or $\Delta_3(\rr)$ can then be written as
\begin{equation}
\Delta_I(\rr)=\sum_{\q{I}{a}}\Delta(\q{I}{a})e^{2i\q{I}{a}\cdot\rr}\;.
\label{EqB1}
\end{equation}
Translating $\Delta_2(\rr)$ in the $\hat{\bf n}$ direction by a distance $s$
corresponds  to the transformation
$\Delta_2(\rr)\rightarrow\Delta_2(\rr-s\hat{\bf n})$
which multiplies each $\Delta(\q{2}{a})$ in the sum in (\ref{EqB1})
by a different phase factor 
$\exp[-2is\q{2}{a}\cdot \hat{\bf n}]$. This is not just
an (irrelevant) overall phase multiplying $\Delta_2(\rr)$ because it
depends on the momentum component. The gap equation for the $\Delta_2$
components, as in (\ref{gl 4}),  is now given by
\begin{widetext}
\begin{equation}
\begin{split}
\Delta^*(\q{2}{a})e^{2is\q{2}{a}\cdot \hat{\bf n}}
 &=-\frac{2{\mu}^2\lambda}{\pi^2}  \Biggl[
 \Delta^*(\q{2}{a})e^{2is\q{2}{a}\cdot\hat{\bf n}}
 \Pi_{31}(\q{2}{a},\q{2}{a})\\
&\!\!\!\!\!\!\!\!\!\!\!\!\!\!\!\!\!\!\!\!\!\!\!
 +\sum_{\q{2}{b}\q{2}{c}\q{2}{d}}
 \Delta^*(\q{2}{b})\Delta(\q{2}{c})\Delta^*(\q{2}{d})
 e^{2is(\q{2}{b}-\q{2}{c}+\q{2}{d})\cdot\hat{\bf n}}
 \,J_{3131}(\q{2}{b},\q{2}{c},\q{2}{d},\q{2}{a})\delta_{\q{2}{b}-\q{2}{c}+\q{2}{d}-\q{2}{a}}\\
&\!\!\!\!\!\!\!\!\!\!\!\!\!\!\!\!\!\!\!\!\!\!\!
 +\ha\sum_{\q{3}{b}\q{3}{c}\q{2}{d}}
 \Delta^*(\q{3}{b})\Delta(\q{3}{c})\Delta^*(\q{2}{d})
 e^{2is\q{2}{d}\cdot\hat{\bf n}}
 \,J_{1213}(\q{3}{b},\q{3}{c},\q{2}{d},\q{2}{a})\delta_{\q{3}{b}-\q{3}{c}+\q{2}{d}-\q{2}{a}}\Biggr]\;.
\end{split}
\end{equation}
\end{widetext}
where we have worked only to cubic order.
Using $\q{2}{b}-\q{2}{c}+\q{2}{d}=\q{2}{a}$ we conclude that the phase factor in
front of the $J_{3131}(\q{2}{b},\q{2}{c},\q{2}{d},\q{2}{a})$ term is simply 
$\exp[2is(\q{2}{a})\cdot\hat{\bf n}]$. In addition, we saw that for
${\q{3}{b}-\q{3}{c}+\q{2}{d}=\q{2}{a}}$ to hold we need to have
${\q{3}{b}=\q{3}{c}}$ and ${\q{2}{d}=\q{2}{a}}$.  This makes
the phase factor in front of $J_{1213}(\q{3}{b},\q{3}{c},\q{2}{d},\q{2}{a})$
also $\exp[2is(\q{2}{a})\cdot\hat{\bf n}]$.  We conclude that (up to cubic order)
the gap equation for 
each $\Delta(\q{2}{a})$ simply picks
up an overall phase.  The same is true for the gap equation for each $\Delta(\q{3}{a})$.
We therefore conclude that the free energy is unchanged up to quartic order
when $\Delta_2(\rr)$ is translated relative to $\Delta_3(\rr)$. This guarantees
that such a translation cannot alleviate the large $\bar\beta_{32}$ arising
from antipodal (or near antipodal) pairs of momenta occurring in $\setq{2}{}$
and $\setq{3}{}$. This argument can easily be extended to include
the sextic terms in the free energy; they too are unchanged when
$\Delta_2(\rr)$ is translated relative to $\Delta_3(\rr)$.

\end{document}